\documentclass[final,3p,times]{elsarticle}

\usepackage{amssymb}
\biboptions{comma,sort&compress}
\usepackage{amsmath}
\usepackage{xcolor}
\usepackage{hyperref}       
\usepackage{url}            
\usepackage{booktabs}       
\usepackage{nicefrac}       
\usepackage{microtype}      
\usepackage{comment}
\usepackage{caption}
\usepackage{subcaption}
\usepackage{multicol}
\usepackage{graphicx}
\usepackage{tabularx}
\usepackage{latexsym}
\usepackage{amstext,amsfonts,amssymb,amsmath}
\usepackage{todonotes}
\usepackage[version=3]{mhchem}
\usepackage{epstopdf}
\AppendGraphicsExtensions{.tif}
\usepackage{color}
\usepackage{bm}
\usepackage{algorithm}
\usepackage{algpseudocode}
\usepackage{multirow}
\usepackage{tikz}
\usepackage{breqn}
\usepackage{etoolbox}
\usepackage[thinc]{esdiff}

\usepackage[normalem]{ulem}

\usepackage{tikz}
\usepackage{breqn}
\usepackage{etoolbox}

\newrobustcmd*{\myVtriangle}[2]{\tikz{\filldraw[draw=#1,fill=#2] (0cm,0.2cm) --
(0.2cm,0.2cm) -- (0.1cm,0cm) -- (0cm,0.2cm);}}

\newrobustcmd*{\mythickVtriangle}[2]{\tikz{\filldraw[line width=0.3mm,draw=#1,fill=#2] (0cm,0.2cm) --
(0.2cm,0.2cm) -- (0.1cm,0cm) -- (0cm,0.2cm);}}

\newrobustcmd*{\mythickErrorVtriangle}[2]{\tikz{\filldraw[line width=0.3mm,draw=#1,fill=#2] (-0.05cm,0.05cm) --
(0.05cm,0.05cm) -- (0cm,-0.05cm) -- (-0.05cm,0.05cm);  \draw[draw=#1] (0.0cm, -0.12cm) -- (0.0cm, 0.12cm) ; \draw[draw=#1] (-0.06cm, 0.12cm) -- (0.06cm, 0.12cm); \draw[draw=#1] (-0.06cm, -0.12cm) -- (0.06cm, -0.12cm)    }}

\newrobustcmd*{\mytriangle}[2]{\tikz{\filldraw[draw=#1,fill=#2] (0.0cm,0.0cm) --
(0.2cm,0cm) -- (0.1cm,0.2cm) -- (0cm,0cm);}}

\newrobustcmd*{\mysquare}[2]{\tikz{\draw[draw=#1,fill=#2] (0cm,0cm)
rectangle (0.2cm,0.2cm)}}

\newrobustcmd*{\mythicktriangle}[2]{\tikz{\filldraw[line width=0.3mm,draw=#1,fill=#2] (0.0cm,0cm) --
(0.2cm,0cm) -- (0.1cm,0.2cm) -- (0.0cm,0cm);}}

\newrobustcmd*{\mythicksquare}[2]{\tikz{\draw[line width=0.3mm,draw=#1,fill=#2] (0cm,0cm)
rectangle (0.2cm,0.2cm)}}

\newrobustcmd*{\mybarredtriangle}[2]{\tikz{\draw[draw=#1,fill=#2] (0,0) --
(0.2cm,0) -- (0.1cm,0.2cm) -- (0cm,0cm); \draw[draw=#1] (-0.1cm, 0.07cm) -- (0.3cm, 0.07cm)}}

\newrobustcmd*{\mythickbarredtriangle}[2]{\tikz{\draw[line width=0.3mm,draw=#1,fill=#2] (0,0) --
(0.2cm,0) -- (0.1cm,0.2cm) -- (0cm,0cm); \draw[draw=#1] (-0.1cm, 0.07cm) -- (0.3cm, 0.07cm)}}

\newrobustcmd*{\mybarredsquare}[2]{\tikz{\draw[draw=#1,fill=#2] (0,0)
rectangle (0.2cm,0.2cm); \draw[draw=#1] (-0.1cm, 0.1cm) -- (0.3cm, 0.1cm)}}

\newrobustcmd*{\mythickbarredsquare}[2]{\tikz{\draw[line width=0.3mm,draw=#1,fill=#2] (0,0)
rectangle (0.2cm,0.2cm); \draw[draw=#1] (-0.1cm, 0.1cm) -- (0.3cm, 0.1cm)}}

\newrobustcmd*{\mybarredcircle}[2]{\tikz{\draw[draw=#1,fill=#2] (0,0)
circle (0.1cm); \draw[draw=#1] (-0.2cm, 0.0cm) -- (0.2cm, 0.0cm)}}

\newrobustcmd*{\mythickbarredcircle}[2]{\tikz{\draw[line width=0.3mm,draw=#1,fill=#2] (0,0)
circle (0.1cm); \draw[draw=#1] (-0.2cm, 0.0cm) -- (0.2cm, 0.0cm)}}

\newrobustcmd*{\mythickErrorcircle}[2]{\tikz{\draw[line width=0.3mm,draw=#1,fill=#2] (0,0)
circle (0.06cm); \draw[draw=#1] (0.0cm, -0.12cm) -- (0.0cm, 0.12cm) ;   \draw[draw=#1] (-0.06cm, 0.12cm) -- (0.06cm, 0.12cm); \draw[draw=#1] (-0.06cm, -0.12cm) -- (0.06cm, -0.12cm)    }}

\newrobustcmd*{\mydashedline}[1]{\tikz{\draw[draw=#1] (-0.2cm, 0.2cm) -- (-0.1cm, 0.2cm); \draw[draw=#1] (-0.0cm, 0.2cm) -- (0.1cm, 0.2cm)}}

\newrobustcmd*{\mythickcross}[1]{\tikz{\draw[line width=0.3mm,draw=#1] (0,0) --
(0.2cm,0); \draw[line width=0.3mm,draw=#1] (0.1cm,-0.1cm) -- (0.1cm,0.1cm);}}

\newrobustcmd*{\mybarredcross}[1]{\tikz{\draw[line width=0.3mm,draw=#1] (0,0) --
(0.2cm,0); \draw[line width=0.3mm,draw=#1] (0.1cm,-0.1cm) -- (0.1cm,0.1cm); \draw[draw=#1] (-0.1cm,0) -- (0.3cm,0);}}

\newrobustcmd*{\myline}[1]{\tikz{\draw[draw=#1] (-0.15cm, 0.1cm) -- (0.15cm, 0.1cm);\draw[line width=0.3mm,draw=#1] (-0.0cm, 0.0cm);}}

\newrobustcmd*{\mythickline}[1]{\tikz{\draw[line width=0.3mm,draw=#1] (-0.15cm, 0.1cm) -- (0.15cm, 0.1cm);\draw[line width=0.3mm,draw=#1] (-0.0cm, 0.0cm);}}

\newrobustcmd*{\mythickdashedline}[1]{\tikz{\draw[line width=0.3mm,draw=#1] (-0.2, 0.1cm) -- (-0.1cm, 0.1cm); \draw[line width=0.3mm,draw=#1] (-0.0cm, 0.1cm) -- (0.1cm, 0.1cm); \draw[line width=0.3mm,draw=#1] (-0.0cm, 0.0cm);}}

\newrobustcmd*{\mythickdasheddottedline}[1]{\tikz{\draw[line width=0.3mm,draw=#1] (-0.22, 0.1cm) -- (-0.13cm, 0.1cm); \draw[line width=0.3mm,draw=#1] (-0.085cm, 0.1cm) -- (-0.055cm, 0.1cm); \draw[line width=0.3mm,draw=#1] (-0.01cm, 0.1cm) -- (0.08cm, 0.1cm); \draw[line width=0.3mm,draw=#1] (-0.0cm, 0.0cm);}}

\newrobustcmd*{\mycircle}[2]{\tikz{\draw[draw=#1,fill=#2] (0,0)
circle (0.1cm);}}

\newrobustcmd*{\mythickcircle}[2]{\tikz{\draw[line width=0.3mm,draw=#1,fill=#2] (0,0)
circle (0.1cm);}}

\newrobustcmd*{\mydot}[1]{\tikz{\draw[line width=0.3mm,draw=#1] (0,0)
circle (0.025cm);}}
\newcommand{\revone}[1]{\textcolor{black}{{#1}}}
\newcommand{\revtwo}[1]{\textcolor{black}{{#1}}}
\newcommand{\revtwobarred}[1]{{}}
\newcommand{\ftexttt}[1]{\texttt{\frenchspacing#1}}

\journal{TBD}

\usepackage{glossaries}
\newacronym{amr}{AMR}{adaptive mesh refinement}
\newacronym{pmf}{PMF}{pre-mixed flame}
\newacronym{cfd}{CFD}{computational fluid dynamics}
\newacronym{ci}{CI}{continuous integration}
\newacronym{gpu}{GPU}{graphics processing unit}
\newacronym{cpu}{CPU}{central processing unit}
\newacronym{hpc}{HPC}{high performance computing}
\newacronym{ppm}{PPM}{piecewise parabolic method}
\newacronym{mol}{MOL}{method of lines}
\newacronym{dof}{DoF}{degrees of freedom}
\newacronym{les}{LES}{large eddy simulation}
\newacronym{dns}{DNS}{direct numerical simulations}
\newacronym{mpi}{MPI}{Message Passing Interface}
\newacronym{ode}{ODE}{ordinary differential equation}
\newacronym{rk}{RK}{Runge-Kutta}
\newacronym{egr}{EGR}{exhaust gas recirculation}
\newacronym{mms}{MMS}{Method of Manufactured Solutions}
\newacronym{ornl}{ORNL}{Oak Ridge National Laboratory}
\newacronym{nrel}{NREL}{National Renewable Energy Laboratory}
\newacronym{anl}{ANL}{Argonne National Laboratory}
\newacronym{gcc}{GCC}{GNU Compiler Collection}
\newacronym{gmg}{GMG}{Geometric Multigrid}
\newacronym{ceptr}{CEPTR}{Chemistry Evaluation for Pele Through Recasting}
\newacronym{qssa}{QSSA}{Quasi-Steady State Approximation}
\newacronym{sdc}{SDC}{spectral deferred correction}
\newacronym[longplural={embedded boundaries}]{eb}{EB}{embedded boundary}
\newacronym{rcci}{RCCI}{reactivity-controlled compression ignition}
\newacronym{lrf}{LRF}{low-reactivity fuel}
\newacronym{tdc}{TDC}{top-dead-center}
\newacronym{ecp}{ECP}{Exascale Computing Project}
\newacronym{uvm}{UVM}{Unified Virtual Memory}
\newacronym{cfl}{CFL}{Courant-Friedrichs-Lewis}
\newacronym{hmom}{HMOM}{Hybrid Method of Moments}
\makeglossaries
\glsdisablehyper

\begin{document}

\begin{frontmatter}

\title{Symbolic construction of the chemical Jacobian of quasi-steady state (QSS) chemistries for Exascale computing platforms}

\author[NREL_CompSci]{Malik Hassanaly}
\ead{malik.hassanaly@nrel.gov}
\author[NREL_CompSci]{Nicholas T. Wimer}
\author[LBNL]{Anne Felden}
\author[NREL_CompSci]{Lucas Esclapez}
\author[FSU]{Julia Ream}
\author[NREL_CompSci]{Marc T. Henry de Frahan}
\author[NREL_CompSci]{Jon Rood}
\author[NREL_CompSci]{Marc Day}

\address[NREL_CompSci]{Computational Science Center, National Renewable Energy Laboratory (NREL), Golden, CO 80401}
\address[LBNL]{Process and Energy Department, Delft University of Technology, 2628 CB Delft, The Netherlands}
\address[FSU]{Department of Mathematics, Florida State University, Tallahassee, FL 32306}

\cortext[cor1]{Corresponding author:}

\begin{abstract}

The Quasi-Steady State Approximation (QSSA) can be an effective tool for reducing the size and stiffness of chemical mechanisms for implementation in computational reacting flow solvers.
\revtwo{However, for many applications, the resulting model still requires implicit methods for efficient time integration.}  In this paper, we outline an approach to formulating the QSSA reduction that is coupled with a strategy to generate C++ source code to evaluate the net species production rates, and the chemical Jacobian. The code-generation component employs a symbolic approach enabling a simple and effective strategy to analytically compute the chemical Jacobian. For computational tractability, the symbolic approach needs to be paired with common subexpression elimination which can negatively affect memory usage. Several solutions are outlined and successfully tested on a 3D multipulse ignition problem, thus allowing portable application across chemical model sizes and GPU capabilities. The implementation of the proposed method is available at \hyperlink{https://github.com/AMReX-Combustion/PelePhysics}{https://github.com/AMReX-Combustion/PelePhysics} under an open-source license.  
\end{abstract}


\end{frontmatter}


\section*{Novelty and Significance}
A symbolic method is proposed to write analytical chemical Jacobians. The benefit of the symbolic method is that it is easy to implement and flexible to any elementary reaction type. Its benefit is shown in the context of QSS-reduced chemistries: there, constructing an analytical chemical Jacobian is complex since one must include the effect of traditional elementary reactions and algebraic closure for the QSS species. To the authors' knowledge, there is no open-source package available to construct analytical Jacobians of QSS-reduced chemistries. We expect this work to facilitate the use of analytical Jacobians in arbitrarily complex chemical mechanisms. The proposed method was integrated into an open-source suite of reacting flow solvers \hyperlink{https://github.com/AMReX-Combustion/PelePhysics}{https://github.com/AMReX-Combustion/PelePhysics} to facilitate its dissemination. 

\section*{CRediT Authorship Contribution Statement}
MH: Methodology, Investigation, Software, Writing - original draft. NTW: Methodology, Investigation, Software, Visualization, Writing - review \& editing. AF: Methodology, Software, Writing - review \& editing. LE: Software, Writing - review \& editing. JR: Methodology, Software, Writing - review \& editing. MTHdF: Software, Writing - review \& editing. JRo: Software, Writing - review \& editing. MD: Funding acquisition, Conceptualization, Software, Writing - review \& editing.

\section{Introduction}

Exascale simulations of turbulent reacting flows have recently become possible~\cite{wimer2023visualizations,alexander2020exascale,malaya2023} due to a sustained investment in the co-design of modern high performance GPU-based computing architectures and performance-portable simulation software \cite{Esclapez2023, henrydefrahan2022, HenrydeFrahan2024, treichler2017s3d, mira2023hpc}. As simulation capabilities grow, so does the complexity of the research questions that can be addressed computationally. It is now possible to consider detailed interactions between turbulence and chemistry in the context of multicomponent and alternative fuels ~\cite{Pignatelli2023,felden2018including}, novel ignition strategies \cite{Chung2023,tang2021probabilistic}, and pollutant formation~\cite{Grader2023,Jaravel2017,mathieu2015experimental,attili2014formation}. 
Enabling these detailed investigations, the chemical kinetic models employed have also grown in size and complexity, exhibiting an increasingly broad range of time scales and ultimately becoming extremely ``stiff'' relative to the turbulent flow dynamics. As discussed in Ref.~\cite{balos2024}, operator-splitting has proven to be an effective framework for coupling stiff chemical kinetics to flow solvers, but even when using the most recent chemistry integration approaches of this type (e.g., \cite{Shi2012,Kodavasal2016,wu2019pareto,Mao2023}), reacting flow simulations employing a finite rate chemistry description can spend half or more of the computing resources required for the full simulation in the chemistry component of the solve alone~\cite{Lu2009}. \revone{To mitigate the cost of the chemistry integration, several approaches have been developed. From a computational standpoint, it was shown that the integration of the chemical system can be greatly accelerated through preconditioning approaches \cite{mcnenly2015faster,walker2023generalized}. In addition, the matrices used to represent the chemical systems can be strategically sparsified thereby allowing to leverage efficient sparse-linear algebra methods \cite{curtis2018using}. In addition, the chemical mechanisms integrated can also be simplified \cite{turanyi2014analysis}. Among the latter strategies, reducing the size of chemical mechanisms based on a-priori knowledge remains a key strategy to mitigate the computational burden of reacting flow simulations \cite{lu2006linear}.}

Many chemistry reduction techniques have been developed that identify and retain key kinetic information exhibited in an arbitrary set of idealized combustion scenarios (such as strain-free flame propagation speeds, extinction strain rates, etc.), that are asserted to be relevant to the model's intended application (e.g., see Table 1 in ~\cite{Felden2019}). The first step usually results in a problem-specific ``skeletal'' mechanism with a smaller number of chemical species and reactions, and one that typically also exhibits reduced stiffness~\cite{LU2008,PEPIOTDESJARDINS2008,Tomlin1992}. A common strategy for additional reduction of the size and stiffness of these models invokes the Quasi-Steady State Approximation (QSSA)~\cite{bodenstein1913theorie,Fraser1988}, i.e.\ that the production and destruction of well-chosen chemical species are assumed to be everywhere in balance. As discussed in detail later in Section~\ref{sec:QSSintro}, the QSSA reduces the dimension of the chemical system; the molar concentrations of the removed species become algebraic constraints on the remaining system.
Critically, the QSSA can result in complex nonlinear algebraic relations between species for which either cumbersome numerical procedures must be designed, or simplified approximations should be formulated~\cite{lu2006systematic}. In the context of simulating turbulent combustion applications, QSSA-reduced chemistries also complicate the construction and evaluation of the chemical Jacobian, which is often required by time-implicit chemistry integration algorithms.

The evaluation of analytic chemical Jacobians has received significant recent attention~\cite{safta2011tchem,niemeyer2017pyjac,perini2012analytical,bisetti2012integration}, and efforts geared towards the way it is assembled have been recognized as beneficial from a computational standpoint~\cite{niemeyer2017pyjac,dijkmans2014gpu}. However, fewer efforts have been dedicated to adapting analytic Jacobian construction methods to QSSA-reduced chemistries. Recently, Sharma~\cite{sharma2022acceleration} discussed the construction of analytical Jacobian of QSSA-reduced chemistry and how to approximate it, but did not explain how to encode the dependence of QSSA species on non-QSSA species, which is a significant hurdle we address here. To the authors' knowledge, there are also no currently available analytic chemical Jacobian construction tools for QSSA-reduced chemistry models.

Recent work in the design of exascale-ready codes \cite{Esclapez2023,henrydefrahan2022} has also highlighted that software maintainability is crucial for long-term sustainability. Therefore, the software implementing the analytical Jacobian of QSS-reduced chemistry models should induce minimal additional complexity in the reacting flow solvers that use them. The novel contributions of this work are as follows:
\begin{enumerate}
    \item A framework for the generation of the analytical chemical Jacobian of QSS-reduced mechanisms is developed. This capability is openly available \revone{in PelePhysics \cite{HenrydeFrahan2024,doecode_5574}}. Jacobians of the chemical models are assembled via a symbolic method that significantly simplifies the logic involved in the resulting C++ implementation.
    \item In general, evaluating a chemical Jacobian is computationally intensive and requires large amounts of memory. The problem is shown to be even worse when a symbolic method is used. Strategies to mitigate memory and computing time are proposed and shown to perform well and only mildly affect the overall computational cost associated with generating the evaluation routines.
\end{enumerate}

The remainder of this paper is organized as follows. 
Section~\ref{sec:QSSimpl} discusses the QSSA in depth and describes how interdependencies and nonlinear coupling between QSS species are handled. Section~\ref{sec:aj_qss} presents an approach to compute and validate the analytic evaluation of the Jacobian for QSSA-reduced chemistry models, and demonstrates increased stability and computational efficiency of implicit integration schemes over alternative approaches for stiff systems. In Section~\ref{sec:perfOpt}, we address the key practical issue of efficient memory use when constructing the chemical Jacobian with a symbolic approach. Section~\ref{sec:exascale} demonstrates the application and scalability of our combined generation and implementation approach on a test problem that models a pulsed-injection ignition. Conclusions are provided in Section~\ref{sec:conclusions}.

\section{Implementation of the QSS-reduced chemistries}
\label{sec:QSSimpl}
In this section, a brief overview of the QSSA is provided. The difficulties caused by QSS species couplings in the construction of a tractable equation set are also discussed.

\subsection{Rate laws and the Quasi-Steady State Approximation}
\label{sec:QSSintro}
A set of $N$ species interacting through $M$ reactions will usually be expressed via a reaction mechanism of the form:
\begin{equation}
    \label{eq:chemMech}
    \sum_{i=1,N} \nu'_{i,j}S_i \leftrightarrow \sum_{i=1,N} \nu''_{i,j}S_i,  \text{  for j=}1,M
\end{equation}
where $S_i$ denotes any species $i$ and $\nu'_{i,j}$ and $\nu''_{i,j}$ are the forward and backwards stoichioimetric coefficients of species $i$ in reaction $j$. The progress of reaction $j$ ($\dot{\mathcal{Q}_j}$) is the sum of the forward ($q_{f,j}$) and backward ($q_{b,j}$) contributions, which are a function of the species molar concentrations ($C_i$) as well as to the forward and reverse reaction rates ($k_{f/b}$):
\begin{equation}
    \label{eq:progRates}
    \dot{\mathcal{Q}_j} = q_{f,j} - q_{b,j} = k_{f,j}\prod_{i=1,N}C_i^{\nu'_{i,j}} - k_{b,j}\prod_{i=1,N}C_i^{\nu''_{i,j}},
\end{equation}
Thus, the evolution of species molar concentrations is the result of contributions from a subset of the $M$ reactions, and is typically expressed via a series of ordinary differential equations (ODEs):
\revtwo{
\begin{equation}
\label{eq:ReacMech}
    \frac{\mathrm dC_i}{ \mathrm dt} = \sum_{j=1,M} \nu_{i,j} \dot{\mathcal{Q}_j} = \dot{\omega}_i(\mathbf{C}, \mathbf{k}), \text{  for i=}1,N 
\end{equation}}
where $\nu_{i,j} = \nu''_{i,j} - \nu'_{i,j}$, $\mathbf{C}$ is the $N$ vector of species molar concentrations, $\mathbf{k}$ is a $2M$ vector of reaction rate functions and $\dot{\omega}_i$ is the net production rate of species $i$. 

To reduce the computational burden and stiffness of reactive flow simulations, QSSA can be formulated on a set of $N_{\rm QSS}$ strategically chosen species. Good candidates are species with very short characteristic timescales and low molar concentration throughout the process of interest~\cite{lu2006systematic}. The QSS approximation for a species $i$ is expressed as~\cite{Turanyi1993}:
\revtwo{
\begin{equation}
\label{eq:QSSi}
    \frac{\mathrm dC_i}{\mathrm dt} = 0,  
\end{equation}}
 
The molar concentrations of QSS species can then be obtained via algebraic relations constructed by inverting this newly constructed system of $N_{\rm QSS}$ equations. Combining Eqs.~\ref{eq:progRates}\&\ref{eq:ReacMech} and substituting into Eq.~\ref{eq:QSSi} can lead to a set of nonlinear algebraic relations, which can involve a large number of species and reaction rates. In particular, a situation that typically arises is when two or more QSS species are on the same side of an elementary reaction: this situation is referred, hereafter, as a quadratic coupling. Quadratic couplings require complex solution strategies that can dominate the implementation and eventually reduce the interest of using QSSA in the first place~\cite{law2003development}, especially if the number of reactions is large. One linearization approach has been discussed in the past \cite{lu2006systematic}, but other linearization methods can be envisioned. For the sake of completeness, different linearization methods are described and compared in the following section. All linearization methods are compatible with a symbolic construction of the analytic Jacobian described in Sec.~\ref{sec:aj_qss}.

\subsection{Nonlinear QSS couplings}
\label{sec:lin}
\paragraph{Linearization approaches}
One approach to avoid quadratic couplings would be to restrict the choice of QSS species to the ones that are not involved in quadratic couplings. A species is said to be involved in a quadratic coupling if it is a reactant in the elementary reaction where a quadratic coupling between QSS species concentrations is found. Here, this approach is referred to as the \textit{species linearization} method. A simple but systematic method to identify the set of QSS species to turn into non-QSS species will be outlined in the following paragraphs. As will be illustrated, this approach often can end up being too restrictive. A second approach to eliminate quadratic couplings is to ignore chemical reactions involving intra-QSS collisions. This approach is rooted in the fact that since QSS species are present in lower molar concentrations than transported species, the probability of observing the reactions that induce quadratic coupling should also be very low. Lu and Law~\cite{lu2006systematic} proposed a linearization technique for irreversible reactions based on this observation. In the case when reversible reactions are considered, and following a similar approach as~\cite{lu2006systematic}, one could ignore the complete reaction to conserve chemical balance, even if only the forward or the backward reaction induces the quadratic coupling. This approach is referred to as the \textit{reaction linearization} method. Finally, a third approach, the \textit{one-sided reaction linearization} method which can be seen as a softer version of the second method, removes only the reaction direction that induces the quadratic coupling. As discussed below, all three of these methods have been implemented in an open-source software distribution available at \hyperlink{https://github.com/AMReX-Combustion/PelePhysics}{https://github.com/AMReX-Combustion/PelePhysics}.

\begin{figure}[th!]
	\centering
    \includegraphics[width=0.99\textwidth]{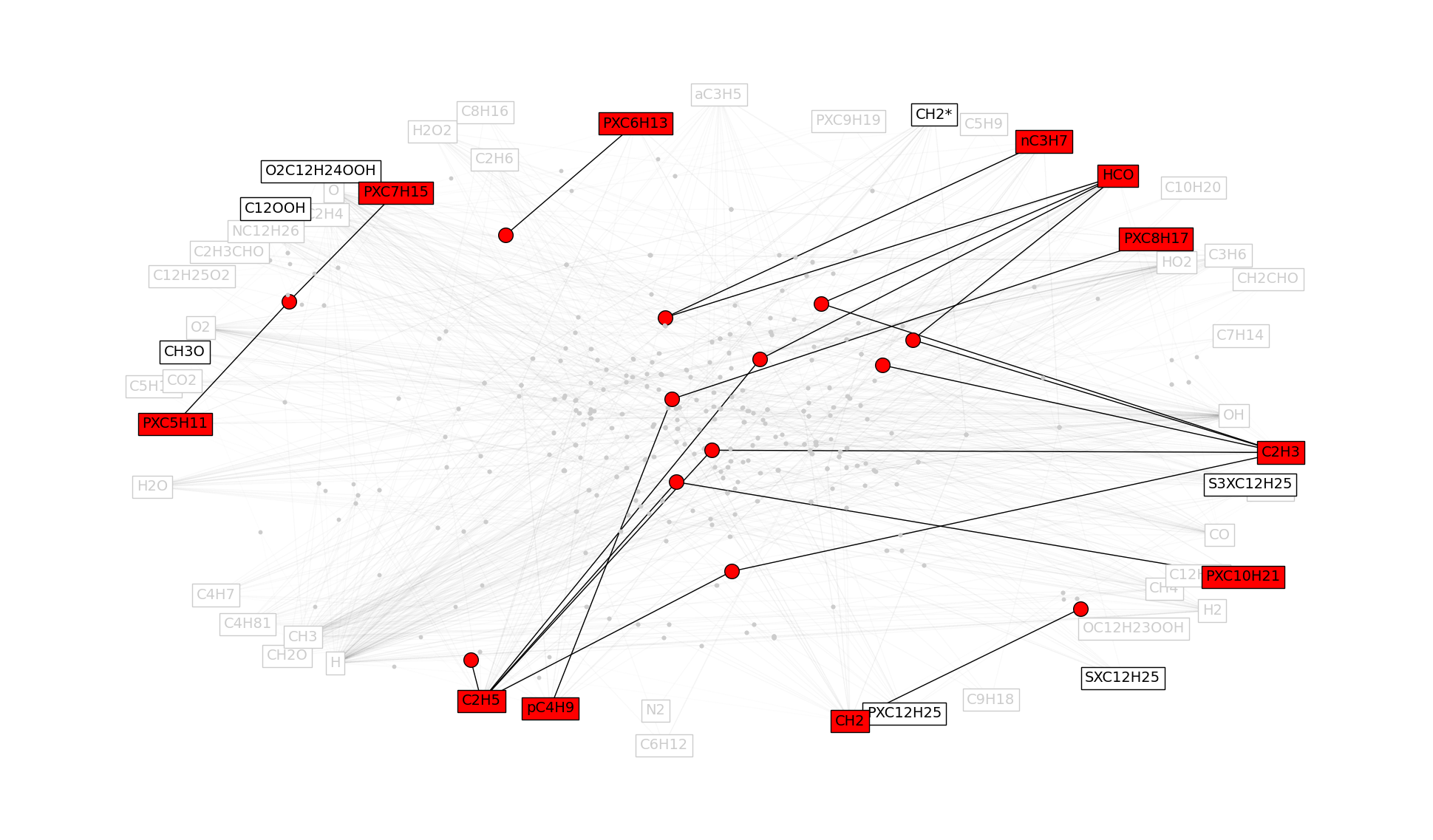}
	\caption{Graph of the 35-species, QSSA reduced, n-dodecane mechanism~\cite{borghesi2018direct}. Textboxes correspond to species, and circles correspond to elementary reactions. The 35-non-QSSA are greyed out, and 11 QSS species involved in quadratic couplings are highlighted in red. Red circles highlight the elementary reactions that induce the quadratic couplings. Links between reaction nodes and species nodes are grayed out except for inducing couplings (i.e., links between red-highlighted reaction nodes and red-highlighted species nodes).}
	\label{fig:linQSS}
\end{figure}

\paragraph{The example of n-dodecane}
The three different linearization approaches discussed above are used to reduce a 53-species n-dodecane skeletal mechanism~\cite{borghesi2018direct}, assumed to contain 18 QSS species. This model is represented graphically in Fig.~\ref{fig:linQSS}. The reaction mechanism is represented as a graph with 2 types of nodes: species (as textboxes) and elementary reactions (as circles). Representing elementary reactions with nodes allows to link more than 2 species to a given reaction. Non-QSS species are greyed out and QSS species that are involved in at least one quadratic coupling are highlighted in red. Quadratic couplings are observed in specific reactions that are also highlighted in red. Other reactions are greyed out. The species linearization method acts upon the red-highlighted QSS species, and both reaction linearization methods act upon the set of red-highlighted reactions.

In the first linearization approach (species linearization method), the objective is to identify a set of QSS species that can eliminate quadratic couplings, if turned into a set of non-QSS species. This set of QSS species must be as small as possible given that every new non-QSS species requires solving an additional transport equation in a reacting flow simulation. A naive but sufficiently fast procedure to identify this set of QSS species is outlined in Algo.~\ref{algo:specieslin}.

\begin{algorithm}
\caption{Identification of QSS species to eliminate in the species linearization method.}
\label{algo:specieslin}
\begin{algorithmic}[1]
\State Identify the set $\mathcal{A}$, the set of QSS species involved in at least one quadratic coupling, where $n \equiv \operatorname{card}(\mathcal{A})$
\State Initialize an empty list $\mathcal{B}$, that holds the candidate sets of QSS species that could eliminate quadratic couplings if turned into non-QSS species
\For{i = 1, n}
\State Create an ensemble $\mathcal{P}_i$ of subsets of $i$ species from $\mathcal{A}$ without repetition, i. e. \State $\mathcal{P}_i= \{ \mathcal{A}_j \}$, where $\mathcal{A}_j \subset \mathcal{A}$,  and $\operatorname{card}(\mathcal{A}_j) = i, \forall j$.
\For{$\mathcal{A}_j \in \mathcal{P}_i$}
\State Turn the set of QSS species $\mathcal{A}_j$ into non-QSS species
\If{No quadratic couplings}
    \State Append $\mathcal{A}_j$ to $\mathcal{B}$.
\EndIf
\State Turn back the set of non-QSS species $\mathcal{A}_j$ into QSS species

\EndFor
\If{$\mathcal{B}$ non empty}
    \State Return $\mathcal{B}$
\EndIf
\EndFor
\end{algorithmic}
\end{algorithm}

In the case of the n-dodecane mechanism illustrated in Fig.~\ref{fig:linQSS}, an initial set of candidate of QSS species to eliminate ($\mathcal{A}$ in Algo.~\ref{algo:specieslin}) contains 11 QSS species. Those species are the red-highlighted species in Fig.~\ref{fig:linQSS}. If all 11 QSS species are turned into non-QSS species, quadratic couplings are all eliminated. However, not all 11 species need to be turned into non-QSS species to eliminate quadratic couplings. For instance, \ce{PXC7H15} and \ce{PXC5H11} and involved in a quadratic coupling through a single elementary reaction \ce{PXC7H15 + PXC5H11 <=> NC12H26}. In this case, either \ce{PXC7H15} or \ce{PXC5H11} need to be turned into non-QSS species. Algorithm~\ref{algo:specieslin} shows a simple method to identify the minimal sets of QSS species that need to be turned into non-QSS species. Following this procedure, a set of 7 QSS species is identified as the smallest set of QSS species that should be turned into non-QSS species ($\mathcal{B}$ in Algo.~\ref{algo:specieslin}). Note that while this strategy is simple to implement, it can induce a significant computational overhead on the CFD solvers given that the original 35-species reduced mechanism has now become a 42-species mechanism.

The other two linearization approaches (the reaction linearization method and the one-sided reaction linearization method) are more straightforward, and rely on eliminating the red-highlighted reactions shown in Fig.~\ref{fig:linQSS}.

The mechanisms obtained with the three linearization methods are compared to the 53-species skeletal version of the n-dodecane mechanism~\cite{Yao2017} by computing ignition delays ($t_{\rm ign}$) for 108 initial chemical states. The 108 initial conditions correspond to 9 initial temperatures evenly distributed in the set $T = [800-1600]$K, 3 equivalence ratio from the set $\phi = \{ 0.5, 1, 2\}$, and 4 pressures from the set $P = \{ 1, 5, 10, 50\}$atm. These conditions span the validity range of the QSSA for this mechanism \cite{borghesi2018direct}. The ignition delay $t_{\rm ign}$ was defined as the earliest time after which the initial temperature increased by at least $400$K. In the absence of ignition, $t_{\rm ign}$ was assigned the arbitrary value of $3$s. To eliminate any considerations regarding the analytic Jacobian construction for now, the results of this Section are obtained by using a Newton method with a Jacobian computed via finite differences~\cite{hindmarsh2005sundials}. Results are reported in terms of the ignition delay error ($\varepsilon_{\rm ign}$), defined as: 
\begin{equation}
    \label{eq:ignErr}
    \varepsilon_{\rm ign,i-j} = \mathbb{E}_k\left(\frac{|t_{\rm ign,i,k} - t_{\rm ign,j,k}|}{t_{\rm ign,i,k}}\right),
\end{equation}
where the error between results from method $i$ and $j$ are compared for the initial condition $k$, and $\mathbb{E}_k$ denotes the averaging operation over all 108 initial conditions considered. In the present case, $i=\rm{SK}$ since results obtained with the skeletal mechanism~\cite{Yao2017} are taken as a reference, and $j=\rm{QSS}$ denotes the use of one of the linearized QSS mechanisms. Values for $\varepsilon_{\rm ign,SK-QSS}$ are shown in Tab.~\ref{tab:linQSS}.

\begin{table}[h!]
\centering
\begin{tabular}{ |c|c| } 
          \hline
          Linearization method &   $\varepsilon_{\rm ign,SK-QSS}$ \\
          \hline
           Species linearization &  $0.848$  \%  \\ 
          \hline
           Reactions linearization & $13.3$ \%  \\
        \hline
           One-sided reaction linearization  & $8.97$ \%  \\ 
        \hline
         \end{tabular}
\caption{Ignition delay errors obtained with different QSS linearization methods of a reduced n-dodecane mechanism~\cite{Yao2017,borghesi2018direct}.}
\label{tab:linQSS}
\end{table} 

It can be seen that the species linearization strategy offers the highest accuracy at the expense of increasing the computational cost (by about 30\% compared to the other two mechanisms). Additionally, the one-sided reaction linearization is 33\% more accurate over the 108 initial conditions than the more stringent reaction linearization. Given these results, the one-sided reaction linearization option is the preferred strategy for the rest of this manuscript. The analytical Jacobian construction method described in Sec.~\ref{sec:aj_qss} could, however, easily adapt to any of these linearization approaches. All three methods are available in the companion repository.

\subsection{Interdepencies between QSS species} 
\label{sec:interdependencies}

After linearization, the system that results from applying Eq.~\ref{eq:ReacMech} and Eq.~\ref{eq:QSSi} must be inverted to formulate the algebraic closure equations for the QSS species molar concentrations. 
In the case of linear couplings between QSS species, the interdependencies can be represented by a block-diagonal matrix~\cite{lu2006systematic}, allowing the use of standard solution approaches. \revtwo{The coupling between interdependent QSS species is a key complication that prevents one from simply aggregating the effect of each elementary reaction one by one into the chemical Jacobian. Instead, one needs to account for all the elementary reactions that appear in the algebraic relation function that maps non-QSS species concentration to QSS species concentration.} The identification and organization of QSS interdependencies to facilitate this representation is summarized next.

The linear system for the $N_{\rm QSS}$ QSS species molar concentrations, stored in the vector $\mathbf{C}_{\rm QSS}$, can be written as follows:
\begin{equation}
\label{eq:full_QSSsystem}
\begin{split}
&\mathbf{A} \mathbf{C}_{\rm QSS} = \mathbf{b}, \\
&A_{\rm m,n} = \frac{\sum_{j=1,M} \nu_{\rm m,j} \left( \delta'_{n,j} q_{f,j} - \delta''_{n,j}q_{b,j} \right)}{C_n} \quad \text{ for } \quad m,n=1,N_{\rm QSS}, \\ &b_m =  \sum_{j=1,M} \nu_{m,j} \left( \prod_{i=1,N_{\rm QSS}} \left(\delta'_{i,j} - 1 \right) q_{f,j} - \prod_{i=1,N_{\rm QSS}} \left(\delta''_{i,j} - 1 \right) q_{b,j} \right) \quad \text{ for } \quad m=1,N_{\rm QSS},
\end{split}
\end{equation}
where $\delta'_{i,j}$, resp. $\delta''_{i,j}$, equals 1 if (QSS) species $i$ is on the left, resp. right, of reaction $j$. $A_{\rm m,n}$ represents the contribution to the net production rate of QSS species $m$ of those reactions that linearly couple QSS species $n$ and $m$; while $b_m$ contains the remaining non-QSS contributions. The $\mathbf{A}$ matrix becomes more dense as QSS species interdependency increases, introducing the need for a non-trivial solution in order to solve for $\mathbf{C}_{\rm QSS}$.

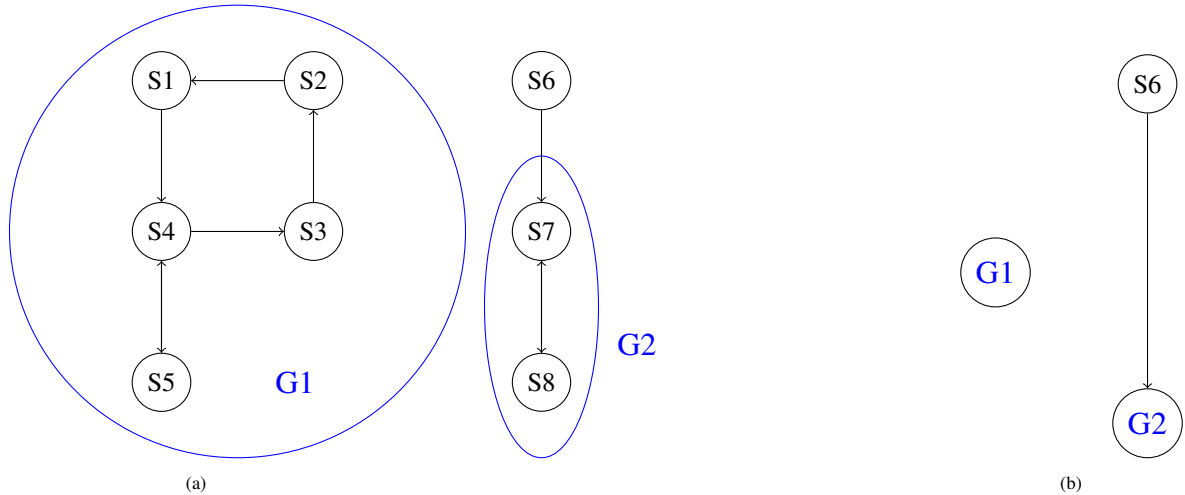
\begin{figure*}[!htb]
    \centering
    \begin{subfigure}[b]{0.3\textwidth}
        \centering 
        \begin{tikzpicture}
    \node[shape=circle,draw=black] (S5) at (0,0) {S5};
    \node[shape=circle,draw=black] (S4) at (0,2) {S4};
    \node[shape=circle,draw=black] (S1) at (0,4) {S1};
    \node[shape=circle,draw=black] (S2) at (2,4) {S2};
    \node[shape=circle,draw=black] (S3) at (2,2) {S3};
    \draw[color=blue] (1,2) circle [radius=3];
    \node[text width=1cm] at (2,0) {\large {\color{blue}G1}};

    \node[shape=circle,draw=black] (S6) at (5,4) {S6};
    \node[shape=circle,draw=black] (S7) at (5,2) {S7};
    \node[shape=circle,draw=black] (S8) at (5,0) {S8};
    \draw[color=blue] (5,1) ellipse (0.75 and 2);
    \node[text width=1cm] at (6.5,0.5) {\large {\color{blue}G2}};

    \path [->] (S1) edge node[left] {} (S4);
    \path [->](S4) edge node[left] {} (S5);
    \path [->](S5) edge node[left] {} (S4);
    \path [->](S4) edge node[left] {} (S3);
    \path [->](S3) edge node[right]{} (S2);
    \path [->](S2) edge node[left] {} (S1);

    \path [->] (S6) edge node[left] {} (S7);
    \path [->](S7) edge node[left] {} (S8);
    \path [->](S8) edge node[left] {} (S7);
\end{tikzpicture}
        \caption{}
        \label{fig:QSSGraph}
    \end{subfigure}
    \hfill
    \begin{subfigure}[b]{0.3\textwidth}
        \centering 
        \begin{tikzpicture}
    \node[shape=circle,draw=black] (G1) at (0,3) {{\color{blue} \large G1}};
    \node[shape=circle,draw=black] (S6) at (2,5.5) {S6};
    \node[shape=circle,draw=black] (G2) at (2,1) {{\color{blue} \large G2}};

    \path [->] (S6) edge node[left] {} (G2);
\end{tikzpicture}
        \caption{}
        \label{fig:QSSGraph_scc}
    \end{subfigure}
    \caption{(a) QSS dependencies through a directed graph: connected components are determined by acyclic paths (e.g., S6 to S7) while strongly connected components (SCC) are determined by cyclic paths (blue, solid groupings). (b) After SCCs are identified, they can be considered as single nodes within a simplified directed graph containing only acyclic paths.}
    \label{fig:GroupsQSS}
\end{figure*}

Following the general procedure outlined by Lu and Law~\cite{Lu2005} the interdependencies between QSS species are expressed as a directed graph. Fig.~\ref{fig:GroupsQSS}(a) shows such a graph for a fictitious kinetic mechanism containing 8 QSS species ($S_1$ to $S_8$). The arrows express connectivity between species and point in the direction of dependence; for example, species $S_2$ is a product in at least one reaction that uses species $S_1$ as a reactant ($S_2$ \textit{needs} $S_1$). Two types of interdependencies can be distinguished: connected components and strongly connected components (SCC). Vertices (the species) within a connected component are related only via unidirectional paths, such as $S_6$ and $S_7$. Vertices within SCCs are related by bi-directional paths, which can be indirect; for every pair of vertices, $S_x$ and $S_y$, within an SCC there exist paths both from $S_x$ to $S_y$ and from $S_y$ to $S_x$~\cite{Tarjan1972}. Species $S_1$ to $S_5$ in Fig.~\ref{fig:GroupsQSS}(a) form a SCC group, labeled $G_1$. Identifying SCCs enables partitioning QSS species into groups, as depicted in Fig.~\ref{fig:GroupsQSS}(b). This information can be used to reorder the species and simplify the expressions that define the entries of the block-diagonal matrix $\mathbf{A}$ \cite{lu2006systematic}. Note also that connected components can be represented by an upper triangular structure, requiring only back substitutions to solve; while SCCs are represented by sub-blocks which require a more complex solution scheme that typically involves pivoting.

To sum-up, the order of algebraic resolution is determined as follows:
\begin{enumerate}
    \item QSS species involved in SCCs are identified and partitioned into groups (Fig.~\ref{fig:GroupsQSS} (a))
    \item The remaining unidirectional dependencies relating groups and individual QSS species are updated (Fig.~\ref{fig:GroupsQSS} (b)) 
    \item Groups and individual QSS species are ordered based on the size and direction of dependencies
\end{enumerate}

The procedure aforementioned requires identifying the set of SCCs and partitioning the set of QSS species appropriately. Several algorithms have been proposed to identify SCCs with various time complexities in terms of the number of graph vertices and edges~\cite{sharir1981strong, Tarjan1972}. The algorithm used herein~\cite{Tarjan1972} is comparable to that used in~\cite{lu2006systematic} in that both are recursive, depth-first search (DFS) algorithms that inherently topologically sort the resulting directed acyclic graph; the two differ in number of DFS traversals needed and storage structure of graph information. Our implementation requires only one DFS traversal of the original graph, as opposed to the multiple traversals needed for both the original and transpose graphs as in~\cite{lu2006systematic}. Additionally, we use an adjacency list to represent graph information, as opposed to the adjacency matrix implementation in~\cite{lu2006systematic}; this eliminates storage of superfluous information regarding non-adjacent vertices. With the use of an adjacency list, the SCC identification has time complexity ${\cal O}$(V + E) where V is the number of vertices and E is the number of edges in the directed graph. Further description of the algorithm is provided in~\ref{sec:appendixQSS}. 

\section{The analytical chemical Jacobian for QSS-reduced chemistries}
\label{sec:aj_qss}

The integration of stiff chemical kinetics in CFD solvers often relies on time-implicit methods to keep the overall timestep size constrained by the fluid dynamics of interest. The implicit time discretization results in a non-linear system generally tackled using a Newton approach. In this section, the shortcomings of using an iterative method to calculate the Newton direction (i.e. methods that do not rely on constructing the system Jacobian) are put forth when larger timesteps are required in CFD solvers. Challenges of constructing analytical Jacobians (AJ) for QSS-reduced chemistries are discussed, typical approximations are investigated and a symbolic approach for the construction of an exact AJ is presented and validated.

\subsection{The need for a chemical Jacobian}
\label{sec:needJac}
At every timestep and in every computational cell of a reacting flow simulation, the instantaneous vector of production rates for the chemical state $\phi$ of the system should be evaluated in order to advance the solution. Typically, $\phi$ is an ($N+1$)-sized vector containing the transported species molar concentrations and an energy variable - here, temperature: $\phi = \{ C_1, C_2, ..., C_N, T\}$. Chemical mechanisms enabling the evaluation of these production rates are typically characterized by a wide range of timescales. This is all the more true as larger, more detailed, chemical mechanisms are being deployed within reacting flow simulations~\cite{niemeyer2017pyjac}. The integration of the resulting stiff system of ODE can require very small timesteps and when an explicit integration of the chemical kinetics over time is unpractical, such as in low-Mach solvers, an implicit strategy is often preferred. 

Implicit time integration schemes usually employ a Newton method to solve the non-linear system resulting from the time discretization of the set of chemical ODEs, relying on forming and inverting the chemical Jacobian of $\phi$~\cite{safta2011tchem,niemeyer2017pyjac, gadalla2022implementation}: 
\begin{equation}
  \label{eq:JacExptr}
  \mathcal{J}_{i,j}=\frac{\partial \dot{\omega}_{i}}{\partial \phi_j}, \text{  for } i,j \in \{1,N+1\},
\end{equation}
to advance the Newton step. Alternatively, matrix-free methods also exist, such as Krylov-based iterative methods, and these can be implemented without knowledge of the full chemical Jacobian~\cite{saad2003iterative}.
Since the computation, storage and inversion of the \textit{full} chemical Jacobian can be prohibitively expensive, these matrix-free and preconditioned Krylov techniques can be attractive. \revone{In what follows, we assess the need for a chemical Jacobian by comparing the robustness and efficiency of a Krylov iterative method without preconditioning, hereafter referred to as IM, with that of Newton strategy that uses a full Jacobian computed using finite differences (FD)~\cite{hindmarsh2005sundials}. Note that while an effective preconditioner could improve the performance of the IM solver \cite{knoll2004jacobian}, it would benefit most large chemical mechanism ($>100$ species \cite{mcnenly2015faster,walker2023generalized}) and its construction would involve computing or approximating elements of $\mathcal{J}$, thus rather reinforcing the argument for needing an approach to compute such elements. While an FD approach to computing $\mathcal{J}$ is robust and simple, the main focus of the paper is the construction of an analytical Jacobian that is considerably more efficient and flexible than its FD counterpart.}

\revone{The comparison between the FD chemical Jacobian and the IM is conducted for the QSS-reduced, 35-species, n-dodecane chemical mechanism using the 0D calculations at constant volume; starting from the same 108 initial chemical states as in Sec.~\ref{sec:lin}.
All 108 0D cases are run for a total time interval of 0.1s, which is divided into substeps over which the chemical state is advanced. The role of these substeps is to mimic timestepping in a fluid simulation.}

\revone{First, a set of reference solutions is constructed using a finite difference chemical Jacobian for each one of the 108 initial states to approximate the true solution. 
The reference simulations use a substep of size $10^{-8}$s and absolute tolerance of $10^{-15}$. Second, the solutions predicted using the FD Jacobian and the Krylov-based IM are generated for absolute tolerances ($\eta$) varied between $10^{-8}$ and $10^{-14}$, and substep size ranging from $3\times 10^{-8}$s to $10^{-4}$s. For each $\eta$ and substep size, all 108 0D calculations are run, and the execution time is recorded, as well as the error in the temperature predictions compared to the reference solutions. The error reported is computed as: 
\begin{equation}
    \varepsilon_{T} = \frac{|| T(t) - \overline{T_{\rm ref}}(t)||_2}{\sqrt{N_{\rm steps}}},
\end{equation} 
where $T$ is the time-dependent temperature history obtained with a given number of substeps and tolerance level, $\overline{T_{ref}}$ is the temperature from the reference solution interpolated to the time instance where $T$ is defined, and $N_{\rm steps}$ is the total number of substeps throughout the total time interval. The scaling factor $\sqrt{N_{\rm steps}}$ ensures that the errors computed with different numbers of substeps are comparable.}

\paragraph{\revone{Convergence at large substep size}} \revone{Hereafter, the error $\varepsilon_{T}$ averaged over the 108 initial conditions is reported, where possible. Results are not reported when any of the 108 calculations took more than $7200$s to complete. This timeout procedure is implemented because, within each substep, the SUNDIALS ODE integrator employs its own variable size time-stepping strategy. Typically, when the ODE integrator uses more than $10^4$ internal steps (threshold typically used in Pele runs \cite{wimer2023visualizations}), the reacting flow simulation would stop and would need to be restarted with a smaller substep size (or timestep in the context of a reacting flow simulation). Here, this threshold is set to $10^9$ internal steps to prevent such early stopping, and allow the integration method to take as many steps as needed to converge. In practice, none of the cases reached the $10^9$ internal steps limit. However, with this approach, the time to solution can become unreasonably long (see also \cite{balos2024}), in which case we declare the simulation unsuccessful.}

\begin{figure}[th!]
	\centering
    \includegraphics[width=0.4\textwidth]{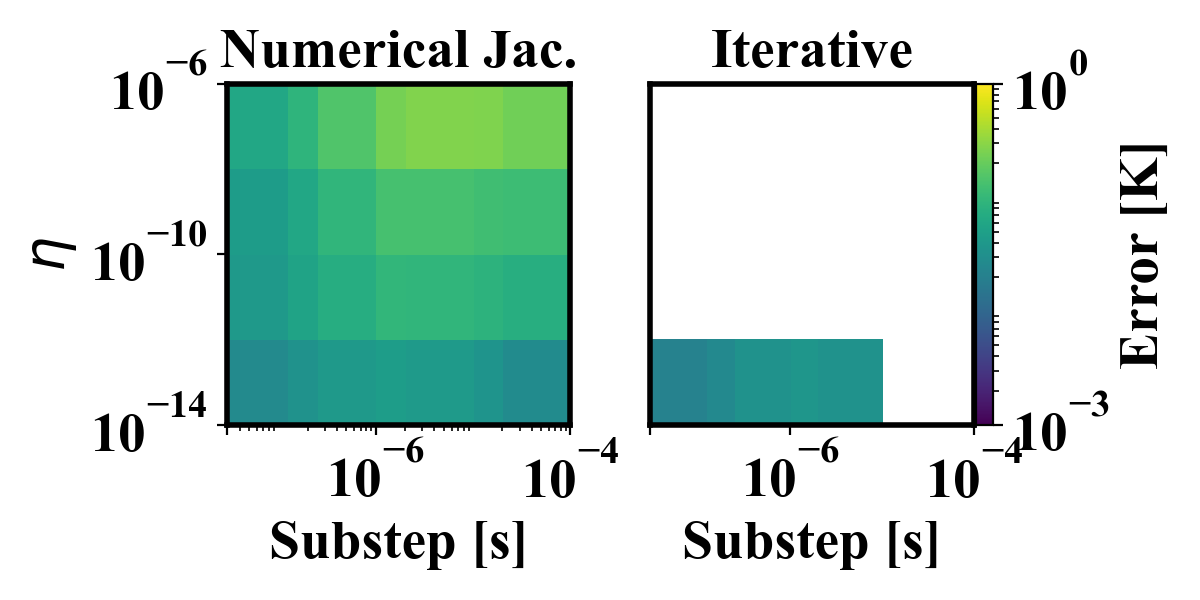}
	\caption{\revone{Average temperature error of 0D constant volume calculations, starting from 108 initial conditions, done with a direct solve requiring a numerical Jacobian (left contour) and a Krylov-based iterative method (right contour). White spaces denote that at least one of the 108 calculations failed to complete in under 7200s.}}
	\label{fig:stab1}
\end{figure} 

\revone{The results shown in Fig.~\ref{fig:stab1} illustrate that, where all 108 calculations successfully complete, the averaged temperature error is similar between the FD Jacobian and the IM. Successful calculations can consistently be obtained with the IM only when the absolute tolerances are sufficiently small ($\leq 10^{-12}$). Moreover, some calculations failed to complete for substep sizes larger than $10^{-5}$s. By contrast, all 108 0D runs successfully complete for all tolerances and substep sizes when using an FD Jacobian. These results emphasize that a chemical Jacobian would be typically needed for low-Mach number solvers~\cite{Esclapez2023,hassanaly2018minimally,desjardins2008high} that use timestep sizes of the order $10^{-6}-10^{-4}$s. For compressible solvers, where timesteps are constrained by the acoustic timescales, the Krylov-based iterative method may however be an acceptable method if an explicit integration is not feasible.}

\paragraph{\revone{Computational cost at large substep size}} \revone{The wall time was recorded for each one of the 0D calculations and was averaged over the 108 initial conditions. The calculations were conducted on the Kestrel HPC cluster at NREL which uses Intel Xeon Sapphire Rapids. Hereafter, the ratio $Wt_{\rm IM}/Wt_{\rm FDJ}$ is reported, where $Wt_{\rm IM}$ is the average wall time over the 108 initial conditions for the Krylov-based IM and $Wt_{\rm FDJ}$ is the average wall time over the 108 initial conditions when using an FD Jacobian.} \revone{The wall time ratio is computed for the smallest absolute tolerance level, ensuring successful completion of all computations.}  
\begin{figure}[th!]
	\centering
    \includegraphics[width=0.4\textwidth]{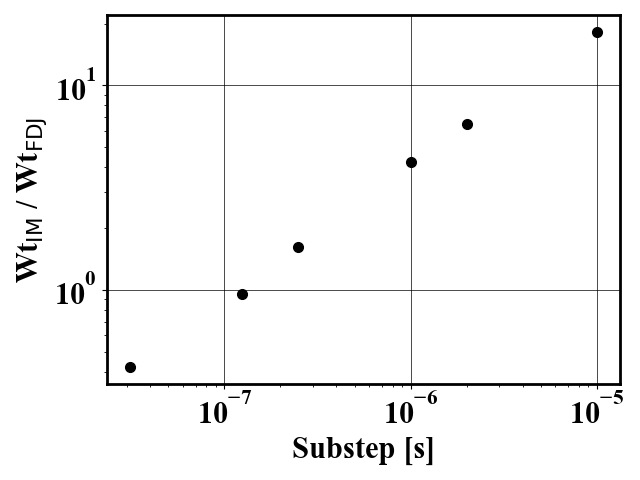}
	\caption{\revone{Average wall time ratio of 0D constant volume calculations starting from 108 initial conditions between a Krylov-based iterative method and a numerical Jacobian for absolute tolerance of $10^{-14}$.}}
	\label{fig:stab}
\end{figure}
\revone{It can be observed from Fig.~\ref{fig:stab}, that as the substep size (i.e. the simulation timestep size) increases, using a chemical Jacobian becomes computationally more efficient than using a GMRES IM. This result also suggests that a chemical Jacobian would be advantageous in low-Mach number simulations.}


\subsection{Construction of analytical chemical Jacobian}
\label{sec:jac_cons}

Finite differences evaluations of the chemical Jacobian can be a very important contribution to the total computational time in CFD simulations, especially if updates are frequently needed during the solve (as is often the case when relying on the BDF methods implemented in CVODE~\cite{hindmarsh2005sundials} for example). One FD evaluation of the chemical Jacobian is typically more expensive than an analytical evaluation, because of the multiple reaction rate evaluations. Additionally, analytical evaluations can leverage the sparsity exhibited by most chemical Jacobians.

Although the computation of an AJ can be an involved process, due to the complexity of certain reaction types, several numerical methods have been successfully developed in the past~\cite{safta2011tchem,niemeyer2017pyjac}. \revtwo{Here, it is chosen to formulate the AJ in terms of molar concentrations (instead of mass fractions) in order to generate a sparser Jacobian \cite{curtis2018using,walker2023generalized}.} Without QSSA, the AJ can be computed by accumulating the contributions of each reaction to entries in the AJ, one after the other (see, e.g., Eq.12 in \cite{perini2012analytical}). In the presence of QSSA, a different approach is needed. One way to proceed is to differentiate between the set of $N_{\rm NQSS}$ non-QSS transported species, and the set of $N_{\rm QSS}$ QSS species which are algebraically related to the set of transported species. As a result, the production rates in Eq.~\ref{eq:JacExptr} indirectly involve the contribution of QSS species, and the AJ now reads:

\begin{equation}
    \label{eq:JQSS}
    \mathcal{J}_{i,j} = \sum_{k=1,N_{\rm NQSS}} \frac{\partial \dot{\omega}_i}{\partial C_k} \frac{\partial  C_k}{\partial C_j} = \sum_{k=1,N_{\rm NQSS}} \left( \sum_{l=1,M_{\rm NQSS}} \frac{ \partial \nu_{i,l} \dot{\mathcal{Q}}_l} {\partial C_k} + 
    \sum_{l=1,M_{\rm QSS}} \frac{ \partial \nu_{i,l} \dot{\mathcal{Q}}_l} {\partial C_k} \right)\frac{\partial  C_k}{\partial C_j},
\end{equation}

A distinction is made between contributions from reactions that do not involve QSS species (the first term in parenthesis in Eq.~\ref{eq:JQSS}) and those who do (the second term in parenthesis in Eq.~\ref{eq:JQSS}).
The non-QSS portion can be computed using classical AJ derivations reported in earlier work (e.g. \cite{perini2012analytical}). The evaluation of the second term in the expression is more difficult because it entails resolving the dependencies of all QSS species (involved in the $\dot{\mathcal{Q}}_l$) with respect to the non-QSS species. Note that a hybrid analytical-numerical strategy could be envisioned here, where one would numerically approximate the $\frac{\partial \mathcal{Q}_l}{\partial C_k}$ terms for each reaction $l$ and non-QSS species $k$. However, the numerical approximation would involve additional production rate evaluations, which should be avoided for computational efficiency.

In the rest of this Section, the construction of AJ for QSS-reduced chemistries is described. First, common simplifications are investigated, before an exact version is constructed and tested. The validation of the approach is done on 0D integrations, as was done previously in Sections~\ref{sec:lin} and~\ref{sec:needJac}, by comparing results against that obtained with FD. Three different QSS-reduced mechanisms are used for this purpose, spanning the range of fuels typically considered in reacting flow simulations. The following nomenclature is adopted: X-Y-\textit{fuel}, where X is the number of non-QSS species considered in the reduced mechanism, Y is the number of QSS species and \textit{fuel} identifies the molecule for which the kinetic mechanism was designed. The first chemistry adopted is a reduced methane mechanism~\cite{sankaran2007structure} referred to as 13-4-\ce{CH4}; the second chemical mechanism is a reduced n-dodecane mechanism~\cite{borghesi2018direct} referred to as 35-18-\ce{N-C12H26}; the third chemical mechanism is a reduced n-heptane mechanism~\cite{yoo2011direct} referred to as 55-33-\ce{N-C7H16}. 

\subsubsection{Simplified construction}
\label{sec:jac_simp}
The easiest simplifying strategy~\cite{sharma2022acceleration} has been suggested for QSS mechanisms, and consists of nullifying the second term of the right-hand side of Eq.~\ref{eq:JQSS}, thus falling back on the classical formulation of non-QSS AJ. This approximation is termed the~\textit{constant QSS AJ}. Another approximation consists of nullifying the QSS species molar concentrations altogether in the \textit{constant QSSA AJ} approximation by invoking the fact that QSS species molar concentrations are usually very low~\cite{lu2006systematic}. This second method is termed the \textit{zero QSS AJ} approximation.

\begin{figure}
     \centering
     \begin{subfigure}[b]{0.35\textwidth}
         \centering
         \begin{tabular}{ |c|c|c| } 
         \hline
          & Constant QSS AJ &  Zero QSS AJ\\
         \hline
         13-4-\ce{CH4} &  \revone{$8.3$ \%} & \revone{$9.9$ \%} \\ 
         \hline
         35-18-\ce{N-C12H26}  & \revone{$22.7$ \%} & \revone{$25.1$ \%} \\
         \hline
         55-33-\ce{N-C7H16}  & \revone{$65.7$ \%} & \revone{$72.1$ \%} \\ 
         \hline
         \end{tabular}
           \vspace{0.75cm}
     \caption{$\varepsilon_J$ computed for three different QSS-reduced chemistries with simplified AJ for 100 randomly sampled chemical states.}
      
      \label{fig:nosymb_stab_stats}
     \end{subfigure}
     \hfill
     \begin{subfigure}[b]{0.4\textwidth}
         \centering
           \includegraphics[width=\textwidth]{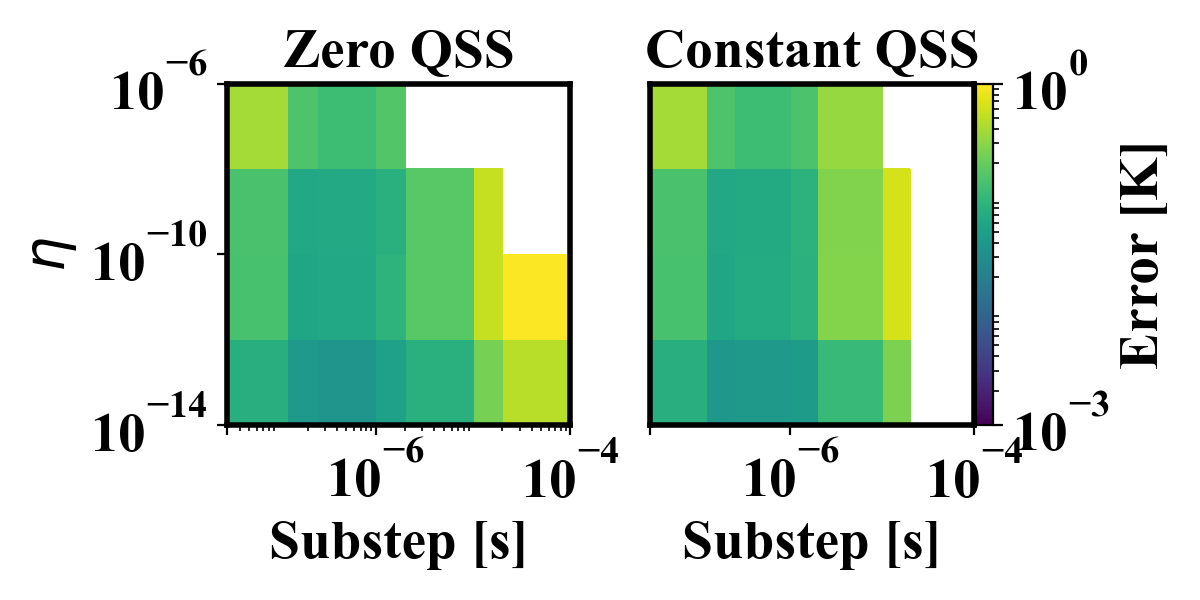}
         \caption{\revone{Mean temperature error profiles obtained with the 0D integration, done with either a Constant QSS AJ (left) or a Zero QSS AJ (right). White spaces denote that at least one of the 108 calculations failed to complete in under 7200s.}}
         \label{fig:nosymb_stab}
     \end{subfigure}
        \caption{Effect of simplified analytical Jacobian construction on QSS-reduced chemistries integration.}
        \label{fig:nosymb}
\end{figure}

To evaluate these strategies, the resulting AJs are compared to an FD chemical Jacobian by computing 0D integrations of 100 initial chemical states. Note that these are different from the 0D integrations presented in Sec.~\ref{sec:needJac}; the only objective here is to interrogate the function that maps temperature and species molar concentrations to the chemical Jacobian. The molar concentrations of non-QSS species are uniformly sampled from the interval $[10^{-4}, 1]$ mol/m$^3$ and initial temperatures are uniformly sampled from the interval $[300, 1300]$K. The finite difference approximation was done with perturbations of size $10^{-4}$ and using a central difference scheme. The error between the analytical approximation of the chemical Jacobian ($\mathcal{J}_{\rm A}$) and the FD one ($\mathcal{J}_{\rm FD}$) is computed using the Frobenius norm over all rows and columns: 
\begin{equation}
    \label{eq:frobErr}
    \varepsilon_J = \mathbb{E}\left(\frac{|| \mathcal{J}_{\rm A} - \mathcal{J}_{\rm FD} ||_F}{ ||\mathcal{J}_{\rm FD} ||_F }\right),
\end{equation}
$\mathbb{E}$ denotes the average taken over the 100 random samples. Results are presented in \revone{Tab}.~\ref{fig:nosymb_stab_stats}. By considering the numerical chemical Jacobian as the ground truth, it can be observed that the constant QSS and zero QSS approximations lead to similar error levels. Errors observed are relatively low, considering how aggressive the approximation employed is. Note that the error increases with the number of QSS species, which suggests that both approximations may be appropriate for mechanisms with a small $N_{\rm NQSS}/N_{\rm QSS}$ ratio. 

Figure~\ref{fig:nosymb_stab} shows the results of the same experiment as reported in Fig.~\ref{fig:stab} \revone{(for the 35-18-N-\ce{C12H26} mechanism only)}. Even though both AJ errors are small (about \revone{25\%} relative error), the ODE integration \revone{fails to converge in under 7200s, especially at large substep size. For the Constant QSS AJ  assumption, this is likely a consequence of the fact that the QSS concentrations cannot be assumed constant throughout the substep. For the Zero QSS AJ, the underlying system is less diagonally dominant as the substep increases and the Zero QSS AJ is too inaccurate to solve the linear system.} \revone{Comparing Fig.~\ref{fig:nosymb_stab} and Fig.~\ref{fig:stab} (left), it is also clear that the average error in the temperature predicted is higher than when using a finite difference Jacobian, especially for timesteps greater than $10^{-6}s$}. However, the errors quickly drop as the timestep and tolerance reduce. Overall simplified constructions of the chemical Jacobian are appropriate only if small timesteps and low tolerances are being used. \revtwo{In the following sections, it is proposed to not rely on the aforementioned approximations, but instead to obtain a Jacobian that can be used with a wider range of tolerances and timesteps.}

\subsubsection{Symbolic construction}
\label{sec:symbCons}
For chemical mechanisms of moderate size (from 30 to 50 transported species) with more than 10 QSS species, which are of interest in reacting flow simulations, it was shown in Sec.~\ref{sec:jac_simp} that neglecting the second right-hand-side term of Eq.~\ref{eq:JQSS} may adversely affect accuracy and stability. \revtwo{Including this term requires knowledge of the concentration of QSS species, which could in principle be obtained via a numerical inversion using a GPU-aware linear algebra library such as MAGMA~\cite{Dongarra:2014}. However, the small size of the chemical system is too limited to exploit the high throughput of the GPU for such a subsystem. Grouping the linear systems of multiple cells into a single batched solve available in MAGMA could fully exploit the GPU, but this approach is incompatible with the function call stack of PelePhysics, which distributes individual cells across GPU threads, and uses a single GPU kernel launch for the entire Jacobian.} 

In this section, a method to construct the AJ of QSS-reduced mechanisms is described and demonstrated. A symbolic method is used whereby the QSS and non-QSS species molar concentrations are symbolically tracked. This can be simply achieved using any symbolic math package such as the~\verb|sympy| package~\cite{10.7717/peerj-cs.103}. When the AJ needs to be written, it can be obtained using symbolic differentiation tools. In practice, the differentiation steps can be expensive but need only to be done once as a pre-processing step, and takes less than 1~min for the 55-33-N-\ce{C7H16}. An advantage of this approach is its flexibility with respect to the implementation of new reaction types. Any differentiable closure equation between transported and modeled species can be included in the chemical Jacobian. The symbolic approach therefore greatly simplifies the maintainability and the readability of the code, since the construction of the Jacobian is done based on the construction of reaction rates. While the manuscript describes the symbolic Jacobian construction for QSS-reduced chemistries, other types of algebraic reductions/simplifications could also benefit from the same approach~\cite{rein1992partial}.

\begin{table}[h!]
\centering
\begin{tabular}{ |c|c|c| } 
    \hline
          & Symbolic AJ with CSE (\# of subexpressions) & Symbolic AJ \\
    \hline
    13-4-\ce{CH4} & 375Kb (2327) & 98Mb  \\ 
    \hline
    35-18-\ce{N-C12H26} & 2.9Mb (16487) & (65Gb+)  \\
    \hline
    55-33-\ce{N-C7H16}  & 5.8Mb (10960) & 1.5Gb  \\ 
    \hline
    \end{tabular}
\caption{Size of the mechanism file generated when using a symbolic AJ computation with and without common subexpression elimination.}
\label{tab:fileSizeQSS}
\end{table} 

The main drawback of the symbolic construction of the Jacobian is that a naive symbolic differentiation can be slow to compute and can result in verbose expressions. In the chemical Jacobian, several matrix entries may involve the contribution of the same set of reaction rates albeit scaled with a different factor. Without additional constraints, a symbolic differentiation would spell the same term many times, resulting in overly verbose expressions. 
Table~\ref{tab:fileSizeQSS} (second column) gives the size of the file \ftexttt{mechanism.H} generated by the code provided in the companion repository and that contains the function that assembles the AJ. It can be seen that, without further treatment, the symbolic differentiation can generate several Gb of code. In the case of the 35-18-N-\ce{C12H26} mechanism \cite{borghesi2018direct}, the code generation could not reach completion and the size indicated is a lower bound of the actual file size that should be expected. Note that the larger the function, the longer the compilation, and the larger the compiled object. Compilation typically failed on HPC clusters using mechanisms files that did not use further treatment, due to memory errors. 

The function that assembles the AJ can be made significantly less verbose by first identifying common subexpressions in the Jacobian entries and precomputing them before assembly. \revtwo{In~\ref{app:subexpbenefit}, an example is provided to illustrate the practical benefit of common subexpression elimination strategies.} Fortunately, common subexpression elimination (CSE) is available in modern symbolic math packages~\cite{10.7717/peerj-cs.103}. Despite the number of species, the CSE can be achieved in only 30s \revtwo{on a single Apple M1 Pro chip} for the 35-18-N-\ce{C12H26} mechanism. The first column of Table~\ref{tab:fileSizeQSS} shows that the code generated with CSE is also several orders of magnitude smaller, which allows compilation on HPC clusters. The number of common subexpressions does not necessarily increase with the mechanism size but with the complexity of the QSS species interdependencies. It can also be observed that the number of common subexpressions identified is a monotonic but sublinear function of mechanism file size without common subexpression elimination. 

Note that the symbolic procedure is only used for the first $N$ rows and columns of the AJ. The contribution of the heat release rate can be simply obtained from the species production rates~\cite{niemeyer2017pyjac}. However the gradients of the production rates with respect to temperature, $\frac{\partial \dot{\omega}_i}{\partial T}$, need a specific and compute-intensive treatment~\cite{sharma2022acceleration}. The strategy adopted here is to approximate these terms via finite difference by computing the species' production rate twice at different temperatures. A finite difference strategy is useful here given that it avoids symbolically encoding the thermodynamics polynomials which can change for each species. In addition, differentiating with respect to temperature is less prone to numerical instability than differentiating with respect to species molar concentrations since the temperature is usually far from 0. \revone{The benefit of a numerical differentiation for gradients with respect to temperature was also recognized elsewhere \cite{mcnenly2015faster,walker2023generalized}.}    

\subsubsection{Validation}
\label{sec:val}
\begin{figure}
     \centering
     \begin{subfigure}[b]{0.35\textwidth}
         \centering
         \begin{tabular}{ |c|c| } 
          \hline
          & Symbolic AJ \\
          \hline
           13-4-\ce{CH4} & \revone{$6.5 \times 10^{-8}$ \%}  \\ 
          \hline
         35-18-\ce{N-C12H26} & \revone{$1.35\times 10^{-6}$ \%}  \\
        \hline
        55-33-\ce{N-C7H16}  & \revone{$4.33\times 10^{-4}$ \%}  \\ 
        \hline
         \end{tabular}
         \vspace{0.75cm}
     \caption{$\varepsilon_J$ computed for three different QSS-reduced chemistries with symbolic AJ.}
      \label{fig:symb_stats}
     \end{subfigure}
     \hfill
     \begin{subfigure}[b]{0.6\textwidth}
         \centering
           \includegraphics[width=\textwidth]{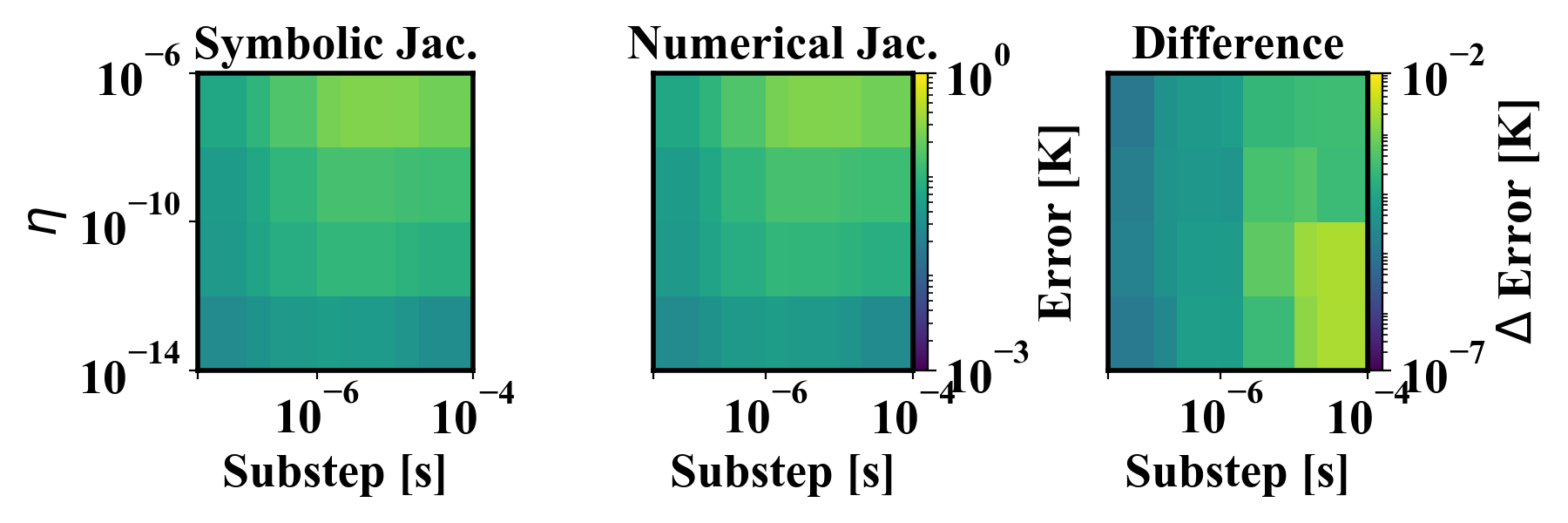}     
         \caption{\revone{Mean squared error of the temperature profiles obtained with the 108 0D ignitions, done with the symbolic analytic chemical Jacobian (left) and the Jacobian computed via finite differences (middle). Right: temperature error differences between the analytic and numerical Jacobian methods.}}
         \label{fig:symb_stab}
     \end{subfigure}
        \caption{Effect of symbolic analytical Jacobian construction on QSS-reduced chemistries integration.}
        \label{fig:symb}
\end{figure}

The symbolic AJ is first validated using the 0D tests starting from the same 100 random initial conditions as the ones reported in Sec.~\ref{sec:jac_simp}. Figure~\ref{fig:symb_stats} shows the accuracy of the symbolic AJ evaluation in terms of the relative Frobenius error (Eq.~\ref{eq:frobErr}). The Frobenius error is several orders of magnitude lower than the one shown for the simplified AJ (Fig.~\ref{fig:nosymb_stab_stats}). \revone{A separate study in~\ref{sec:conv} shows that the error between the symbolic and the finite difference Jacobian can be attributed to the discretization errors in the construction of the finite difference Jacobian.}

Next, the same 108 0D ignitions as in Sec.~\ref{sec:jac_simp} are reproduced with the symbolic AJ. The stability domain (domain over which the ODE integrator converged in less than 7200s) of the symbolic construction is the same as that of the finite differences one, and the errors incurred follow the same trend with respect to substep sizes and tolerances (Fig.~\ref{fig:symb_stab}).

Finally, the accuracy of the symbolic AJ is evaluated by estimating ignition delay values for the same $108$ initial conditions that were described and used in Sec.~\ref{sec:lin}. 
The ignition delay error is reported in terms of the rescaled mean absolute error as defined in Eq.~\ref{eq:ignErr}. Table~\ref{tab:ignErr} shows the ignition delay error between the skeletal mechanism and the reduced mechanism integrated with an FD Jacobian ($\varepsilon_{\rm ign,SK-QSS}$, first column), as well as the ignition delay error between the reduced mechanism integrated with an FD Jacobian and the symbolic AJ ($\varepsilon_{\rm ign,QSS-QSSAJ}$, second column). The QSSA affects the ignition delay by $8.97$\%. Meanwhile, the symbolic AJ exactly reproduces the ignition delay of that obtained with the FD Jacobian. Note that the ignition delay is accurate up to the size of the timestep used which ranges from $10^{-10}$s to $10^{-4}$s. The 108 experiments were reproduced each 5 times on a single CPU, and timed. The computational speedup induced by the use of an AJ is denoted by $Wt_{\rm QSS} / Wt_{\rm QSSAJ}$ in Tab.~\ref{tab:ignErr}. The error reported is the standard deviation of the speedup across the 108 realizations. It can be observed that the AJ strategy provides a significant speedup over an FD Jacobian as also noted elsewhere \cite{rouhi2011development}. The speedup is also consistent across all the realizations considered. The speedup appears to increase with the number of species which dictates the number of production rate evaluations needed to compute the numerical Jacobian. However, the speedup observed is nearly the same between the 13-4-\ce{CH4} and the 35-18-\ce{N-C12H26} mechanisms despite the fact that the latter mechanism contains three times as many non-QSS species. This observation points to the fact that the cost of \revtwo{evaluating} the analytic chemical Jacobian can vary depending on its density, and the computational complexity of each term.

\begin{table}[h!]
\centering
\begin{tabular}{ |c|c|c|c| } 
    \hline
          & $\varepsilon_{\rm ign,SK-QSS}$ & $\varepsilon_{\rm ign,QSS-QSSAJ}$ & $Wt_{\rm QSS} / Wt_{\rm QSSAJ}$ \\
    \hline
     13-4-\ce{CH4} & $4.02 \times 10^-3$ \% & $0$ \% & $1.35 \pm 0.118$ \\ 
    \hline
     35-18-\ce{N-C12H26} & $8.97$ \% & $0$ \% & $1.35 \pm 0.0665$ \\
    \hline
    55-33-\ce{N-C7H16} & $4.91$ \% & $0$ \% & $1.99 \pm 0.195$ \\ 
    \hline
    \end{tabular}
\caption{Ignition delay errors between the skeletal and the QSS mechanisms integrated with an FD Jacobian (first column); and the QSS mechanisms integrated with an FD or a symbolic AJ (second column). The last column denotes the speedup when using a symbolic AJ against a FDJ on a CPU.}
\label{tab:ignErr}
\end{table} 
    
In terms of memory usage, which can be critical for memory-limited computing hardware, the use of an AJ for reduced mechanisms can induce significant memory pressure. Table~\ref{tab:memQSS} shows the fraction of non-zero entries in the first N rows and N columns of the AJ, as obtained over 100 randomly sampled chemical states. For the 35-18-\ce{N-C12H26} mechanism and the 55-33-\ce{N-C7H16} mechanism, the second term in the right-hand side of Eq.~\ref{eq:JQSS} increases the number of non-zero entries by a factor of about 2. This loss in sparsity is due to the coupling of each reaction with possibly many other reactions that can be related to the molar concentration of a single QSS species (see Sec.~\ref{sec:interdependencies}). 

\begin{table}[h!]
\centering
\begin{tabular}{ |c|c|c|c| } 
    \hline
          & Symbolic AJ & Constant QSS AJ  &  Zero QSS AJ\\
    \hline
    13-4-\ce{CH4} & $0.489$ & $0.468$ & $0.38$ \\ 
    \hline
    35-18-\ce{N-C12H26} & $0.447$ & $0.241$ & $0.176$ \\
    \hline
    55-33-\ce{N-C7H16}  & $0.357$ & $0.183$ & $0.159$ \\ 
    \hline
    \end{tabular}
\caption{Fraction of non-zero entries in the AJ using an exact symbolic method, or approximated as in Sec.~\ref{sec:jac_simp}.}
\label{tab:memQSS}
\end{table}

\section{Performance improvements for the chemical Jacobian}
\label{sec:perfOpt}
Assembling a chemical Jacobian is computationally expensive in general, irrespective of whether the QSS approximation is used. In terms of computational intensity, each term of the AJ requires evaluating the derivative of several species' production rates. In terms of memory requirements, a matrix of size $(N+1) \times (N+1)$ needs to be stored in memory for every computational cell in the domain. The use of the QSSA can further result in a relatively dense Jacobian (Tab.~\ref{tab:memQSS}) which increases the number of entries to compute. As mentioned in Sec.~\ref{sec:symbCons}, the use of the symbolic method needs to be paired with CSE. However, the precomputed expressions are additional scalars that must be stored in memory. In the case of the 35-18-\ce{N-C12H26} mechanism, $16487$ scalars need to be stored in memory which is nearly 13 times more than the number of entries in the chemical Jacobian matrix.

In this section, several strategies to mitigate memory use and reduce the computational of the AJ construction are outlined. Their effect is then quantified in Sec.~\ref{sec:exascale}.

\subsection{Memory footprint reduction}
\label{sec:perfmem}
In the case of a symbolic AJ that uses CSE, the memory use is primarily impacted by the number of common subexpressions that need to be precomputed. Reducing the number of common subexpressions can be done by simply replacing these expressions in the final Jacobian entry computation. However, if the common subexpressions are naively replaced, one will simply negate the benefits of CSE (Sec.~\ref{sec:symbCons}). Instead, the common subexpressions must be strategically replaced depending on the amount of added computation that they will incur.

\subsubsection{Replacement based on operation count}
It is proposed to select the common subexpressions to replace based on the number of additional operations incurred if the common subexpression is replaced. The replacement procedure is shown in Algo.~\ref{alg:subexp_rep}.

\begin{algorithm}
\caption{Algorithm of common subexpression replacement}\label{alg:subexp_rep}
\begin{algorithmic}[1]
\State User inputs a threshold of maximum operation count $n_{\rm op}$
\For{$\rm{opThreshold} = 1:n_{\rm op}$} (Loop 1)
    \For{$\rm{subexp} \in \rm{allSubexpressions}$} (Loop 2)
         \State $\rm{totalOp}_{\rm subexp} = (\rm{nUse}_{\rm subexp} - 1) \times \rm{nOp}_{\rm subexp}$ (number of operations added if $\rm{subexp}$ is eliminated)
         \If{$\rm{totalOp}_{\rm subexp} < \rm{opThreshold}$}
            \State Replace $\rm{subexp}$
         \EndIf
    \EndFor
\EndFor
\end{algorithmic}
\end{algorithm}

In Loop 2 of Algo.~\ref{alg:subexp_rep}, for each common subexpression "$\rm{subexp}$", one first counts the number of times the common subexpression is used ($\rm{nUse}_{\rm subexp}$) and counts the number of operations (addition, multiplication, exponential, power, etc.) in the common subexpression $\rm{nOp}_{\rm subexp}$. By replacing the common subexpression in the rest of the AJ assembling function, one would trade the precomputation of one scalar with an added $(\rm{nUse}_{\rm subexp} - 1) \times \rm{nOp}_{\rm subexp}$ operations. If the number of additional operations is less than the threshold $\rm{opThreshold}$, the common subexpression is replaced.

The role of Loop 1 of Algo.~\ref{alg:subexp_rep}, is to ensure that the number of common subexpressions eliminated grows as operation count threshold  $n_{\rm op}$ inputted by the user increases. In our experience, this loop does not necessarily lead to a higher number of common subexpression eliminated but provides an intuitive control for memory footprint.

\subsubsection{Recycling of common subexpression}
The symbolic expressions constructed with the symbolic method can be divided into two categories: 1) common subexpressions, 2) AJ entries. The common subexpressions are computed sequentially and the role of some of the earlier CSEs is to simplify the computation of future CSEs. In this case, there is an opportunity to reuse precomputed common subexpressions' memory space once they are not useful for subsequent calculations. This method is termed~\textit{subexpression recycling} and is described in Algo.~\ref{alg:subexp_recycle}. 

\begin{algorithm}
\caption{Algorithm of subexpression recycling}\label{alg:subexp_recycle}
\begin{algorithmic}[1]
\For{$\rm{subexp} \in \rm{allSubexpressions}$}
    \If{$\rm{subexp}$ not in Jacobian terms}
    \State Find $\rm{last_{\rm{subexp}}}$ the last subexpression that uses $\rm{subexp}$
    \State Reuse $\rm{subexp}$ and eliminate $\rm{last_{\rm{subexp}}}$
    \EndIf
\EndFor
\end{algorithmic}
\end{algorithm}

Algorithm~\ref{alg:subexp_recycle} simply consists in looping through each common subexpression $\rm{subexp}$ and identifying the ones that are not directly used in the AJ entries. By finding the last common subexpression that uses $\rm{subexp}$, one can reuse $\rm{subexp}$ instead of allocating an additional scalar.

\subsection{Computational cost reduction on GPU}
\label{sec:perfcomp}
Assembling the chemical Jacobian is a computationally intensive process that requires several non-elementary operations, such as using power operations or exponential. Several best practices for writing GPU-ready code include transforming some non-elementary operations \cite{guide2013cuda}. In particular, instead of using a \verb|pow| operation for small integer powers (2 or 3), it can be more efficient to use explicit multiplication. Likewise, transforming powers of $10$ with base-$10$ exponentiation was listed in Ref.~\cite{guide2013cuda} as possibly advantageous. In Sec.~\ref{sec:exascale}, the effect of these two optimizations on computing cost is explored.

\section{Deployment on Exascale machines}
\label{sec:exascale}
In previous sections, the validation of the symbolic approach for chemical Jacobian generation was done using 0D ignition. Here the focus is on exercising the symbolic chemical Jacobian constructed on a problem that can run on exascale-ready HPC clusters. 

\subsection{Configuration}

\begin{figure}
    \centering
    \begin{subfigure}[t]{0.24\textwidth}
        \centering
        \includegraphics[width=\textwidth]{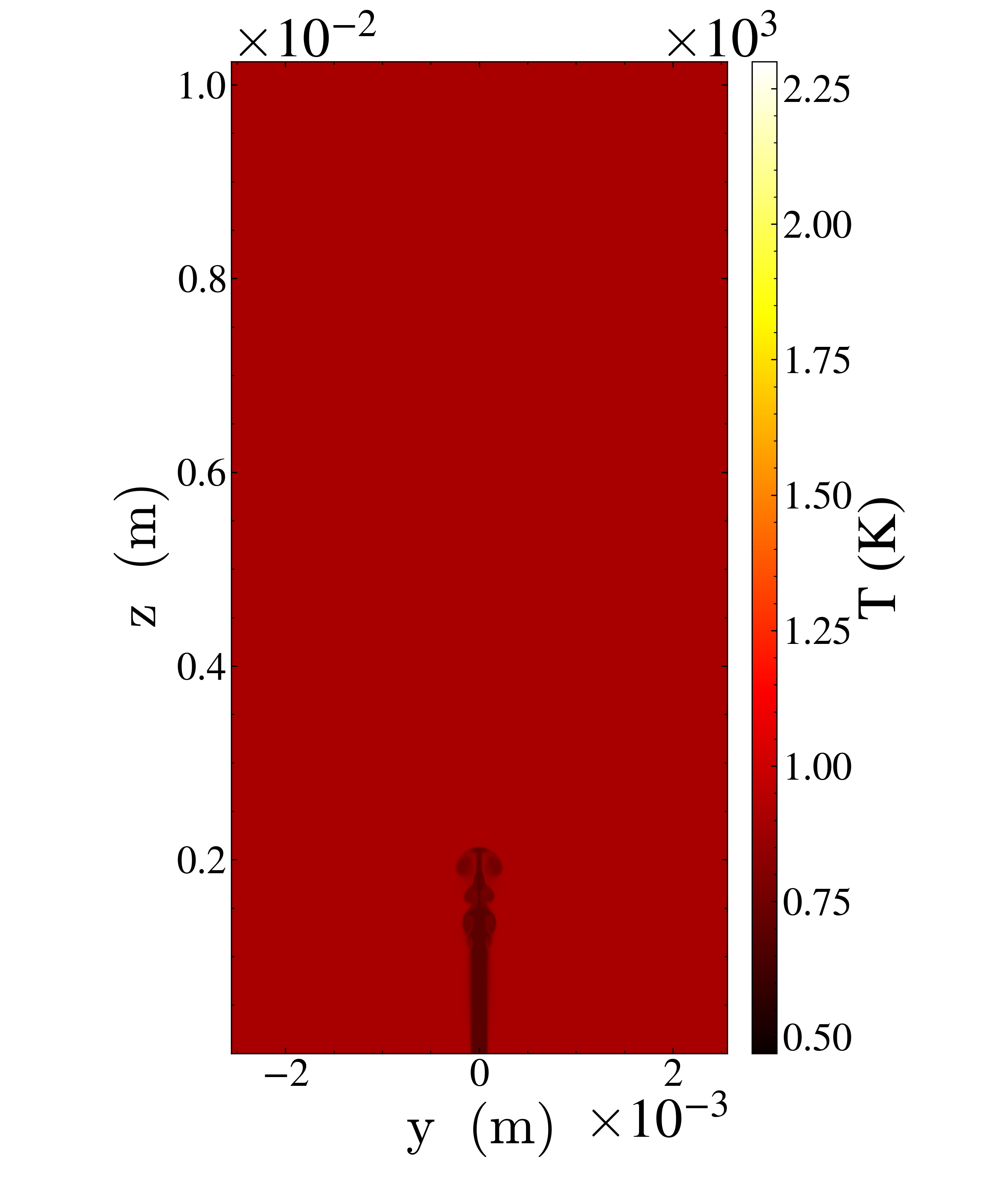}
        \caption{$t = 0.25$ms. 470K Dodecane jet enters methane-air background mixture at 900K.}
        \label{subfig:aj_25}
    \end{subfigure}
    \hfill
    \begin{subfigure}[t]{0.24\textwidth}
        \centering
        \includegraphics[width=\textwidth]{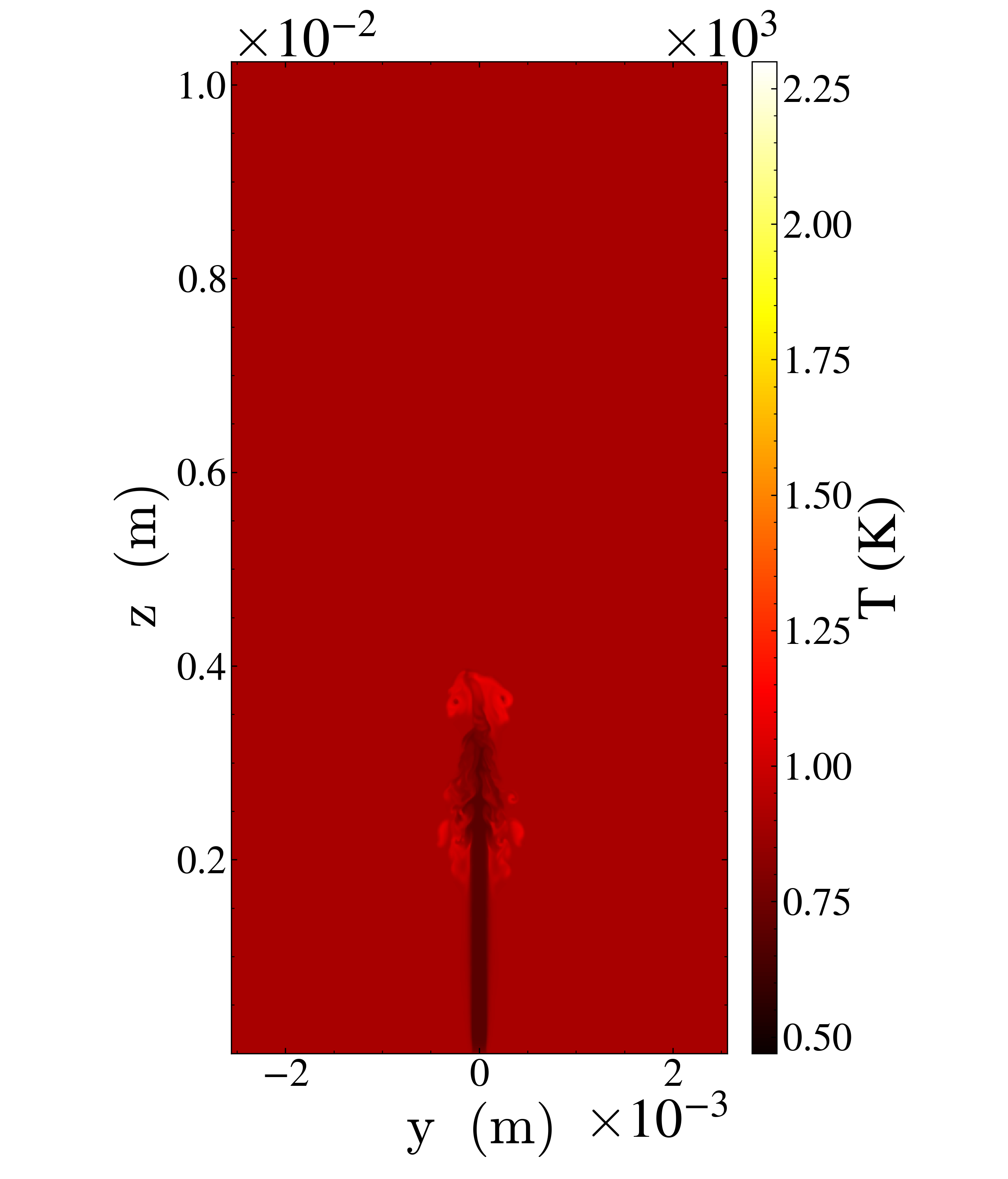}
        \caption{$t = 0.5$ms. The dodecane jet forms a shear mixing layer with the background.}
        \label{subfig:aj_50}
    \end{subfigure}
    \hfill
    \begin{subfigure}[t]{0.24\textwidth}
        \centering
        \includegraphics[width=\textwidth]{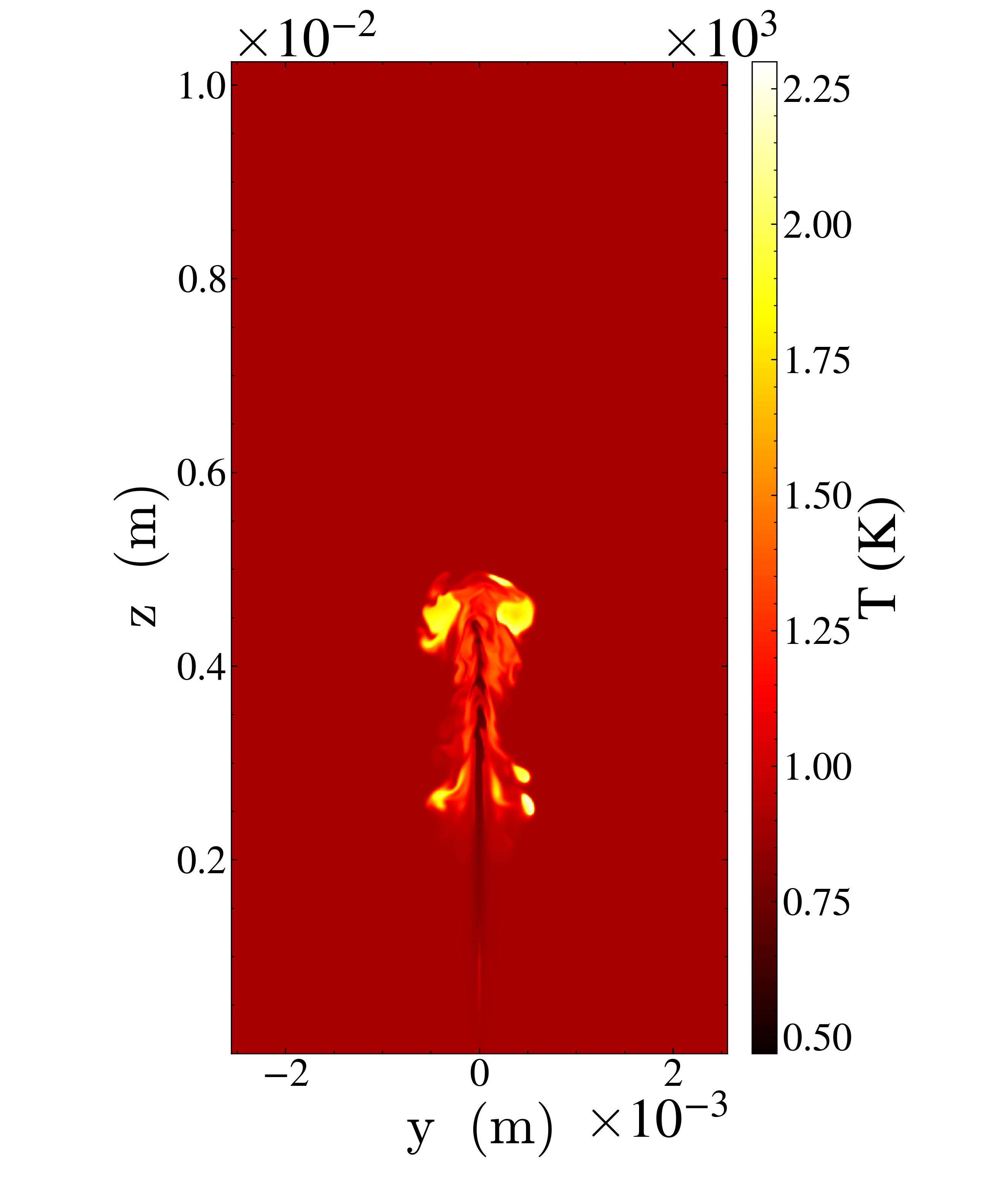}
        \caption{$t = 0.75$ms. Onset of auto-ignition along the shear interface.}
        \label{subfig:aj_75}
    \end{subfigure}
    \hfill
    \begin{subfigure}[t]{0.24\textwidth}
        \centering
        \includegraphics[width=\textwidth]{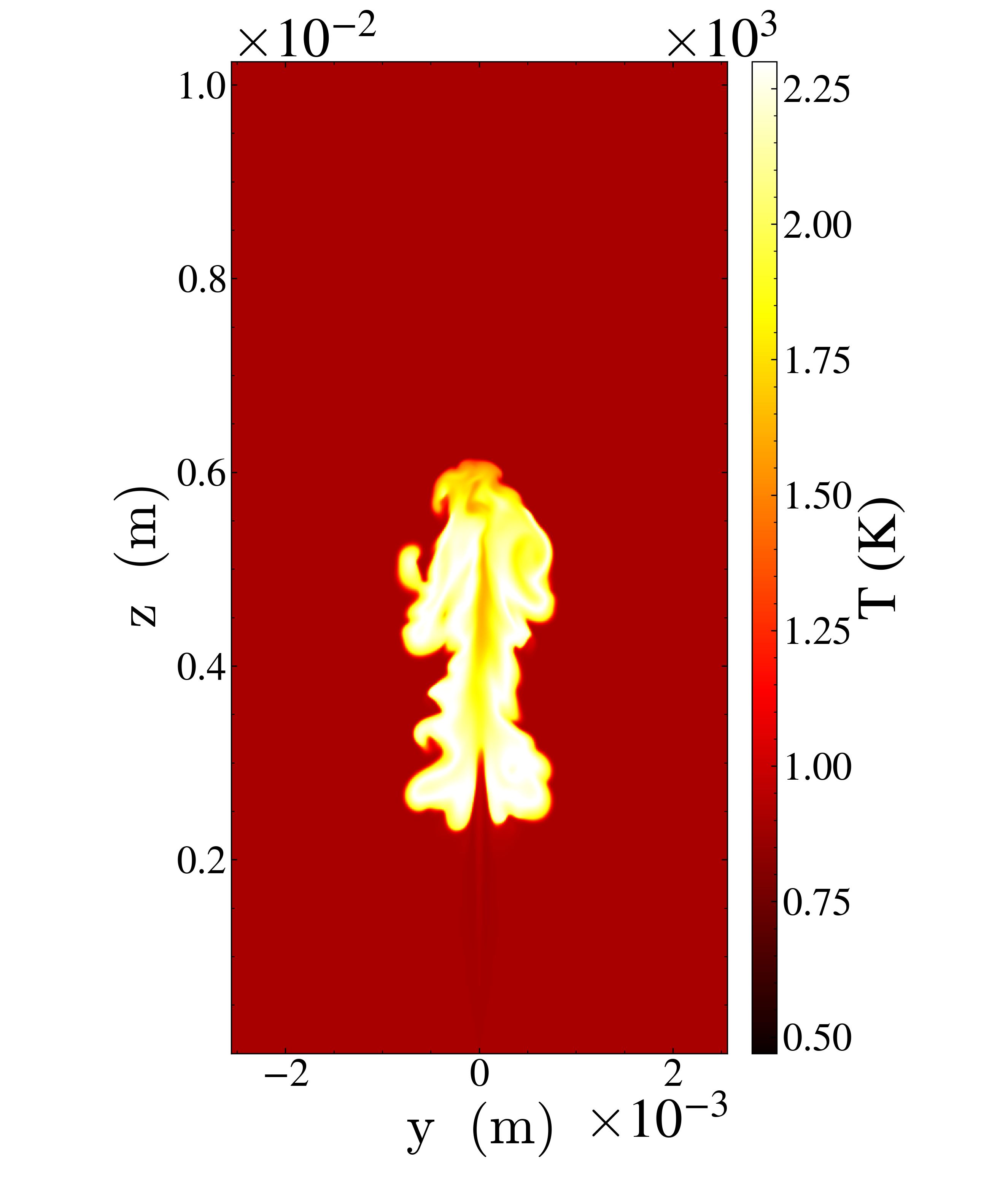}
        \caption{$t = 1.0$ms. Fully-developed flame-front propagating into the domain.}
        \label{subfig:aj_100}
    \end{subfigure}
    \hfill
    \begin{subfigure}[t]{0.24\textwidth}
        \centering
        \includegraphics[width=\textwidth]{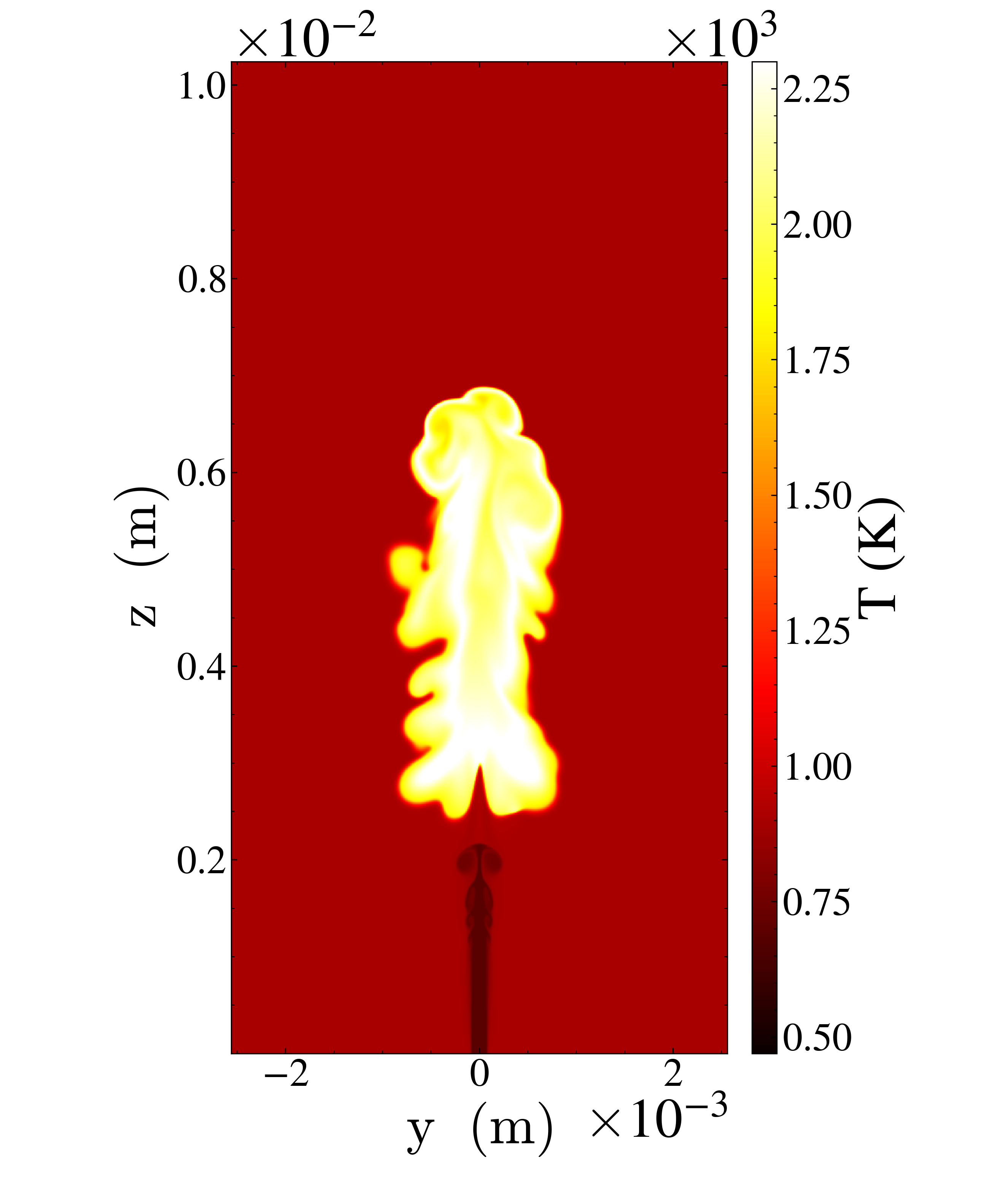}
        \caption{$t = 1.25$ms. The second dodecane jet pulse forms behind the first flame front.}
        \label{subfig:aj_125}
    \end{subfigure}
    \hfill
    \begin{subfigure}[t]{0.24\textwidth}
        \centering
        \includegraphics[width=\textwidth]{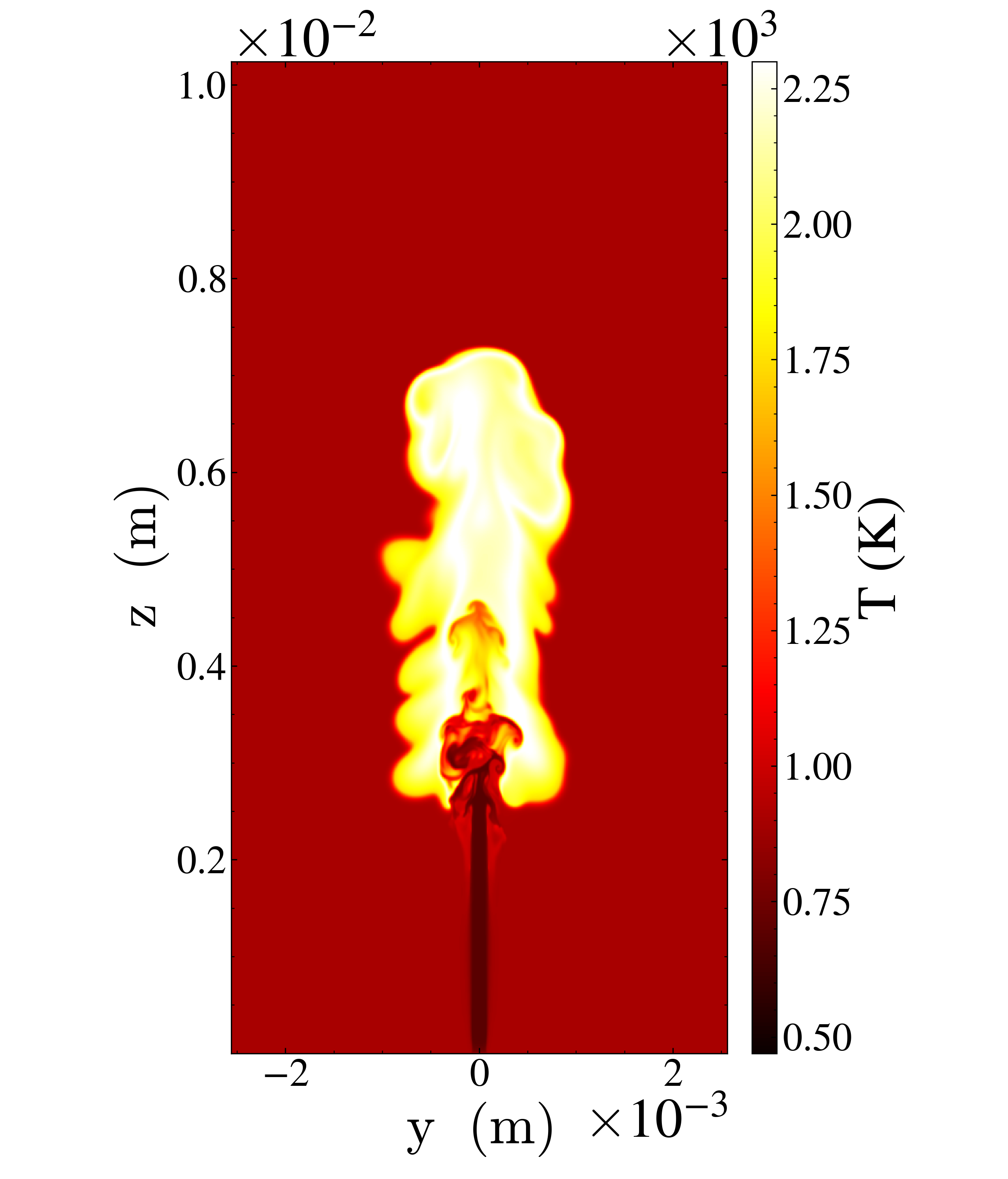}
        \caption{$t = 1.5$ms. The second fuel stream mixes with the first flame and ignites.}
        \label{subfig:aj_150}
    \end{subfigure}
    \hfill
    \begin{subfigure}[t]{0.24\textwidth}
        \centering
        \includegraphics[width=\textwidth]{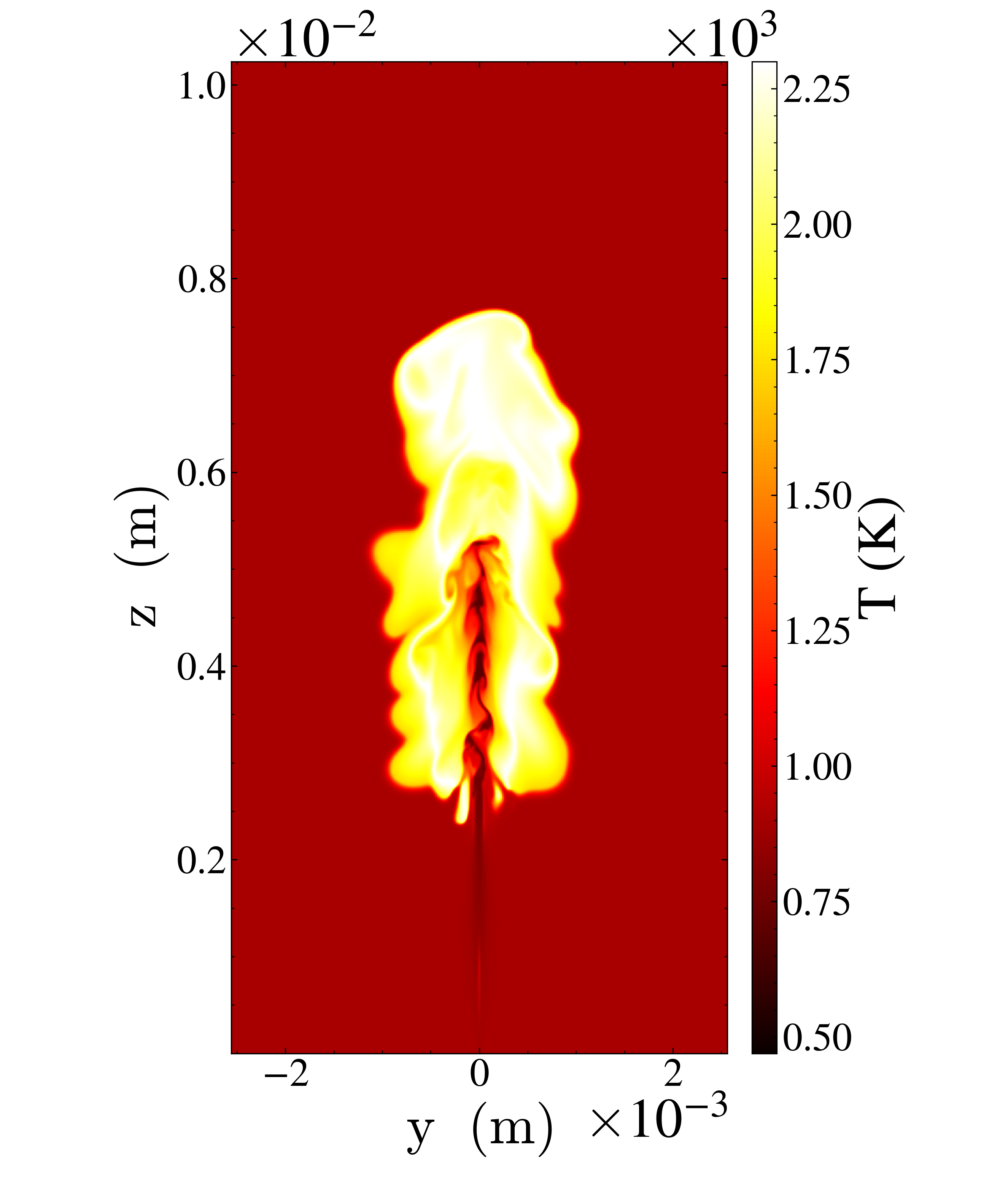}
        \caption{$t = 1.75$ms. Further development of secondary ignition process.}
        \label{subfig:aj_175}
    \end{subfigure}
    \hfill
    \begin{subfigure}[t]{0.24\textwidth}
        \centering
        \includegraphics[width=\textwidth]{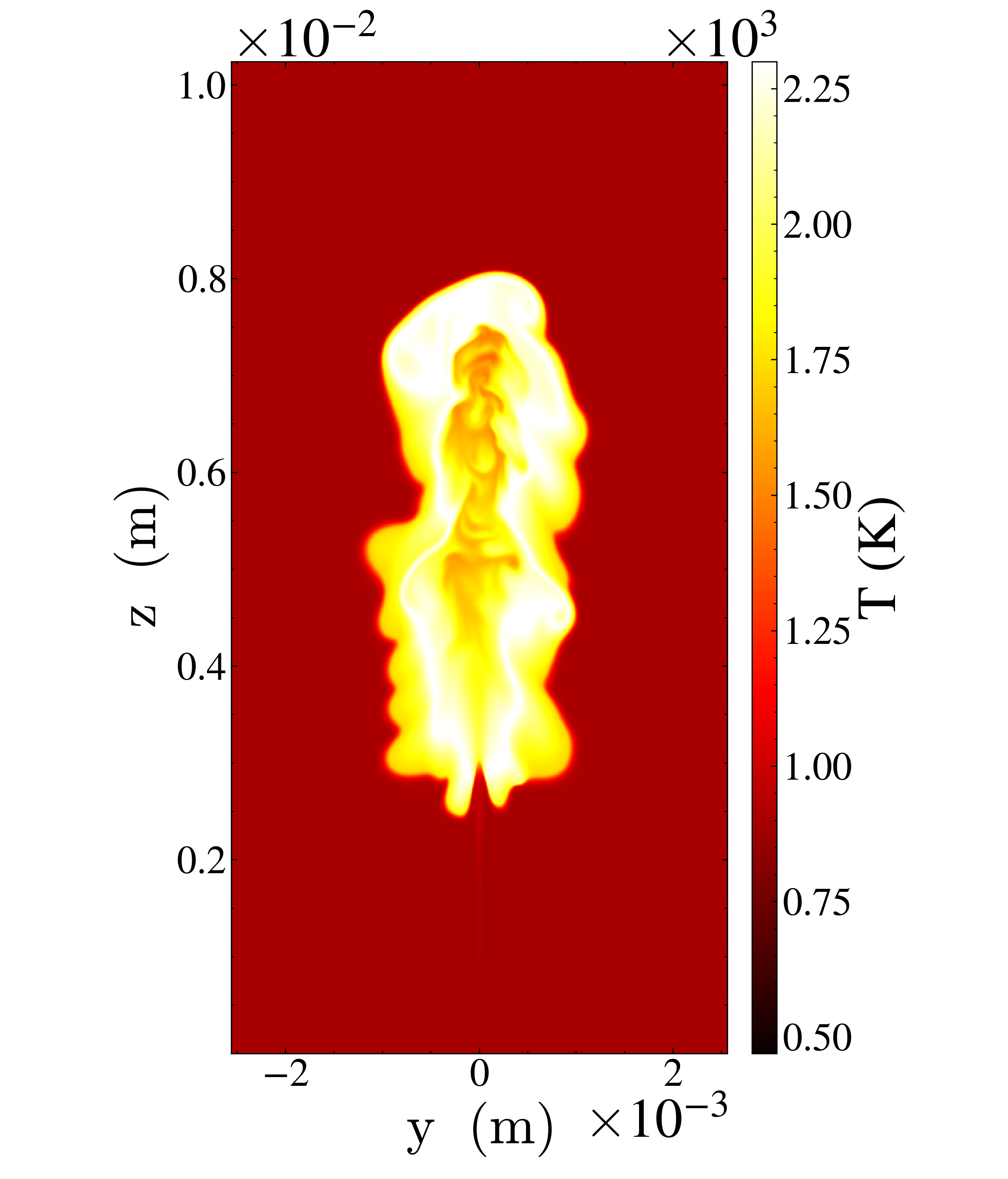}
        \caption{$t = 2.0$ms. The fully developed combined flame fronts propagate through the domain and continue the combustion process.}
        \label{subfig:aj_200}
    \end{subfigure}
    \caption{PeleLMeX multipulse n-dodecane fuel jet simulation using the analytic Jacobian QSS formulation at 0.25ms increments during the simulation.}
    \label{fig:schematic_multipulse}
\end{figure}

The test-case simulated is a 3D multi-pulse fuel injection of pure N-\ce{C12H26} fuel into a premixed oxidizing air-fuel mixture with an equivalence ratio of $\phi = 0.5$ that consists of air and \ce{CH4} at 900K and 60 bars. The multi-pulse scenario includes a first short pulse of the N-\ce{C12H26} into the prefilled chamber. As the jet develops, it entrains the preheated \ce{CH4}-air mixture and the first pulse ignites. After 0.5 ms, the first pulse injection stops. After a dwell time of an additional 0.5 ms, a second pulse of pure N-\ce{C12H26} fuel (same composition as the first pulse) enters the domain at 1 ms. The second pulse lasts 0.5 ms and interacts with the burning gases from the first pulse combustion process and ignites in turn. The simulation proceeds until there is significant combustion of both jet pulses after about 2 ms. The entire ignition sequence is illustrated in Fig.~\ref{fig:schematic_multipulse}. This test case was chosen since it is representative of many engineering combustion problems that include turbulent mixing of fuel and oxidizer on length scales that are typical of internal combustion engines.

The case was simulated on the Crusher exascale-ready HPC cluster located at Oak Ridge National Laboratory. Crusher is a testbed for the DOE first Exascale platform, Frontier. introduced in 2023. Each of the compute nodes on the Crusher machine consists of one 64-core AMD ``Optimized 3rd Gen EPY'' CPU and four AMD MI250X, each with 2 graphics compute dies that are equivalent to 8 standard GPUs per node. The case was simulated using the PeleLMeX solver \cite{Esclapez2023} which is an exascale capable \gls{amr} based low-Mach solver for reacting flow simulations. The simulations presented in this study used a total of 4 levels of \gls{amr} that refine based upon the vorticity and temperature gradient in order to capture the development of the shear layer of the incoming fuel stream as it mixes with the background oxidizer as well as adequately capture the start of the ignition process. Because the simulations are all performed using \gls{amr}, the total number of grid cells increases as the simulation progresses with the number of cells approximately 250 million during the onset of the ignition process. All of the cases in this study were simulated using between 8 and 30 Crusher nodes (64 -- 240 GPUs) with fewer nodes required during the start of the simulation and the number of nodes increasing as the flow developed and during the ignition process. No subgrid-scale model is used in any of the simulations and the chemistry is modeled using a detailed N-\ce{C12H26} mechanism. In the following, either the 53-species skeletal N-\ce{C12H26} mechanism \cite{Yao2017} or the QSS reduced 35-species \ce{N-C12H26} mechanisms \cite{borghesi2018direct} are used.

\subsection{Results}
\label{sec:exa_results}

Three cases are run to ensure that the symbolic Jacobian does not affect the simulation accuracy. All three cases rely on SUNDIALS CVODE integrator, but use different linear solvers within the Newton method CVODE employs to solve the discretized nonlinear system of ODEs. Case 1 uses the skeletal mechanism \cite{Yao2017} and its integration rely on an iterative Jacobian-free method (GMRES); Case 2 uses the QSS-reduced mechanism \cite{borghesi2018direct} combined with an iterative Jacobian-free method (GMRES); Case 3 uses the QSS-reduced mechanism \cite{borghesi2018direct} combined with a dense direct linear solve from the MAGMA library \cite{tomov2010dense,tomov2010towards,dongarra2014accelerating} which uses an analytic Jacobian constructed symbolically.

\begin{figure}
    \centering
    \begin{subfigure}[t]{0.33\textwidth}
        \centering
        \includegraphics[width=\textwidth]{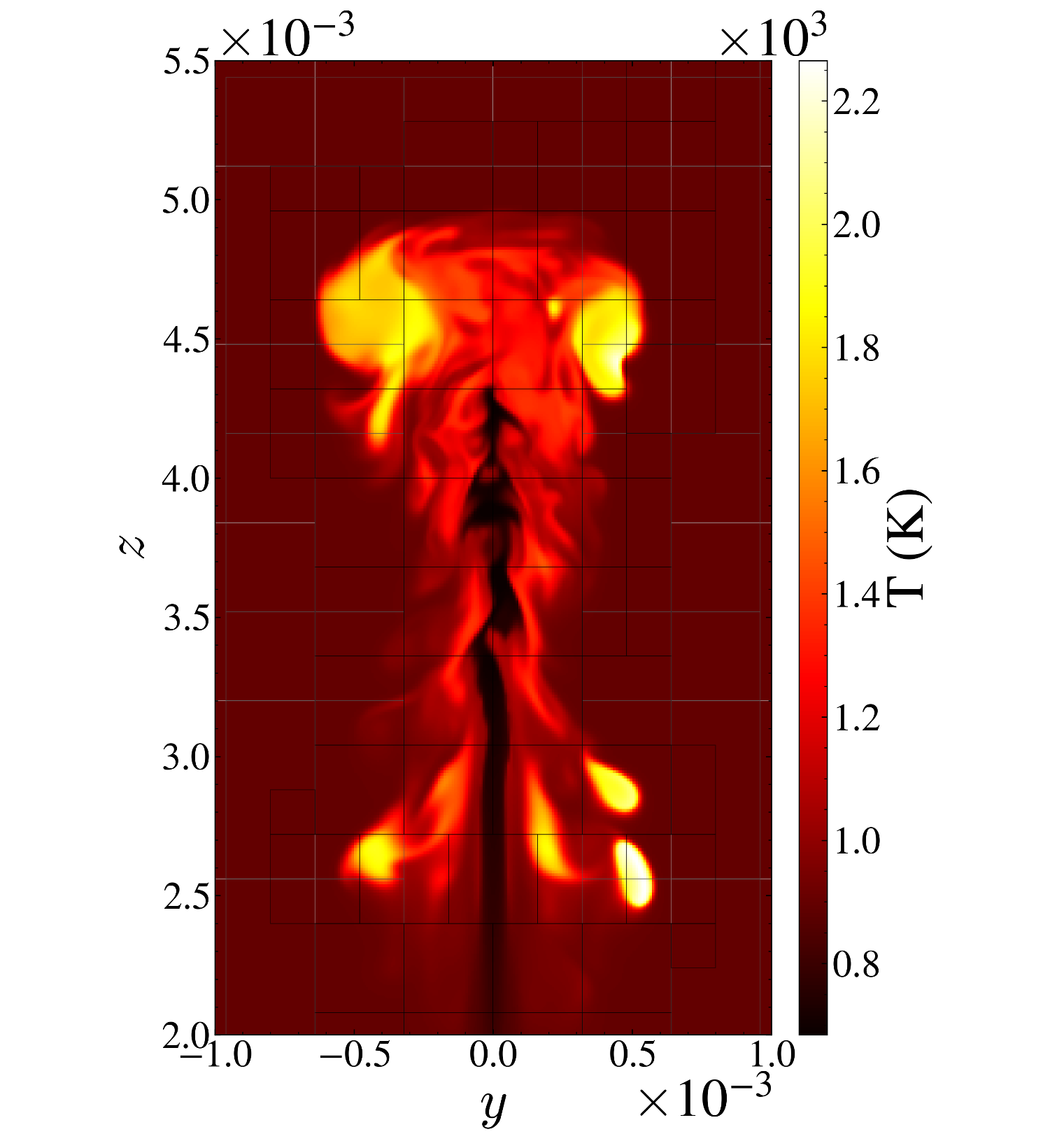}
        \caption{$t = 0.75$ms. Skeletal mechanism using a Jacobian-free iterative integration (Case 1).}
        \label{subfig:close_sk_75}
    \end{subfigure}
    \hfill
    \begin{subfigure}[t]{0.33\textwidth}
        \centering
        \includegraphics[width=\textwidth]{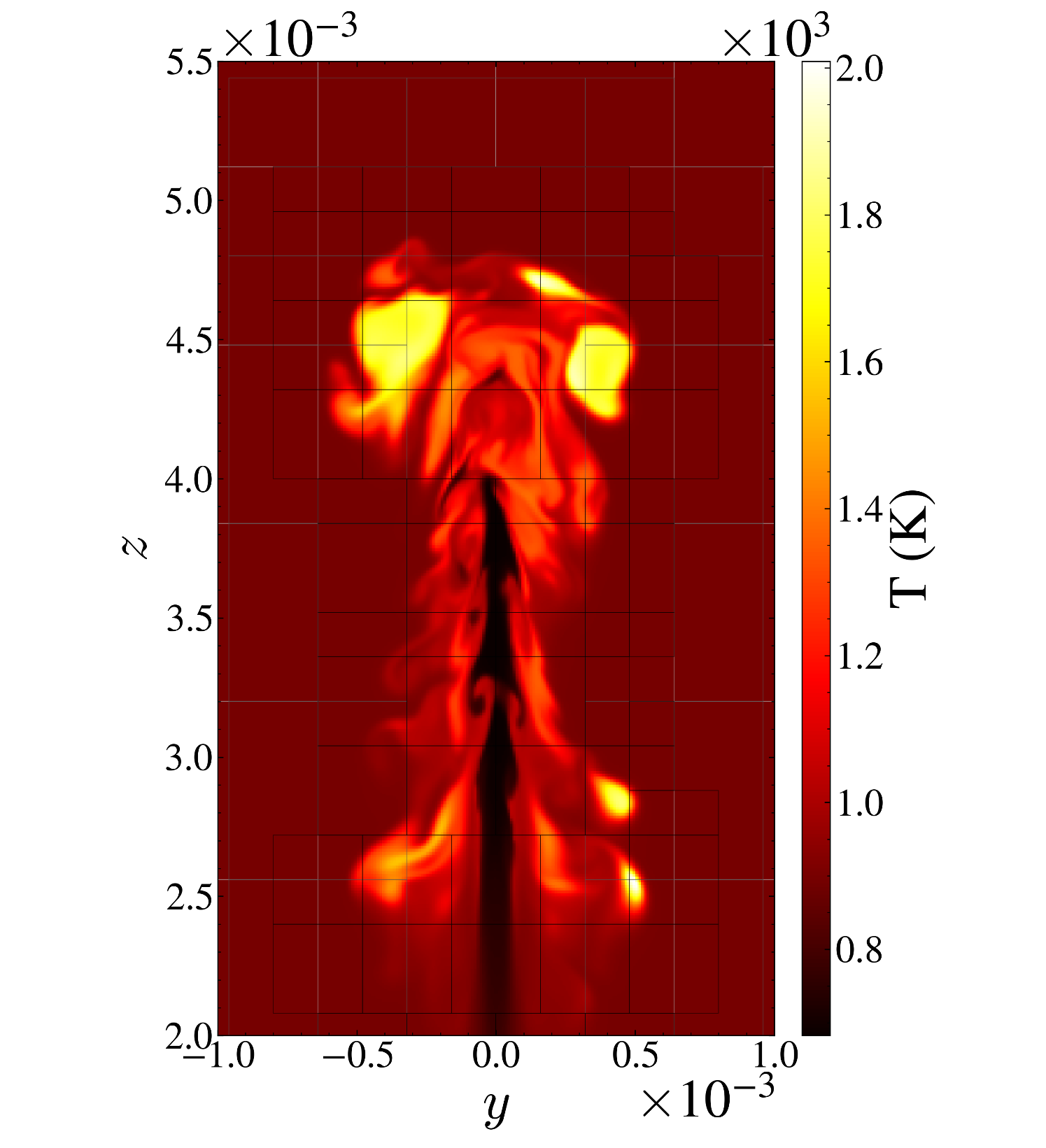}
        \caption{$t = 0.75$ms. QSS mechanism using a Jacobian-free iterative integration (Case 2).}
        \label{subfig:close_gmres_75}
    \end{subfigure}
    \begin{subfigure}[t]{0.33\textwidth}
        \centering
        \includegraphics[width=\textwidth]{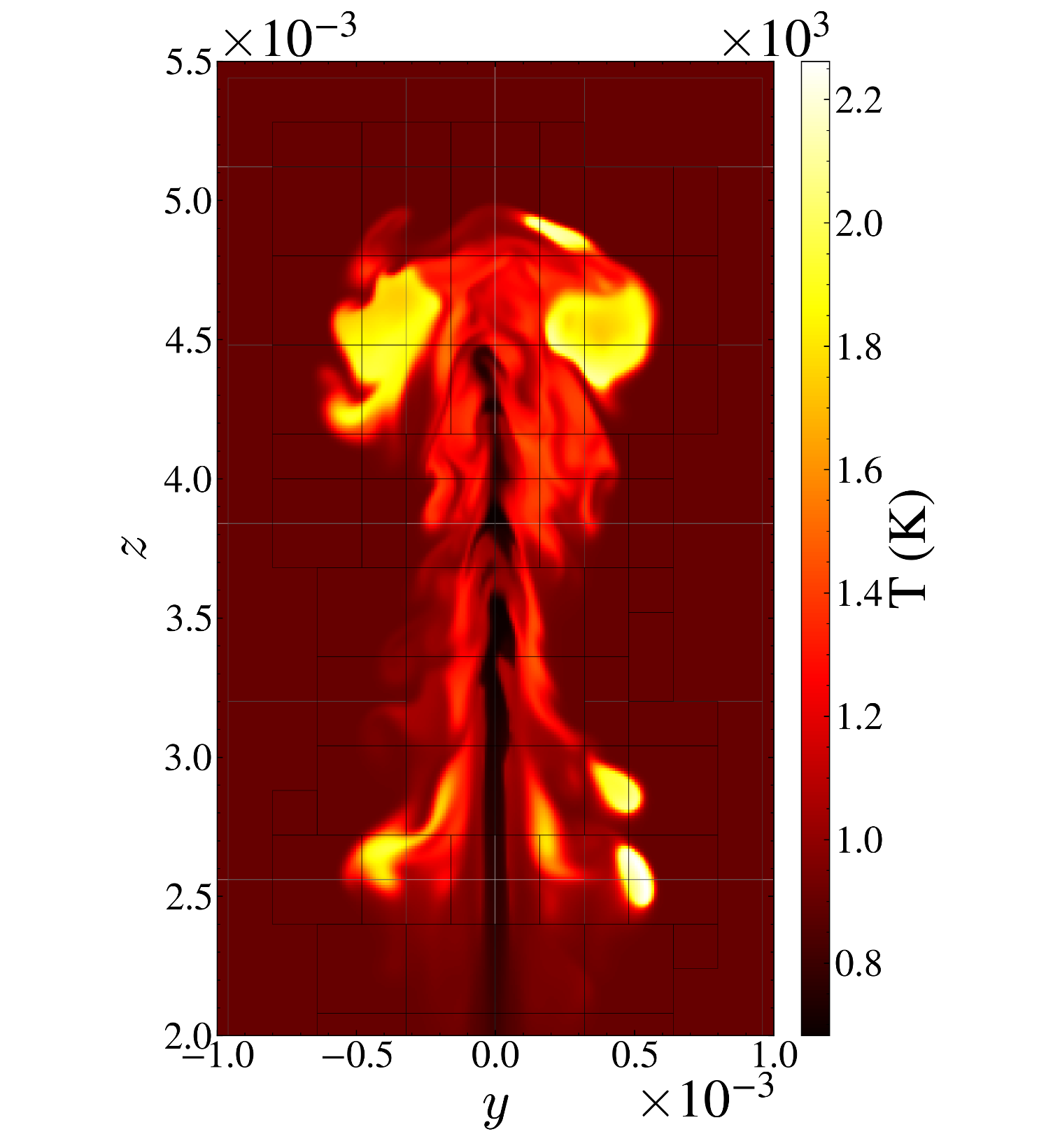}
        \caption{$t = 0.75$ms. QSS mechanism using a symbolic analytic Jacobian (Case 3).}
        \label{subfig:close_aj_75}
    \end{subfigure}
    \caption{Close-up comparison of the start of ignition for the three different solution methods.}
    \label{fig:mech_comparison_ignition_close}
\end{figure}

Figure~\ref{fig:mech_comparison_ignition_close} shows a close-up view of the start of ignition for each of the three cases after 0.75 ms. \revone{Given the chaotic nature of the system investigated, any infinitesimal perturbations even due to rounding errors \cite{senoner2008growth} will exponentially amplify over time \cite{hassanaly2019ensemble, hassanaly2019lyapunov}. Therefore, it is not possible to guarantee an exact agreement of instantaneous snapshots. The snapshots in Fig.~\ref{fig:mech_comparison_ignition_close} can only be used for qualitative assessment and show reasonable agreement between all three cases in terms of the location and sizes of the onset of the ignition kernels}. A key observation in these images is that both of the QSS mechanisms' run develop an ignition kernel at the leading edge of the penetrating jet, whereas the skeletal mechanism only develops two ignition kernels along the sheer vortex ring and three ignition kernels at the lower portions of the shear jet layer. Overall, the use of the analytic Jacobian formulation leads to accuracy levels on par with Jacobian-free iterative methods.

\begin{figure}
    \centering
    \includegraphics[width=\textwidth]{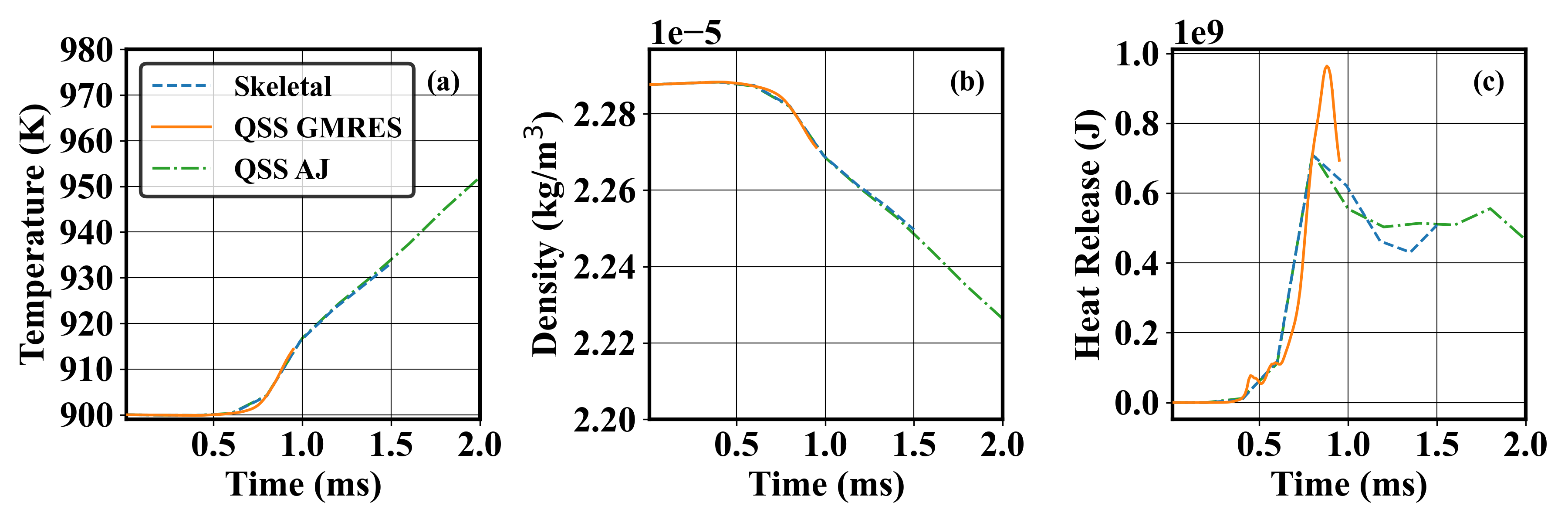}
    \caption{Comparison of the domain averaged (a) temperature, (b) density, and (c) heat release for the skeletal mechanism using GMRES, the QSS mechanism using GMRES, and the QSS mechanism using the analytic Jacobian formulation.}
    \label{fig:mech_comp}
\end{figure}

Quantitatively, Figure~\ref{fig:mech_comp} shows the comparison of the domain average fields for: (a) temperature, (b) density, and (c) heat release for the three different mechanisms examined. The temperature and density averages are consistent among the different chemical mechanisms as these quantities are primarily dependent on the mixing between the fuel stream and background mixture. The integrated heat release tracks the ignition sequence. While all cases ignite similarly, Case 2 (QSS mechanism integrated with a Jacobian-free method) shows an initial spike in the heat release that is larger than the other cases. Case 3 (QSS mechanism with analytic Jacobian formulation) shows results that are closer to the original skeletal mechanism which suggests that the initial heat release spike may be the results of a numerical inaccuracy during the chemistry integration. 

For both mechanisms integrated with a Jacobian-free method (Case 1 and Case 2), the simulation was stopped before reaching 2 ms because of numerical instabilities that occurred during the ignition process. In practice, the cases would need to be restarted with a smaller timestep, or with a higher number of GMRES iterations. These instabilities were not observed when using the analytic Jacobian formulation and the solution was able to progress smoothly up to 5 ms without failure. This observation echos the 0D integration results shown in Fig.~\ref{fig:stab} which motivated the construction of the AJ.

In terms of computational time, significant differences were observed between all three cases. After the onset of ignition, Case 1 took approximately 162 seconds per timestep, Case 2 took 80 seconds per timestep, and Case 3 took 44 seconds per timestep. As could be expected, Case 2 was approximately twice as fast as Case 1 primarily because a smaller number of species needed to be transported. \revtwobarred{and because the QSS assumption reduced the stiffness of the mechanism which allowed for fewer GMRES iterations for a convergent solution. }Case 3 was approximately 1.8 times faster than Case 2 while leading to higher robustness as discussed in the previous paragraph. Note that the robustness of Case 2 could be improved by increasing the number of iterations for the chemistry integration allowed per timestep (capped at 10000), at the cost of further widening the computational cost between Case 2 and Case 3. The computational speedup and robustness of Case 3 are particularly clear in the case simulated. However, they can be expected to be less clear when using a compressible solver where the fluid timestep size is small enough to make the chemical Jacobian diagonally dominant.

\subsection{Performance optimization}
\label{sec:perf}

In this section, the effect of the performance optimization methods described in Sec.~\ref{sec:perfOpt} is evaluated using the multipulse ignition case described in Sec.~\ref{sec:exascale}. In total, 24 different formulations of the chemical Jacobian function were tested by varying the operation count threshold on common subexpression replacement ($n_{\rm op}$ in Algo.~\ref{alg:subexp_rep}) in the set \{ 0, 10, 20 \}, by either recycling or not recycling subexpressions (Algo.~\ref{alg:subexp_recycle}), by either replacing small integer power with multiplications or not, by either replacing powers of 10 with base-10 exponentials or not.

For each of the 24 mechanisms, Case 3 (see Sec.~\ref{sec:exa_results}) was simulated for 10 timesteps on 30 nodes (240 GPUs) starting from t=0.75 ms. The criteria used to evaluate each of the 24 cases are 1) the total computational time needed by the chemistry integration module (which dominates the computational cost) and 2) the number of common subexpressions precomputed. The effect of each optimization method is obtained by computing the relative compute time and the relative memory use between cases that are the same, except for the one optimization parameter being investigated.

\subsubsection{Effect of memory footprint optimization}

\begin{figure}[h]
    \centering
    \includegraphics[width=0.4\textwidth]{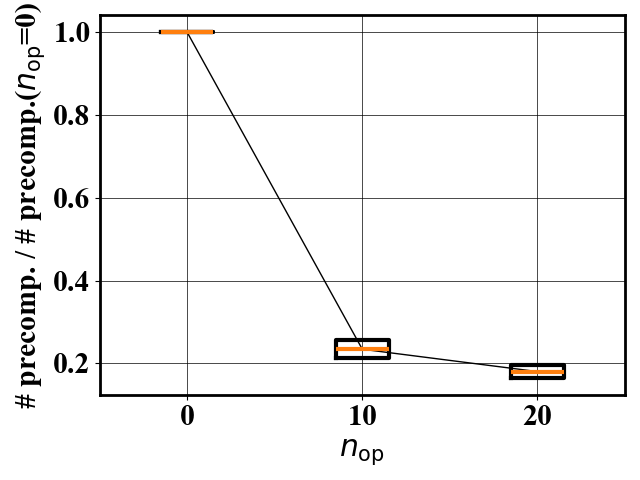}
    \includegraphics[width=0.4\textwidth]{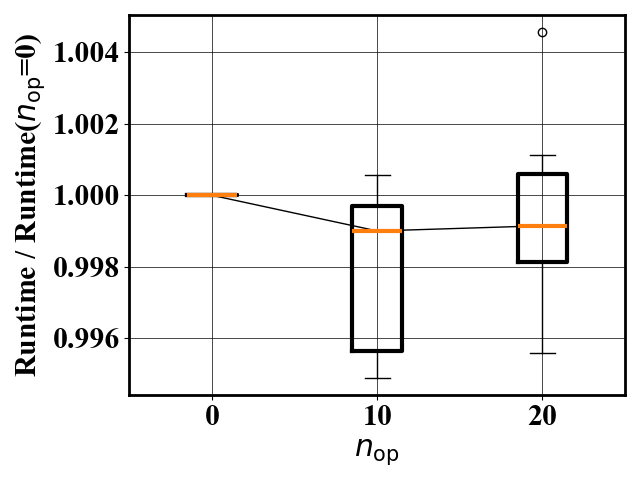}
    \caption{Effect of $n_{\rm op}$ in Algo.~\ref{alg:subexp_rep} on the number of precomputed subexpressions (left) and the computing time of the chemistry integration (right). The orange line denotes the median, and the box plot denotes the 95\% confidence interval.}
    \label{fig:effect_subexp_rep}
\end{figure}

The effect of Algo.~\ref{alg:subexp_rep} is shown in Fig~.\ref{fig:effect_subexp_rep} in terms of memory use (left) and computing time (right). It can be observed that using $n_{\rm op}=20$ can reduce the number of precomputed common subexpressions by a factor of 5 to 10 which brings the number of precomputed subexpressions close to the number of entries in the Jacobian matrix. While small values of $n_{\rm op}$ drastically reduce the number of precomputed subexpressions, the number of precomputed expressions decreased sublinearly as $n_{\rm op}$ further increases. In terms of computing time, $n_{\rm op}$ has almost no effect on chemistry integration computational intensity for the time range considered. This result suggests that the highest value of $n_{\rm op}$ investigated can be used when deploying QSS-reduced chemistries on exascale-ready HPC platforms. 

\begin{figure}[h]
    \centering
    \includegraphics[width=0.4\textwidth]{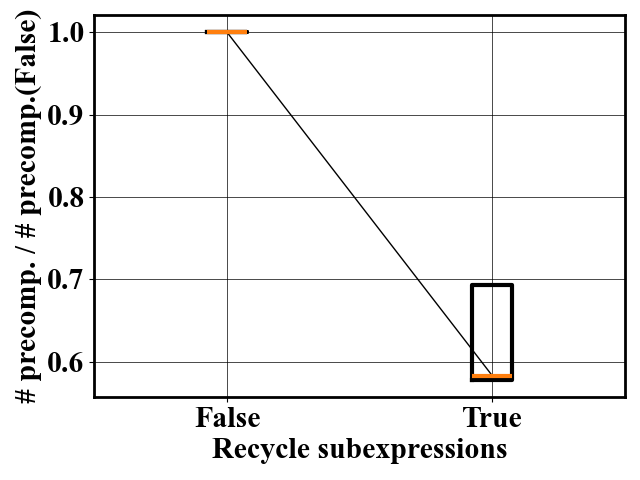}
    \includegraphics[width=0.4\textwidth]{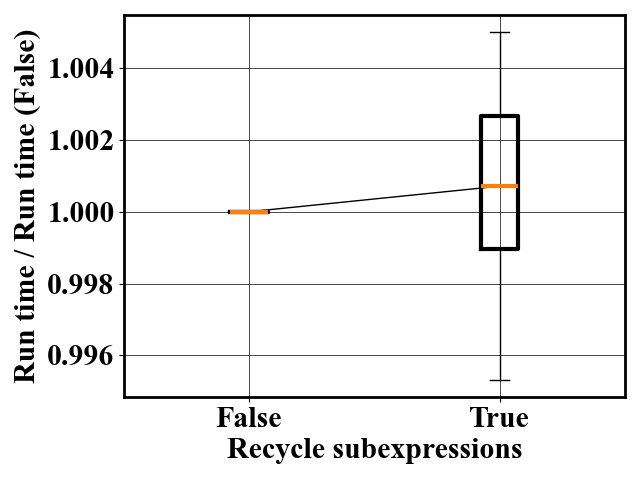}
    \caption{Effect of Algo.~\ref{alg:subexp_recycle} on the number of precomputed subexpressions (left) and the computing time of the chemistry integration (right). The orange line denotes the median, and the box plot denotes the 95\% confidence interval.}
    \label{fig:effect_subexp_rec}
\end{figure}

Figure~\ref{fig:effect_subexp_rec} suggests that Algo.~\ref{alg:subexp_recycle} also has a significant effect on memory use since it tends to reduce the number of precomputed expressions almost by 45\%. Similar to Algo.~\ref{alg:subexp_rep}, it has almost no effect on the computational time. This result suggests that subexpression recycling should be used in general on exascale-ready HPC platforms.

\subsubsection{Effect of computing optimization}

\begin{figure}[h]
    \centering
    \includegraphics[width=0.4\textwidth]{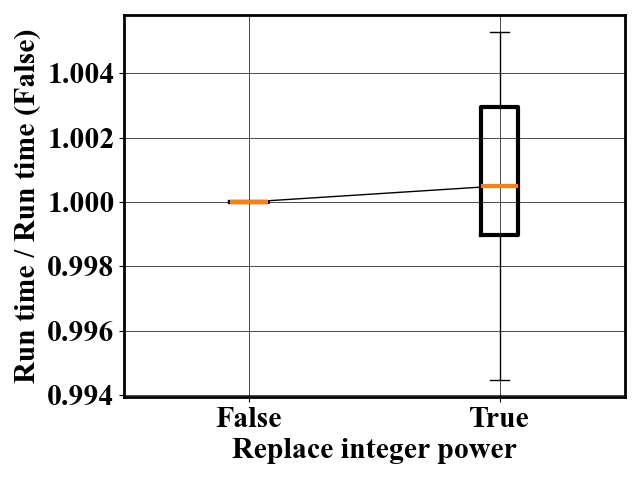}
    \includegraphics[width=0.4\textwidth]{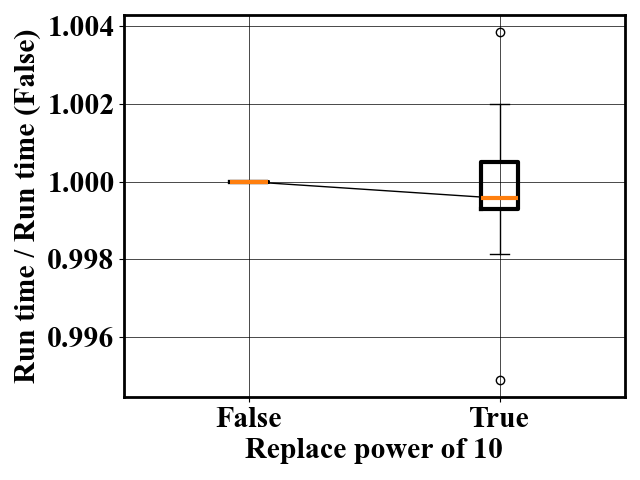}
    \caption{Effect of replacing small integer powers with multiplications (left) and powers of 10 with base-10 exponential (right) on the computing time of the chemistry integration (right). The orange line denotes the median, and the box plot denotes the 95\% confidence interval.}
    \label{fig:effect_gpu_opt}
\end{figure}

The effects of mathematical operations optimizations for GPU described in Sec.~\ref{sec:perfcomp} are shown in Fig.~\ref{fig:effect_gpu_opt}. Neither the small integer power replacement nor the power of 10 replacement has a significant effect on the computing time. This result may be due to modern compiler optimization that already performs operation optimization \cite{guide2013cuda}. Overall, this result suggests that the mathematical operation optimizations listed in Sec.~\ref{sec:perfcomp} are not necessary on exascale-ready HPC platforms.

\subsection{Scaling}

\begin{figure}
\centering
    \includegraphics[width=0.5\textwidth]{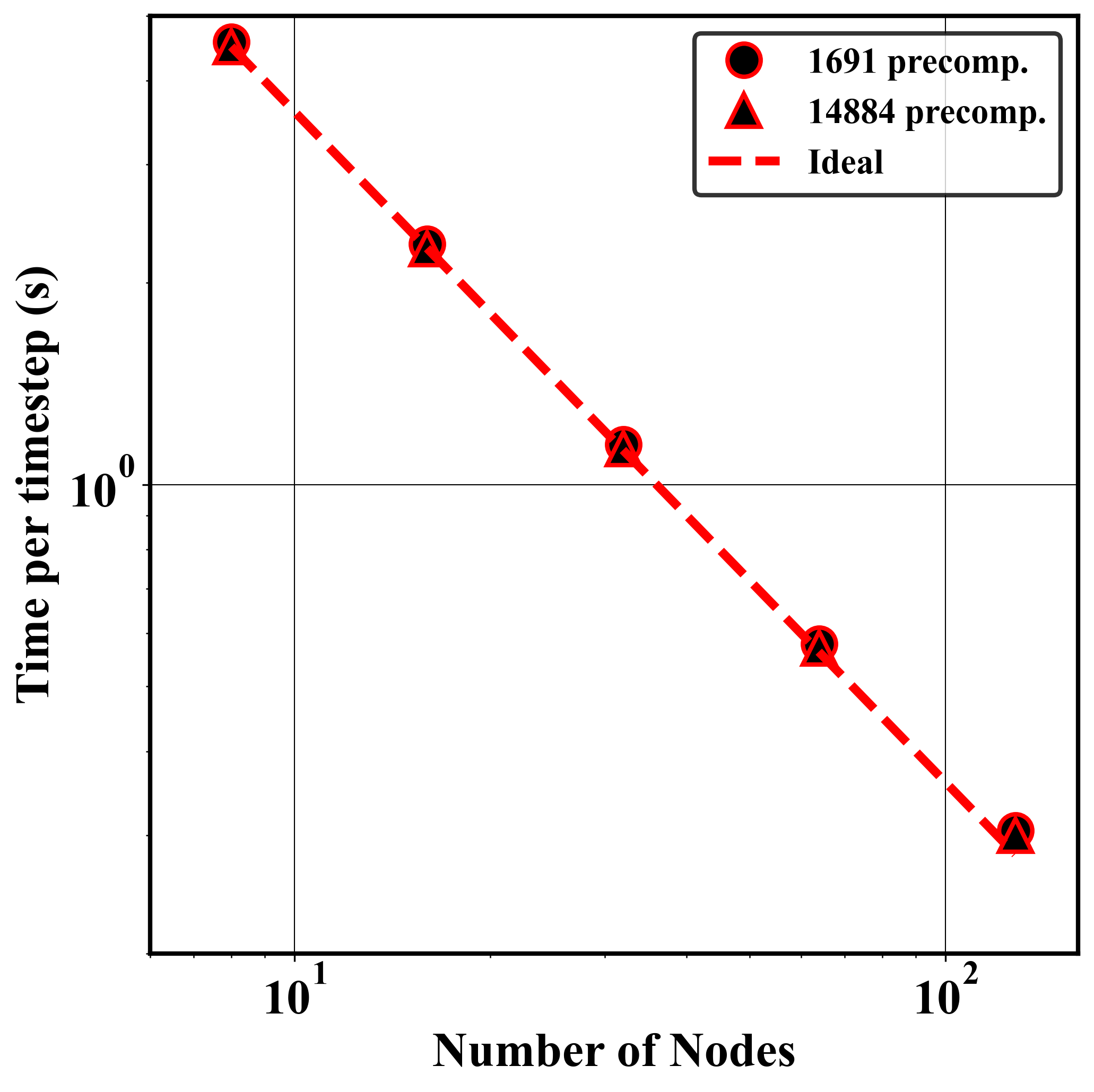}
    \caption{\revone{Strong scaling of the AJ routines for a QSS mechanism without memory optimization (\mythicktriangle{red}{black}) and with memory optimization (\mythickcircle{red}{black}) shown against ideal scaling (\mythickdashedline{red}).}}
    \label{fig:strong_scaling}
\end{figure}

Finally, to ensure that the Jacobian optimizations do not negatively impact the scalability of the solver, strong scaling of the analytic Jacobian evaluation was performed for \revone{8 random mechanisms out of the 24 mechanisms described in Sec.~\ref{sec:perf}. None of the mechanisms show significant deviation from the ideal scaling (not shown here) which can be expected because the modifications in the Jacobian formulation are local to each computing unit. The scaling results are shown for two QSS mechanisms: one with all memory optimization turned on, resulting in 1691 precomputed terms; and one with all the memory optimization turned off, resulting in 14884 precomputed terms. Note that the latter number of precomputed terms departs from the original value of 16487: in all the 24 mechanisms tested, precomputed expressions that did not include any operation were replaced.}.  Figure~\ref{fig:strong_scaling} shows the results of the strong scaling study from 4--128 nodes (32--1024 GPUs) on the Frontier exascale-ready cluster (owing to its larger size than Crusher). The scaling was conducted by simulating Case 3 for 10 timesteps, starting at $t=2$ms. The results show that despite the differences in the formulation of the Jacobian, no adverse scaling issue was introduced. The consequence of the above results is that the analytic Jacobian can be written in a variety of ways that each preserve the computational runtime and scalability of the solver while still providing memory optimizations.

\section{Conclusions}
\label{sec:conclusions}
A comprehensive analysis of chemistry integration of QSS-reduced chemistries using an analytic Jacobian was performed. Several QSS coupling linearization methods were discussed and compared and a 0D analysis of different ODE integration methods suggested that the construction of a Jacobian was needed, especially for low-Mach number solvers. The construction of an analytic Jacobian can provide significant speedup compared to a finite difference Jacobian and a Jacobian-free method. However, constructing an analytic Jacobian of QSS-reduced chemistries can be fairly involved. In this work, a symbolic method was used which allowed to generate an analytic Jacobian if algebraic closure for QSS species can be obtained. Without further treatment, a symbolic Jacobian was found to be overly verbose. Using common subexpression eliminations addressed the problem but increased the memory requirements given that precomputed quantities needed to be stored in memory. Several optimization methods were devised that could reduce the number of precomputed terms by an order of magnitude without incurring additional computational costs. 
The method was exercised in a 3D multipulse ignition problem that was run on an Exascale-ready platform. It was verified that the symbolic Jacobian provided significant speedup compared to a Jacobian-free method and that the procedure did not affect scalability.

\section{Acknowledgments} 
This work was authored in part by the National Renewable Energy Laboratory, operated by Alliance for Sustainable Energy, LLC, for the U.S. Department of Energy (DOE) under Contract No. DE-AC36-08GO28308. Funding was provided by U.S. Department of Energy Office of Science and National Nuclear Security Administration. The views expressed in the article do not necessarily represent the views of the DOE or the U.S. Government. The U.S. Government retains and the publisher, by accepting the article for publication, acknowledges that the U.S. Government retains a nonexclusive, paid-up, irrevocable, worldwide license to publish or reproduce the published form of this work, or allow others to do so, for U.S. Government purposes. This research was supported by the Exascale Computing Project (17-SC-20-SC), a collaborative effort of the U.S. Department of Energy Office of Science and the National Nuclear Security Administration. A portion of the research was performed using computational resources sponsored by the Department of Energy’s Office of Energy Efficiency and Renewable Energy and located at the National Renewable Energy Laboratory. This research used resources of the Oak Ridge Leadership Computing Facility, which is a DOE Office of Science User Facility supported under Contract DE-AC05-00OR22725.

\appendix

\section{Automated code generation}
\label{sec:appendixA}

Automated generation of the mechanism files to be included in application codes is a critical step because of the large variety of possible mechanisms and the importance of being able to switch mechanism types rapidly to explore different cases. We have written a Python-based code generator, called \gls{ceptr}, which is included in PelePhysics \cite{HenrydeFrahan2024,doecode_5574}. \gls{ceptr} is used to generate C++ code for the evaluation of chemical reaction mechanisms. It is particularly formulated to generate efficient code for heterogeneous computing architectures, e.g., \glspl{gpu}. The input file for \gls{ceptr} is a Cantera mechanism input file, which can easily be obtained from a CHEMKIN mechanism file. This input file is parsed for finite rate chemistry evaluation and transport model specifications. These specifications are used to automatically formulate C++ expressions that are written to C++ header and source code files. The rate expressions are parsed to eliminate redundant computation by precomputing expressions, e.g., the arguments of the exponential functions in the Arrhenius rate equations. Efforts were made to reduce thread private arrays, in favor of constant global arrays to reduce register pressure, with a bias towards the \gls{gpu} programming model.

\section{An overview of Tarjan's algorithm for SCC QSS species identification}
\label{sec:appendixQSS}
At the start, all vertices (QSS species) are considered unvisited. Choose any vertex to initiate the recursion; this will be known as the parent. Add the parent to a potential SCC stack and assign it a low-link value, which should match the discovery order upon initialization (note: position in the stack implicitly stores the discovery order via the use of an ordered dictionary). Loop over the adjacent vertices of the parent; these are considered the parent’s children. If a child has not been visited yet (i.e., the child does not have a low-link value), recurse; the child becomes the next parent, enabling the depth-first search to continue. 

Throughout the recursion, low-link values are updated in two scenarios: (1) the current child has been visited before and exists on the stack (note: if a child has been visited before and is not on the stack, it is part of a previously discovered SCC and can thus be ignored). The parent’s low-link value is updated with the child’s discovery order if this value is lower. (2) the end of a recursion is reached (i.e., all children of a given parent have been visited and updates have commenced as described in (1); the parent then reverts to being a child for its parent). The parent’s low-link value is updated with the child’s low-link value if this value is lower. If the low-link value and discovery order of the parent are still equal to each other after the end-of-recursion update, that vertex is the root of an SCC. Vertices are then popped from the stack until the root vertex is removed, generating the fully sequestered SCC. Vertices that were visited but are part of a different SCC remain on the stack until the recursion they are a part of is completed and the next root is found. 

Once all SCCs are found, edge information is updated in the adjacency list to reflect SCC dependencies, as demonstrated in Fig.~\ref{fig:GroupsQSS} (b), and to determine the order of resolution. A conclusive order of execution for back
substitution will always be possible at this point, as no cyclical paths will occur outside of the already determined SCCs.

With only unidirectional edges remaining, back substitution is implemented for the final dependencies while an in-house Gaussian pivoting routine is used to manage the algebraic relationships within individual SCCs.

\section{\revone{Convergence between the finite difference and the symbolic Jacobian}}
\label{sec:conv}
\revone{In Sec.~\ref{sec:symbCons}, the accuracy of the symbolic Jacobian is assessed by comparing it to a finite difference Jacobian computed with a single discretization size for the species concentrations which allows a direct comparison with the zero QSS and the constant QSS approaches. Here, the error between the numerical Jacobian and the symbolic Jacobian is studied for different discretization sizes in the molar concentration space.}

\begin{figure}
\centering
    \includegraphics[width=0.3\textwidth]{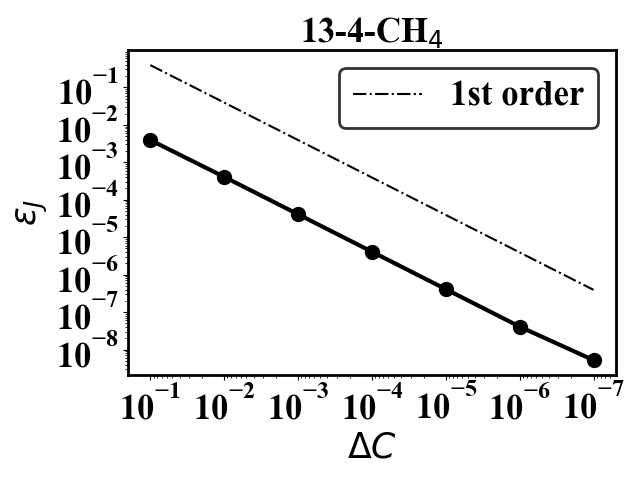}
    \includegraphics[width=0.3\textwidth]{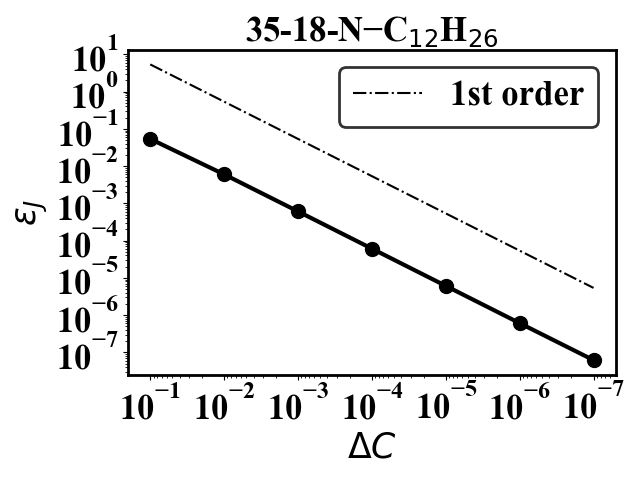}
    \includegraphics[width=0.3\textwidth]{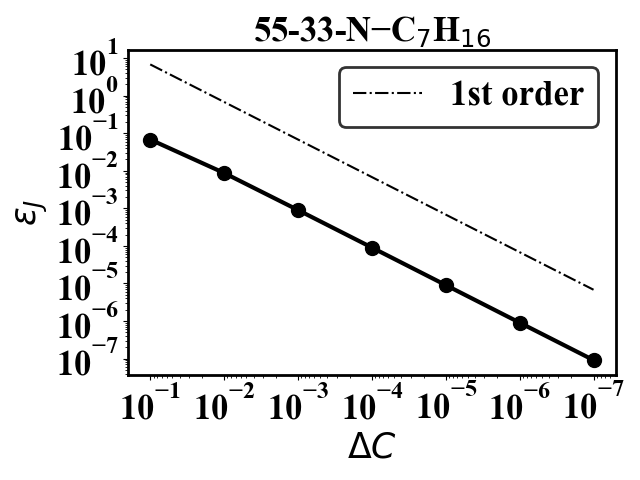}
    \caption{\revone{Convergence of the finite difference Jacobian towards the symbolic Jacobian as a function of the discretization size for the molar concentrations. Left:   13-4-\ce{CH4}. Middle: 35-18-\ce{N-C12H26}. Right: 55-33-\ce{N-C7H16}.}}
    \label{fig:mech_conv}
\end{figure}

\revone{In Fig.~\ref{fig:mech_conv}, the relative error between the finite difference Jacobian and the semi-analytical Jacobian (Eq.~\ref{eq:frobErr}) is computed and averaged over 1000 random samples throughout the composition space (molar concentration of species uniformly sampled from [0.1-1.1] mol/m$^3$ and temperature uniformly sampled from $T = [300-1300]$ K. The FD Jacobian is computed with a one-sided, first-order method, computed with discretization sizes in the molar concentration space $\Delta C \in [10^{-1}, 10^{-7}]$. It can be seen that the FDJ converges towards the symbolic AJ generated, for all three mechanisms considered. The error also converges with a first-order rate which suggests that the error between the symbolic and the finite difference Jacobian is explained by the discretization size.}

\section{\revtwo{Practical benefit of common subexpression elimination}}
\label{app:subexpbenefit}

\revtwo{In Sec.~\ref{sec:symbCons}, the common subexpressions identified are recurrent mathematical expressions that are found to appear several times in several entries of the chemical Jacobian. By precomputing such expressions, one can a) abbreviate the file size and b) reduce the total number of arithmetic operations. Abbreviating the file size is necessary to ensure that the mechanism generated can be compiled.}

\revtwo{As an illustrative example, consider the case of the 13-4-\ce{CH4} mechanisms without memory optimization. One common expression found is {\ftexttt{const amrex::Real x54 = sc[12] * kf_qss[28];}}. Here,  \ftexttt{x54} is the name of the common subexpression, \ftexttt{sc[12]} is one of the non-QSS species concentrations, and \ftexttt{kf_qss[28]} is one of the rates constants. This expression is reused three times when constructing the analytical Jacobian.}

\revtwo{In terms of file size abbreviation: creating the common subexpression requires 44 characters and using the expression later requires 5 characters for each common subexpression occurrence, for a total of 59 characters. In contrast, without the common subexpression, one would need 21 characters for every common subexpression occurrence, hence 63 characters.}

\revtwo{In terms of the number of arithmetic operations: creating the aforementioned common subexpression requires one multiplication. Not creating the expression requires performing one multiplication three times. Creating a common subexpression allows to reduce the total number of arithmetic operations, at the expense of one variable stored in memory.}

\revtwo{Obviously, the larger the common subexpression (in terms of characters and number of arithmetic operations), and the more often it is used, the larger the gains. The question of which common subexpression to keep is tackled in Sec.~\ref{sec:perf} where it is shown that it is possible to reduce the number of precomputations without modifying the computational cost.}




\begin{thebibliography}{67}
\expandafter\ifx\csname natexlab\endcsname\relax\def\natexlab#1{#1}\fi
\providecommand{\url}[1]{\texttt{#1}}
\providecommand{\path}[1]{#1}
\providecommand{\DOIprefix}{doi:}
\providecommand{\ArXivprefix}{arXiv:}
\providecommand{\URLprefix}{URL: }
\providecommand{\Pubmedprefix}{pmid:}
\providecommand{\doi}[1]{\href{http://dx.doi.org/#1}{\path{#1}}}
\providecommand{\Pubmed}[1]{\href{pmid:#1}{\path{#1}}}
\providecommand{\bibinfo}[2]{#2}
\ifx\xfnm\relax \def\xfnm[#1]{\unskip,\space#1}\fi
\bibitem[{Wimer et~al.(2023)Wimer, Esclapez, Brunhart-Lupo, {Henry de Frahan}, Rahimi, Hassanaly, Rood, Yellapantula, Sitaraman, Perry et~al.}]{wimer2023visualizations}
\bibinfo{author}{N.~T. Wimer}, \bibinfo{author}{L.~Esclapez}, \bibinfo{author}{N.~Brunhart-Lupo}, \bibinfo{author}{M.~T. {Henry de Frahan}}, \bibinfo{author}{M.~Rahimi}, \bibinfo{author}{M.~Hassanaly}, \bibinfo{author}{J.~Rood}, \bibinfo{author}{S.~Yellapantula}, \bibinfo{author}{H.~Sitaraman}, \bibinfo{author}{B.~Perry}, et~al.,
\newblock \bibinfo{title}{{Visualizations of a methane/diesel RCCI engine using PeleC and PeleLMeX}},
\newblock \bibinfo{journal}{Phys. Rev. Fl.} \bibinfo{volume}{8} (\bibinfo{year}{2023}) \bibinfo{pages}{110511}.
\bibitem[{Alexander et~al.(2020)Alexander, Almgren, Bell, Bhattacharjee, Chen, Colella, Daniel, DeSlippe, Diachin, Draeger et~al.}]{alexander2020exascale}
\bibinfo{author}{F.~Alexander}, \bibinfo{author}{A.~Almgren}, \bibinfo{author}{J.~Bell}, \bibinfo{author}{A.~Bhattacharjee}, \bibinfo{author}{J.~Chen}, \bibinfo{author}{P.~Colella}, \bibinfo{author}{D.~Daniel}, \bibinfo{author}{J.~DeSlippe}, \bibinfo{author}{L.~Diachin}, \bibinfo{author}{E.~Draeger}, et~al.,
\newblock \bibinfo{title}{Exascale applications: skin in the game},
\newblock \bibinfo{journal}{Philos. T. R. Soc. A.} \bibinfo{volume}{378} (\bibinfo{year}{2020}) \bibinfo{pages}{20190056}.
\bibitem[{Malaya et~al.(2023)Malaya, Messer, Glenski, Georgiadou, Lietz, Gottiparthi, Day, Chen, Rood, Esclapez, White~III, Jansen, Curtis, Nichols, Kurzak, Chalmers, Freitag, Bauman, Fanfarillo, Budiardja, Papatheodore, Frontiere, Mcdougall, Norman, Sreepathi, Roth, Bykov, Wolfe, Mullowney, Eisenbach, {Henry de Frahan}, and Joubert}]{malaya2023}
\bibinfo{author}{N.~Malaya}, \bibinfo{author}{B.~Messer}, \bibinfo{author}{J.~Glenski}, \bibinfo{author}{A.~Georgiadou}, \bibinfo{author}{J.~Lietz}, \bibinfo{author}{K.~Gottiparthi}, \bibinfo{author}{M.~Day}, \bibinfo{author}{J.~Chen}, \bibinfo{author}{J.~Rood}, \bibinfo{author}{L.~Esclapez}, \bibinfo{author}{J.~White~III}, \bibinfo{author}{G.~R. Jansen}, \bibinfo{author}{N.~Curtis}, \bibinfo{author}{S.~Nichols}, \bibinfo{author}{J.~Kurzak}, \bibinfo{author}{N.~Chalmers}, \bibinfo{author}{C.~Freitag}, \bibinfo{author}{P.~Bauman}, \bibinfo{author}{A.~Fanfarillo}, \bibinfo{author}{R.~D. Budiardja}, \bibinfo{author}{T.~Papatheodore}, \bibinfo{author}{N.~Frontiere}, \bibinfo{author}{D.~Mcdougall}, \bibinfo{author}{M.~Norman}, \bibinfo{author}{S.~Sreepathi}, \bibinfo{author}{P.~Roth}, \bibinfo{author}{D.~Bykov}, \bibinfo{author}{N.~Wolfe}, \bibinfo{author}{P.~Mullowney}, \bibinfo{author}{M.~Eisenbach}, \bibinfo{author}{M.~T. {Henry de Frahan}}, \bibinfo{author}{W.~Joubert},
\newblock \bibinfo{title}{{Experiences Readying Applications for Exascale}},
\newblock in: \bibinfo{booktitle}{Proceedings of the International Conference for High Performance Computing, Networking, Storage and Analysis}, SC '23, \bibinfo{publisher}{Association for Computing Machinery}, \bibinfo{address}{New York, NY, USA}, \bibinfo{year}{2023}, pp. \bibinfo{pages}{1--13}.
\bibitem[{Esclapez et~al.(2023)Esclapez, Day, Bell, Felden, Gilet, Grout, {Henry de Frahan}, Motheau, Nonaka, Owen, Perry, Rood, Wimer, and Zhang}]{Esclapez2023}
\bibinfo{author}{L.~Esclapez}, \bibinfo{author}{M.~Day}, \bibinfo{author}{J.~Bell}, \bibinfo{author}{A.~Felden}, \bibinfo{author}{C.~Gilet}, \bibinfo{author}{R.~Grout}, \bibinfo{author}{M.~{Henry de Frahan}}, \bibinfo{author}{E.~Motheau}, \bibinfo{author}{A.~Nonaka}, \bibinfo{author}{L.~Owen}, \bibinfo{author}{B.~Perry}, \bibinfo{author}{J.~Rood}, \bibinfo{author}{N.~Wimer}, \bibinfo{author}{W.~Zhang},
\newblock \bibinfo{title}{{PeleLMeX: an AMR Low Mach Number Reactive Flow Simulation Code without level sub-cycling}},
\newblock \bibinfo{journal}{Journal of Open Source Software} \bibinfo{volume}{8} (\bibinfo{year}{2023}) \bibinfo{pages}{5450}.
\bibitem[{{Henry de Frahan} et~al.(2022){Henry de Frahan}, Rood, Day, Sitaraman, Yellapantula, Perry, Grout, Almgren, Zhang, Bell, and Chen}]{henrydefrahan2022}
\bibinfo{author}{M.~T. {Henry de Frahan}}, \bibinfo{author}{J.~S. Rood}, \bibinfo{author}{M.~S. Day}, \bibinfo{author}{H.~Sitaraman}, \bibinfo{author}{S.~Yellapantula}, \bibinfo{author}{B.~A. Perry}, \bibinfo{author}{R.~W. Grout}, \bibinfo{author}{A.~Almgren}, \bibinfo{author}{W.~Zhang}, \bibinfo{author}{J.~B. Bell}, \bibinfo{author}{J.~H. Chen},
\newblock \bibinfo{title}{{{{PeleC}}: {{An}} Adaptive Mesh Refinement Solver for Compressible Reacting Flows}},
\newblock \bibinfo{journal}{Int. J. High Perform. Comput. Appl.} \bibinfo{volume}{37} (\bibinfo{year}{2022}) \bibinfo{pages}{109434202211211}.
\bibitem[{{Henry de Frahan} et~al.(2024){Henry de Frahan}, Esclapez, Rood, Wimer, Mullowney, Perry, Owen, Sitaraman, Yellapantula, Hassanaly, Rahimi, Martin, Doronina, A., Rieth, Ge, Sankaran, Almgren, Zhang, Bell, Grout, Day, and Chen}]{HenrydeFrahan2024}
\bibinfo{author}{M.~T. {Henry de Frahan}}, \bibinfo{author}{L.~Esclapez}, \bibinfo{author}{J.~Rood}, \bibinfo{author}{N.~T. Wimer}, \bibinfo{author}{P.~Mullowney}, \bibinfo{author}{B.~A. Perry}, \bibinfo{author}{L.~Owen}, \bibinfo{author}{H.~Sitaraman}, \bibinfo{author}{S.~Yellapantula}, \bibinfo{author}{M.~Hassanaly}, \bibinfo{author}{M.~J. Rahimi}, \bibinfo{author}{M.~J. Martin}, \bibinfo{author}{O.~A. Doronina}, \bibinfo{author}{S.~N. A.}, \bibinfo{author}{M.~Rieth}, \bibinfo{author}{W.~Ge}, \bibinfo{author}{R.~Sankaran}, \bibinfo{author}{A.~S. Almgren}, \bibinfo{author}{W.~Zhang}, \bibinfo{author}{J.~B. Bell}, \bibinfo{author}{R.~Grout}, \bibinfo{author}{M.~S. Day}, \bibinfo{author}{J.~H. Chen},
\newblock \bibinfo{title}{{The Pele Simulation Suite for Reacting Flows at Exascale}},
\newblock \bibinfo{journal}{Proceedings of the 2024 SIAM Conference on Parallel Processing for Scientific Computing}  (\bibinfo{year}{2024}) \bibinfo{pages}{13--25}.
\bibitem[{Treichler et~al.(2017)Treichler, Bauer, Bhagatwala, Borghesi, Sankaran, Kolla, McCormick, Slaughter, Lee, Aiken et~al.}]{treichler2017s3d}
\bibinfo{author}{S.~Treichler}, \bibinfo{author}{M.~Bauer}, \bibinfo{author}{A.~Bhagatwala}, \bibinfo{author}{G.~Borghesi}, \bibinfo{author}{R.~Sankaran}, \bibinfo{author}{H.~Kolla}, \bibinfo{author}{P.~S. McCormick}, \bibinfo{author}{E.~Slaughter}, \bibinfo{author}{W.~Lee}, \bibinfo{author}{A.~Aiken}, et~al.,
\newblock \bibinfo{title}{{S3D-Legion: An exascale software for direct numerical simulation of turbulent combustion with complex multicomponent chemistry}},
\newblock in: \bibinfo{booktitle}{Exascale Scientific Applications}, \bibinfo{publisher}{Chapman and Hall/CRC}, \bibinfo{year}{2017}, pp. \bibinfo{pages}{257--278}.
\bibitem[{Mira et~al.(2023)Mira, P{\'e}rez-S{\'a}nchez, Borrell, and Houzeaux}]{mira2023hpc}
\bibinfo{author}{D.~Mira}, \bibinfo{author}{E.~J. P{\'e}rez-S{\'a}nchez}, \bibinfo{author}{R.~Borrell}, \bibinfo{author}{G.~Houzeaux},
\newblock \bibinfo{title}{{HPC}-enabling technologies for high-fidelity combustion simulations},
\newblock \bibinfo{journal}{P. Combust. Inst.} \bibinfo{volume}{39} (\bibinfo{year}{2023}) \bibinfo{pages}{5091--5125}.
\bibitem[{Pignatelli et~al.(2023)Pignatelli, Passad, {\AA}kerblom, Nilsson, Nilsson, and Fureby}]{Pignatelli2023}
\bibinfo{author}{F.~Pignatelli}, \bibinfo{author}{M.~Passad}, \bibinfo{author}{A.~{\AA}kerblom}, \bibinfo{author}{T.~Nilsson}, \bibinfo{author}{E.~Nilsson}, \bibinfo{author}{C.~Fureby},
\newblock \bibinfo{title}{{Predictions of Spray Combustion using Conventional Category A Fuels and Exploratory Category C Fuels}},
\newblock in: \bibinfo{booktitle}{{AIAA} Scitech 2023 Forum}, \bibinfo{publisher}{American Institute of Aeronautics and Astronautics}, \bibinfo{year}{2023}, p. \bibinfo{pages}{1486}.
\bibitem[{Felden et~al.(2018)Felden, Esclapez, Riber, Cuenot, and Wang}]{felden2018including}
\bibinfo{author}{A.~Felden}, \bibinfo{author}{L.~Esclapez}, \bibinfo{author}{E.~Riber}, \bibinfo{author}{B.~Cuenot}, \bibinfo{author}{H.~Wang},
\newblock \bibinfo{title}{{Including real fuel chemistry in LES of turbulent spray combustion}},
\newblock \bibinfo{journal}{Combust. Flame} \bibinfo{volume}{193} (\bibinfo{year}{2018}) \bibinfo{pages}{397--416}.
\bibitem[{Chung et~al.(2023)Chung, Ly, and Ihme}]{Chung2023}
\bibinfo{author}{W.~T. Chung}, \bibinfo{author}{N.~Ly}, \bibinfo{author}{M.~Ihme},
\newblock \bibinfo{title}{{LES} of {HCCI} combustion of iso-octane/air in a flat-piston rapid compression machine},
\newblock \bibinfo{journal}{P. Combust. Inst.} \bibinfo{volume}{39} (\bibinfo{year}{2023}) \bibinfo{pages}{5309--5317}.
\bibitem[{Tang et~al.(2021)Tang, Hassanaly, Raman, Sforzo, and Seitzman}]{tang2021probabilistic}
\bibinfo{author}{Y.~Tang}, \bibinfo{author}{M.~Hassanaly}, \bibinfo{author}{V.~Raman}, \bibinfo{author}{B.~A. Sforzo}, \bibinfo{author}{J.~Seitzman},
\newblock \bibinfo{title}{Probabilistic modeling of forced ignition of alternative jet fuels},
\newblock \bibinfo{journal}{P. Combust. Inst.} \bibinfo{volume}{38} (\bibinfo{year}{2021}) \bibinfo{pages}{2589--2596}.
\bibitem[{Grader and Gerlinger(2023)}]{Grader2023}
\bibinfo{author}{M.~Grader}, \bibinfo{author}{P.~Gerlinger},
\newblock \bibinfo{title}{Influence of operating conditions on flow field dynamics and soot formation in an aero-engine model combustor},
\newblock \bibinfo{journal}{Combust. Flame}  (\bibinfo{year}{2023}) \bibinfo{pages}{112712}.
\bibitem[{Jaravel et~al.(2017)Jaravel, Riber, Cuenot, and Bulat}]{Jaravel2017}
\bibinfo{author}{T.~Jaravel}, \bibinfo{author}{E.~Riber}, \bibinfo{author}{B.~Cuenot}, \bibinfo{author}{G.~Bulat},
\newblock \bibinfo{title}{{Large Eddy Simulation of an industrial gas turbine combustor using reduced chemistry with accurate pollutant prediction}},
\newblock \bibinfo{journal}{P. Combust. Inst.} \bibinfo{volume}{36} (\bibinfo{year}{2017}) \bibinfo{pages}{3817--3825}.
\bibitem[{Mathieu and Petersen(2015)}]{mathieu2015experimental}
\bibinfo{author}{O.~Mathieu}, \bibinfo{author}{E.~L. Petersen},
\newblock \bibinfo{title}{{Experimental and modeling study on the high-temperature oxidation of Ammonia and related NOx chemistry}},
\newblock \bibinfo{journal}{Combust. Flame} \bibinfo{volume}{162} (\bibinfo{year}{2015}) \bibinfo{pages}{554--570}.
\bibitem[{Attili et~al.(2014)Attili, Bisetti, Mueller, and Pitsch}]{attili2014formation}
\bibinfo{author}{A.~Attili}, \bibinfo{author}{F.~Bisetti}, \bibinfo{author}{M.~E. Mueller}, \bibinfo{author}{H.~Pitsch},
\newblock \bibinfo{title}{Formation, growth, and transport of soot in a three-dimensional turbulent non-premixed jet flame},
\newblock \bibinfo{journal}{Combust. Flame} \bibinfo{volume}{161} (\bibinfo{year}{2014}) \bibinfo{pages}{1849--1865}.
\bibitem[{Balos et~al.(2024)Balos, Day, Esclapez, Felden, Gardner, Hassanaly, Reynolds, Rood, Sexton, Sitaraman, Wimer, and Woodward}]{balos2024}
\bibinfo{author}{C.~J. Balos}, \bibinfo{author}{M.~Day}, \bibinfo{author}{L.~Esclapez}, \bibinfo{author}{A.~M. Felden}, \bibinfo{author}{D.~J. Gardner}, \bibinfo{author}{M.~Hassanaly}, \bibinfo{author}{D.~R. Reynolds}, \bibinfo{author}{J.~Rood}, \bibinfo{author}{J.~M. Sexton}, \bibinfo{author}{H.~Sitaraman}, \bibinfo{author}{N.~T. Wimer}, \bibinfo{author}{C.~S. Woodward},
\newblock \bibinfo{title}{{SUNDIALS Time Integrators for Exascale Applications with Many Independent ODE Systems}},
\newblock \bibinfo{journal}{arXiv preprint arXiv:2405.01713}  (\bibinfo{year}{2024}).
\bibitem[{Shi et~al.(2012)Shi, Green, Wong, and Oluwole}]{Shi2012}
\bibinfo{author}{Y.~Shi}, \bibinfo{author}{W.~H. Green}, \bibinfo{author}{H.-W. Wong}, \bibinfo{author}{O.~O. Oluwole},
\newblock \bibinfo{title}{Accelerating multi-dimensional combustion simulations using gpu and hybrid explicit/implicit ode integration},
\newblock \bibinfo{journal}{Combust. Flame} \bibinfo{volume}{159} (\bibinfo{year}{2012}) \bibinfo{pages}{2388–2397}.
\bibitem[{Kodavasal et~al.(2016)Kodavasal, Harms, Srivastava, Som, Quan, Richards, and García}]{Kodavasal2016}
\bibinfo{author}{J.~Kodavasal}, \bibinfo{author}{K.~Harms}, \bibinfo{author}{P.~Srivastava}, \bibinfo{author}{S.~Som}, \bibinfo{author}{S.~Quan}, \bibinfo{author}{K.~Richards}, \bibinfo{author}{M.~García},
\newblock \bibinfo{title}{Development of a stiffness-based chemistry load balancing scheme, and optimization of input/output and communication, to enable massively parallel high-fidelity internal combustion engine simulations},
\newblock \bibinfo{journal}{J. Energ. Resour.-ASME} \bibinfo{volume}{138} (\bibinfo{year}{2016}).
\bibitem[{Wu et~al.(2019)Wu, Ma, Jaravel, and Ihme}]{wu2019pareto}
\bibinfo{author}{H.~Wu}, \bibinfo{author}{P.~C. Ma}, \bibinfo{author}{T.~Jaravel}, \bibinfo{author}{M.~Ihme},
\newblock \bibinfo{title}{{Pareto-efficient combustion modeling for improved CO-emission prediction in LES of a piloted turbulent dimethyl ether jet flame}},
\newblock \bibinfo{journal}{P. Combust. Inst.} \bibinfo{volume}{37} (\bibinfo{year}{2019}) \bibinfo{pages}{2267--2276}.
\bibitem[{Mao et~al.(2023)Mao, Lin, Zhang, Zhang, Xu, and Chen}]{Mao2023}
\bibinfo{author}{R.~Mao}, \bibinfo{author}{M.~Lin}, \bibinfo{author}{Y.~Zhang}, \bibinfo{author}{T.~Zhang}, \bibinfo{author}{Z.-Q.~J. Xu}, \bibinfo{author}{Z.~X. Chen},
\newblock \bibinfo{title}{{DeepFlame}: {A} deep learning empowered open-source platform for reacting flow simulations},
\newblock \bibinfo{journal}{Comput. Phys. Commun.} \bibinfo{volume}{291} (\bibinfo{year}{2023}) \bibinfo{pages}{108842}.
\bibitem[{Lu and Law(2009)}]{Lu2009}
\bibinfo{author}{T.~Lu}, \bibinfo{author}{C.~K. Law},
\newblock \bibinfo{title}{Toward accommodating realistic fuel chemistry in large-scale computations},
\newblock \bibinfo{journal}{Prog. Energ. Combust.} \bibinfo{volume}{35} (\bibinfo{year}{2009}) \bibinfo{pages}{192--215}.
\bibitem[{McNenly et~al.(2015)McNenly, Whitesides, and Flowers}]{mcnenly2015faster}
\bibinfo{author}{M.~J. McNenly}, \bibinfo{author}{R.~A. Whitesides}, \bibinfo{author}{D.~L. Flowers},
\newblock \bibinfo{title}{Faster solvers for large kinetic mechanisms using adaptive preconditioners},
\newblock \bibinfo{journal}{P. Combust. Inst.} \bibinfo{volume}{35} (\bibinfo{year}{2015}) \bibinfo{pages}{581--587}.
\bibitem[{Walker et~al.(2023)Walker, Speth, and Niemeyer}]{walker2023generalized}
\bibinfo{author}{A.~S. Walker}, \bibinfo{author}{R.~L. Speth}, \bibinfo{author}{K.~E. Niemeyer},
\newblock \bibinfo{title}{Generalized preconditioning for accelerating simulations with large kinetic models},
\newblock \bibinfo{journal}{P. Combust. Inst.} \bibinfo{volume}{39} (\bibinfo{year}{2023}) \bibinfo{pages}{5395--5403}.
\bibitem[{Curtis et~al.(2018)Curtis, Niemeyer, and Sung}]{curtis2018using}
\bibinfo{author}{N.~J. Curtis}, \bibinfo{author}{K.~E. Niemeyer}, \bibinfo{author}{C.-J. Sung},
\newblock \bibinfo{title}{{Using SIMD and SIMT vectorization to evaluate sparse chemical kinetic Jacobian matrices and thermochemical source terms}},
\newblock \bibinfo{journal}{Combust. Flame} \bibinfo{volume}{198} (\bibinfo{year}{2018}) \bibinfo{pages}{186--204}.
\bibitem[{Tur{\'a}nyi and Tomlin(2014)}]{turanyi2014analysis}
\bibinfo{author}{T.~Tur{\'a}nyi}, \bibinfo{author}{A.~S. Tomlin}, \bibinfo{title}{Analysis of kinetic reaction mechanisms}, volume~\bibinfo{volume}{20}, \bibinfo{publisher}{Springer}, \bibinfo{year}{2014}.
\bibitem[{Lu and Law(2006)}]{lu2006linear}
\bibinfo{author}{T.~Lu}, \bibinfo{author}{C.~K. Law},
\newblock \bibinfo{title}{{Linear time reduction of large kinetic mechanisms with directed relation graph: n-Heptane and iso-octane}},
\newblock \bibinfo{journal}{Combust. Flame} \bibinfo{volume}{144} (\bibinfo{year}{2006}) \bibinfo{pages}{24--36}.
\bibitem[{Felden et~al.(2019)Felden, Pepiot, Esclapez, Riber, and Cuenot}]{Felden2019}
\bibinfo{author}{A.~Felden}, \bibinfo{author}{P.~Pepiot}, \bibinfo{author}{L.~Esclapez}, \bibinfo{author}{E.~Riber}, \bibinfo{author}{B.~Cuenot},
\newblock \bibinfo{title}{Including analytically reduced chemistry ({ARC}) in {CFD} applications},
\newblock \bibinfo{journal}{Acta Astronautica} \bibinfo{volume}{158} (\bibinfo{year}{2019}) \bibinfo{pages}{444--459}.
\bibitem[{Lu and Law(2008)}]{LU2008}
\bibinfo{author}{T.~Lu}, \bibinfo{author}{C.~K. Law},
\newblock \bibinfo{title}{Strategies for mechanism reduction for large hydrocarbons: n-heptane},
\newblock \bibinfo{journal}{Combust. Flame} \bibinfo{volume}{154} (\bibinfo{year}{2008}) \bibinfo{pages}{153--163}.
\bibitem[{Pepiot and Pitsch(2008)}]{PEPIOTDESJARDINS2008}
\bibinfo{author}{P.~Pepiot}, \bibinfo{author}{H.~Pitsch},
\newblock \bibinfo{title}{An efficient error-propagation-based reduction method for large chemical kinetic mechanisms},
\newblock \bibinfo{journal}{Combust. Flame} \bibinfo{volume}{154} (\bibinfo{year}{2008}) \bibinfo{pages}{67--81}.
\bibitem[{Tomlin et~al.(1992)Tomlin, Pilling, Tur{\'{a}}nyi, Merkin, and Brindley}]{Tomlin1992}
\bibinfo{author}{A.~S. Tomlin}, \bibinfo{author}{M.~J. Pilling}, \bibinfo{author}{T.~Tur{\'{a}}nyi}, \bibinfo{author}{J.~H. Merkin}, \bibinfo{author}{J.~Brindley},
\newblock \bibinfo{title}{{Mechanism reduction for the oscillatory oxidation of hydrogen: Sensitivity and quasi-steady-state analyses}},
\newblock \bibinfo{journal}{Combust. Flame} \bibinfo{volume}{91} (\bibinfo{year}{1992}) \bibinfo{pages}{107--130}.
\bibitem[{Bodenstein(1913)}]{bodenstein1913theorie}
\bibinfo{author}{M.~Bodenstein},
\newblock \bibinfo{title}{Eine theorie der photochemischen reaktionsgeschwindigkeiten},
\newblock \bibinfo{journal}{Z. Phys. Chem.} \bibinfo{volume}{85} (\bibinfo{year}{1913}) \bibinfo{pages}{329--397}.
\bibitem[{Fraser(1988)}]{Fraser1988}
\bibinfo{author}{S.~J. Fraser},
\newblock \bibinfo{title}{{The steady state and equilibrium approximations: A geometrical picture}},
\newblock \bibinfo{journal}{J. Chem. Phys.} \bibinfo{volume}{88} (\bibinfo{year}{1988}) \bibinfo{pages}{4732--4738}.
\bibitem[{Lu and Law(2006)}]{lu2006systematic}
\bibinfo{author}{T.~Lu}, \bibinfo{author}{C.~K. Law},
\newblock \bibinfo{title}{Systematic approach to obtain analytic solutions of quasi steady state species in reduced mechanisms},
\newblock \bibinfo{journal}{J. Phys. Chem. A} \bibinfo{volume}{110} (\bibinfo{year}{2006}) \bibinfo{pages}{13202--13208}.
\bibitem[{Safta et~al.(2011)Safta, Najm, and Knio}]{safta2011tchem}
\bibinfo{author}{C.~Safta}, \bibinfo{author}{H.~N. Najm}, \bibinfo{author}{O.~Knio}, \bibinfo{title}{{TChem-a software toolkit for the analysis of complex kinetic models}}, \bibinfo{type}{Technical Report}, Sandia National Lab.(SNL-CA), Livermore, CA (United States), \bibinfo{year}{2011}.
\bibitem[{Niemeyer et~al.(2017)Niemeyer, Curtis, and Sung}]{niemeyer2017pyjac}
\bibinfo{author}{K.~E. Niemeyer}, \bibinfo{author}{N.~J. Curtis}, \bibinfo{author}{C.-J. Sung},
\newblock \bibinfo{title}{{pyJac: Analytical Jacobian generator for chemical kinetics}},
\newblock \bibinfo{journal}{Comput. Phys. Commun.} \bibinfo{volume}{215} (\bibinfo{year}{2017}) \bibinfo{pages}{188--203}.
\bibitem[{Perini et~al.(2012)Perini, Galligani, and Reitz}]{perini2012analytical}
\bibinfo{author}{F.~Perini}, \bibinfo{author}{E.~Galligani}, \bibinfo{author}{R.~D. Reitz},
\newblock \bibinfo{title}{{An analytical Jacobian approach to sparse reaction kinetics for computationally efficient combustion modeling with large reaction mechanisms}},
\newblock \bibinfo{journal}{Energy. Fuel.} \bibinfo{volume}{26} (\bibinfo{year}{2012}) \bibinfo{pages}{4804--4822}.
\bibitem[{Bisetti(2012)}]{bisetti2012integration}
\bibinfo{author}{F.~Bisetti},
\newblock \bibinfo{title}{{Integration of large chemical kinetic mechanisms via exponential methods with Krylov approximations to Jacobian matrix functions}},
\newblock \bibinfo{journal}{Combust. Theor. Model.} \bibinfo{volume}{16} (\bibinfo{year}{2012}) \bibinfo{pages}{387--418}.
\bibitem[{Dijkmans et~al.(2014)Dijkmans, Schietekat, Van~Geem, and Marin}]{dijkmans2014gpu}
\bibinfo{author}{T.~Dijkmans}, \bibinfo{author}{C.~M. Schietekat}, \bibinfo{author}{K.~M. Van~Geem}, \bibinfo{author}{G.~B. Marin},
\newblock \bibinfo{title}{{GPU based simulation of reactive mixtures with detailed chemistry in combination with tabulation and an analytical Jacobian}},
\newblock \bibinfo{journal}{Comput. Chem. Eng.} \bibinfo{volume}{71} (\bibinfo{year}{2014}) \bibinfo{pages}{521--531}.
\bibitem[{Sharma(2022)}]{sharma2022acceleration}
\bibinfo{author}{P.~Sharma}, \bibinfo{title}{{Acceleration Techniques for Efficient and Accurate Particle PDF Simulations of Large-Scale Turbulent Combustion}}, Ph.D. thesis, Cornell University, \bibinfo{year}{2022}.
\bibitem[{Sitaraman et~al.(2017)Sitaraman, Yellapantula, and Grout}]{doecode_5574}
\bibinfo{author}{H.~Sitaraman}, \bibinfo{author}{S.~Yellapantula}, \bibinfo{author}{R.~Grout}, \bibinfo{title}{Pelephysics [swr-17-36]}, \bibinfo{howpublished}{[Computer Software] \url{https://doi.org/10.11578/dc.20171025.1989}}, \bibinfo{year}{2017}. \URLprefix \url{https://doi.org/10.11578/dc.20171025.1989}. \DOIprefix\doi{10.11578/dc.20171025.1989}.
\bibitem[{Turanyi et~al.(1993)Turanyi, Tomlin, and Pilling}]{Turanyi1993}
\bibinfo{author}{T.~Turanyi}, \bibinfo{author}{A.~S. Tomlin}, \bibinfo{author}{M.~J. Pilling},
\newblock \bibinfo{title}{On the error of the quasi-steady-state approximation},
\newblock \bibinfo{journal}{J. Phys. Chem.} \bibinfo{volume}{97} (\bibinfo{year}{1993}) \bibinfo{pages}{163--172}.
\bibitem[{Law et~al.(2003)Law, Sung, Wang, and Lu}]{law2003development}
\bibinfo{author}{C.~K. Law}, \bibinfo{author}{C.~J. Sung}, \bibinfo{author}{H.~Wang}, \bibinfo{author}{T.~Lu},
\newblock \bibinfo{title}{Development of comprehensive detailed and reduced reaction mechanisms for combustion modeling},
\newblock \bibinfo{journal}{AIAA J.} \bibinfo{volume}{41} (\bibinfo{year}{2003}) \bibinfo{pages}{1629--1646}.
\bibitem[{Borghesi et~al.(2018)Borghesi, Krisman, Lu, and Chen}]{borghesi2018direct}
\bibinfo{author}{G.~Borghesi}, \bibinfo{author}{A.~Krisman}, \bibinfo{author}{T.~Lu}, \bibinfo{author}{J.~H. Chen},
\newblock \bibinfo{title}{Direct numerical simulation of a temporally evolving air/n-dodecane jet at low-temperature diesel-relevant conditions},
\newblock \bibinfo{journal}{Combust. Flame} \bibinfo{volume}{195} (\bibinfo{year}{2018}) \bibinfo{pages}{183--202}.
\bibitem[{Yao et~al.(2017)Yao, Pei, Zhong, Som, Lu, and Luo}]{Yao2017}
\bibinfo{author}{T.~Yao}, \bibinfo{author}{Y.~Pei}, \bibinfo{author}{B.-J. Zhong}, \bibinfo{author}{S.~Som}, \bibinfo{author}{T.~Lu}, \bibinfo{author}{K.~H. Luo},
\newblock \bibinfo{title}{A compact skeletal mechanism for n-dodecane with optimized semi-global low-temperature chemistry for diesel engine simulations},
\newblock \bibinfo{journal}{Fuel} \bibinfo{volume}{191} (\bibinfo{year}{2017}) \bibinfo{pages}{339--349}.
\bibitem[{Hindmarsh et~al.(2005)Hindmarsh, Brown, Grant, Lee, Serban, Shumaker, and Woodward}]{hindmarsh2005sundials}
\bibinfo{author}{A.~C. Hindmarsh}, \bibinfo{author}{P.~N. Brown}, \bibinfo{author}{K.~E. Grant}, \bibinfo{author}{S.~L. Lee}, \bibinfo{author}{R.~Serban}, \bibinfo{author}{D.~E. Shumaker}, \bibinfo{author}{C.~S. Woodward},
\newblock \bibinfo{title}{{SUNDIALS: Suite of nonlinear and differential/algebraic equation solvers}},
\newblock \bibinfo{journal}{ACM T. Math. Software} \bibinfo{volume}{31} (\bibinfo{year}{2005}) \bibinfo{pages}{363--396}.
\bibitem[{Lu and Law(2005)}]{Lu2005}
\bibinfo{author}{T.~Lu}, \bibinfo{author}{C.~K. Law},
\newblock \bibinfo{title}{A directed relation graph method for mechanism reduction},
\newblock \bibinfo{journal}{P. Combust. Inst.} \bibinfo{volume}{30} (\bibinfo{year}{2005}) \bibinfo{pages}{1333–1341}.
\bibitem[{Tarjan(1972)}]{Tarjan1972}
\bibinfo{author}{R.~Tarjan},
\newblock \bibinfo{title}{{Depth-First Search and Linear Graph Algorithms}},
\newblock \bibinfo{journal}{{SIAM} Journal on Computing} \bibinfo{volume}{1} (\bibinfo{year}{1972}) \bibinfo{pages}{146--160}.
\bibitem[{Sharir(1981)}]{sharir1981strong}
\bibinfo{author}{M.~Sharir},
\newblock \bibinfo{title}{A strong-connectivity algorithm and its applications in data flow analysis},
\newblock \bibinfo{journal}{Computers \& Mathematics with Applications} \bibinfo{volume}{7} (\bibinfo{year}{1981}) \bibinfo{pages}{67--72}.
\bibitem[{Gadalla(2022)}]{gadalla2022implementation}
\bibinfo{author}{M.~Gadalla},
\newblock \bibinfo{title}{{Implementation of Analytical Jacobian and Chemical Explosive Mode Analysis (CEMA) in OpenFOAM}},
\newblock \bibinfo{journal}{arXiv preprint arXiv:2205.07416}  (\bibinfo{year}{2022}).
\bibitem[{Saad(2003)}]{saad2003iterative}
\bibinfo{author}{Y.~Saad}, \bibinfo{title}{Iterative methods for sparse linear systems}, \bibinfo{publisher}{SIAM}, \bibinfo{year}{2003}.
\bibitem[{Knoll and Keyes(2004)}]{knoll2004jacobian}
\bibinfo{author}{D.~A. Knoll}, \bibinfo{author}{D.~E. Keyes},
\newblock \bibinfo{title}{{Jacobian-free Newton--Krylov methods: a survey of approaches and applications}},
\newblock \bibinfo{journal}{J. Comput. Phys.} \bibinfo{volume}{193} (\bibinfo{year}{2004}) \bibinfo{pages}{357--397}.
\bibitem[{Hassanaly et~al.(2018)Hassanaly, Koo, Lietz, Chong, and Raman}]{hassanaly2018minimally}
\bibinfo{author}{M.~Hassanaly}, \bibinfo{author}{H.~Koo}, \bibinfo{author}{C.~F. Lietz}, \bibinfo{author}{S.~T. Chong}, \bibinfo{author}{V.~Raman},
\newblock \bibinfo{title}{{A minimally-dissipative low-Mach number solver for complex reacting flows in OpenFOAM}},
\newblock \bibinfo{journal}{Comput. Fluids} \bibinfo{volume}{162} (\bibinfo{year}{2018}) \bibinfo{pages}{11--25}.
\bibitem[{Desjardins et~al.(2008)Desjardins, Blanquart, Balarac, and Pitsch}]{desjardins2008high}
\bibinfo{author}{O.~Desjardins}, \bibinfo{author}{G.~Blanquart}, \bibinfo{author}{G.~Balarac}, \bibinfo{author}{H.~Pitsch},
\newblock \bibinfo{title}{High order conservative finite difference scheme for variable density low mach number turbulent flows},
\newblock \bibinfo{journal}{J. Comput. Phys.} \bibinfo{volume}{227} (\bibinfo{year}{2008}) \bibinfo{pages}{7125--7159}.
\bibitem[{Sankaran et~al.(2007)Sankaran, Hawkes, Chen, Lu, and Law}]{sankaran2007structure}
\bibinfo{author}{R.~Sankaran}, \bibinfo{author}{E.~R. Hawkes}, \bibinfo{author}{J.~H. Chen}, \bibinfo{author}{T.~Lu}, \bibinfo{author}{C.~K. Law},
\newblock \bibinfo{title}{{Structure of a spatially developing turbulent lean methane--air Bunsen flame}},
\newblock \bibinfo{journal}{P. Combust. Inst.} \bibinfo{volume}{31} (\bibinfo{year}{2007}) \bibinfo{pages}{1291--1298}.
\bibitem[{Yoo et~al.(2011)Yoo, Lu, Chen, and Law}]{yoo2011direct}
\bibinfo{author}{C.~S. Yoo}, \bibinfo{author}{T.~Lu}, \bibinfo{author}{J.~H. Chen}, \bibinfo{author}{C.~K. Law},
\newblock \bibinfo{title}{{Direct numerical simulations of ignition of a lean n-heptane/air mixture with temperature inhomogeneities at constant volume: Parametric study}},
\newblock \bibinfo{journal}{Combust. Flame} \bibinfo{volume}{158} (\bibinfo{year}{2011}) \bibinfo{pages}{1727--1741}.
\bibitem[{Dongarra et~al.(2014)Dongarra, Gates, Haidar, Kurzak, Luszczek, Tomov, and Yamazaki}]{Dongarra:2014}
\bibinfo{author}{J.~Dongarra}, \bibinfo{author}{M.~Gates}, \bibinfo{author}{A.~Haidar}, \bibinfo{author}{J.~Kurzak}, \bibinfo{author}{P.~Luszczek}, \bibinfo{author}{S.~Tomov}, \bibinfo{author}{I.~Yamazaki},
\newblock \bibinfo{title}{Accelerating numerical dense linear algebra calculations with {GPU}s},
\newblock \bibinfo{journal}{Numerical Computations with GPUs}  (\bibinfo{year}{2014}) \bibinfo{pages}{1--26}.
\bibitem[{Meurer et~al.(2017)Meurer, Smith, Paprocki, \v{C}ert\'{i}k, Kirpichev, Rocklin, Kumar, Ivanov, Moore, Singh, Rathnayake, Vig, Granger, Muller, Bonazzi, Gupta, Vats, Johansson, Pedregosa, Curry, Terrel, Rou\v{c}ka, Saboo, Fernando, Kulal, Cimrman, and Scopatz}]{10.7717/peerj-cs.103}
\bibinfo{author}{A.~Meurer}, \bibinfo{author}{C.~P. Smith}, \bibinfo{author}{M.~Paprocki}, \bibinfo{author}{O.~\v{C}ert\'{i}k}, \bibinfo{author}{S.~B. Kirpichev}, \bibinfo{author}{M.~Rocklin}, \bibinfo{author}{A.~Kumar}, \bibinfo{author}{S.~Ivanov}, \bibinfo{author}{J.~K. Moore}, \bibinfo{author}{S.~Singh}, \bibinfo{author}{T.~Rathnayake}, \bibinfo{author}{S.~Vig}, \bibinfo{author}{B.~E. Granger}, \bibinfo{author}{R.~P. Muller}, \bibinfo{author}{F.~Bonazzi}, \bibinfo{author}{H.~Gupta}, \bibinfo{author}{S.~Vats}, \bibinfo{author}{F.~Johansson}, \bibinfo{author}{F.~Pedregosa}, \bibinfo{author}{M.~J. Curry}, \bibinfo{author}{A.~R. Terrel}, \bibinfo{author}{v.~Rou\v{c}ka}, \bibinfo{author}{A.~Saboo}, \bibinfo{author}{I.~Fernando}, \bibinfo{author}{S.~Kulal}, \bibinfo{author}{R.~Cimrman}, \bibinfo{author}{A.~Scopatz},
\newblock \bibinfo{title}{Sympy: symbolic computing in python},
\newblock \bibinfo{journal}{PeerJ Computer Science} \bibinfo{volume}{3} (\bibinfo{year}{2017}) \bibinfo{pages}{e103}.
\bibitem[{Rein(1992)}]{rein1992partial}
\bibinfo{author}{M.~Rein},
\newblock \bibinfo{title}{The partial-equilibrium approximation in reacting flows},
\newblock \bibinfo{journal}{Phys. Fluids A-Fluid.} \bibinfo{volume}{4} (\bibinfo{year}{1992}) \bibinfo{pages}{873--886}.
\bibitem[{Rouhi~Youssefi(2011)}]{rouhi2011development}
\bibinfo{author}{M.~Rouhi~Youssefi}, \bibinfo{title}{{Development of Analytic Tools for Computational Flame Diagnostics}}, Ph.D. thesis, University of Connecticut, \bibinfo{year}{2011}.
\bibitem[{Guide(2013)}]{guide2013cuda}
\bibinfo{author}{D.~Guide},
\newblock \bibinfo{title}{{Cuda C best practices guide}},
\newblock \bibinfo{journal}{NVIDIA, July}  (\bibinfo{year}{2013}).
\bibitem[{Tomov et~al.(2010{\natexlab{a}})Tomov, Nath, Ltaief, and Dongarra}]{tomov2010dense}
\bibinfo{author}{S.~Tomov}, \bibinfo{author}{R.~Nath}, \bibinfo{author}{H.~Ltaief}, \bibinfo{author}{J.~Dongarra},
\newblock \bibinfo{title}{Dense linear algebra solvers for multicore with gpu accelerators},
\newblock in: \bibinfo{booktitle}{2010 IEEE International Symposium on Parallel \& Distributed Processing, Workshops and Phd Forum (IPDPSW)}, \bibinfo{organization}{IEEE}, \bibinfo{year}{2010}{\natexlab{a}}, pp. \bibinfo{pages}{1--8}.
\bibitem[{Tomov et~al.(2010{\natexlab{b}})Tomov, Dongarra, and Baboulin}]{tomov2010towards}
\bibinfo{author}{S.~Tomov}, \bibinfo{author}{J.~Dongarra}, \bibinfo{author}{M.~Baboulin},
\newblock \bibinfo{title}{Towards dense linear algebra for hybrid gpu accelerated manycore systems},
\newblock \bibinfo{journal}{Parallel Comput.} \bibinfo{volume}{36} (\bibinfo{year}{2010}{\natexlab{b}}) \bibinfo{pages}{232--240}.
\bibitem[{Dongarra et~al.(2014)Dongarra, Gates, Haidar, Kurzak, Luszczek, Tomov, and Yamazaki}]{dongarra2014accelerating}
\bibinfo{author}{J.~Dongarra}, \bibinfo{author}{M.~Gates}, \bibinfo{author}{A.~Haidar}, \bibinfo{author}{J.~Kurzak}, \bibinfo{author}{P.~Luszczek}, \bibinfo{author}{S.~Tomov}, \bibinfo{author}{I.~Yamazaki},
\newblock \bibinfo{title}{Accelerating numerical dense linear algebra calculations with gpus},
\newblock \bibinfo{journal}{Numerical computations with GPUs}  (\bibinfo{year}{2014}) \bibinfo{pages}{3--28}.
\bibitem[{Senoner et~al.(2008)Senoner, Garc{\'\i}a, Mendez, Staffelbach, Vermorel, and Poinsot}]{senoner2008growth}
\bibinfo{author}{J.-M. Senoner}, \bibinfo{author}{M.~Garc{\'\i}a}, \bibinfo{author}{S.~Mendez}, \bibinfo{author}{G.~Staffelbach}, \bibinfo{author}{O.~Vermorel}, \bibinfo{author}{T.~Poinsot},
\newblock \bibinfo{title}{Growth of rounding errors and repetitivity of large eddy simulations},
\newblock \bibinfo{journal}{AIAA J.} \bibinfo{volume}{46} (\bibinfo{year}{2008}) \bibinfo{pages}{1773--1781}.
\bibitem[{Hassanaly and Raman(2019{\natexlab{a}})}]{hassanaly2019ensemble}
\bibinfo{author}{M.~Hassanaly}, \bibinfo{author}{V.~Raman},
\newblock \bibinfo{title}{Ensemble-{LES} analysis of perturbation response of turbulent partially-premixed flames},
\newblock \bibinfo{journal}{P. Combust. Inst.} \bibinfo{volume}{37} (\bibinfo{year}{2019}{\natexlab{a}}) \bibinfo{pages}{2249--2257}.
\bibitem[{Hassanaly and Raman(2019{\natexlab{b}})}]{hassanaly2019lyapunov}
\bibinfo{author}{M.~Hassanaly}, \bibinfo{author}{V.~Raman},
\newblock \bibinfo{title}{Lyapunov spectrum of forced homogeneous isotropic turbulent flows},
\newblock \bibinfo{journal}{Phys. Rev. Fl.} \bibinfo{volume}{4} (\bibinfo{year}{2019}{\natexlab{b}}) \bibinfo{pages}{114608}.

\end{thebibliography}
\bibliographystyle{elsarticle-num-names}

\end{document}